\documentstyle[twoside,12pt,epsf,aps]{revtex}
\relax
\newcommand{\MET}{\mbox{$\protect \raisebox{.3ex}{$\not$}\et$}}

\def\W{{\em W\/ }}
\def\Z0{${\em Z^0\/}$}

\def\r#1 {$^{#1}$}

\newcommand{\et}{{\rm E}_{\scriptscriptstyle\rm T}}

\newcommand{\met}{\mbox{$\protect \raisebox{.3ex}{$\not$}\et \ $}}

\newcommand{\ppbar}{p\bar{p}}

\newcommand{\ttbar}{t\bar{t}}
\newcommand{\bbbar}{b\bar{b}}

\newcommand{\gbb} { g \rightarrow b\bar{b} }
\newcommand{\gcc} { g \rightarrow c\bar{c}}

\newcommand{\gev}  { {\rm GeV}}

\newcommand{\gevc} { {\rm GeV/c  }}
\newcommand{\gevcc}{ {\rm GeV/c^2}}

\def\gepsfcentered#1{
  \def\testit{#1}
  \def\lbracket{[}
  \ifx\testit\lbracket
    \let\dofilecmd=\gepsfwithopt
  \else
    \let\dofilecmd=\gepsfnoopt
  \fi
  \dofilecmd}

\def\gepsfnoopt#1{
  \begin{center}
  \leavevmode
  \epsffile{#1}
  \end{center}}

\def\gepsfwithopt#1 #2 #3 #4]#5{
  \begin{center}
  \leavevmode
  \gepsfmaxx=0.94\textwidth
  \epsffile[#1 #2 #3 #4]{#5}
  \end{center}}

\newdimen\gepsfmaxx
\def\epsfsize#1#2{
  \ifnum \epsfxsize=0
    \ifnum \epsfysize=0
      \ifnum #1 > \gepsfmaxx
        \gepsfmaxx
      \else
        #1
      \fi
    \else
      \epsfxsize
    \fi
  \else
    \epsfxsize
  \fi
}

\begin{document}
\bibliographystyle{unsrt}
\title{Study of the heavy flavor content of jets produced in association with
       $W$ bosons in $p\bar{p}$ collisions at $\sqrt{s}$ = 1.8 TeV}
\maketitle
\font\eightit=cmti8
\def\r#1{\ignorespaces $^{#1}$}
\hfilneg
\begin{sloppypar}
\noindent
D.~Acosta,\r {12} T.~Affolder,\r {23} H.~Akimoto,\r {44}
A.~Akopian,\r {36}  P.~Amaral,\r 8  D.~Ambrose,\r {31}
D.~Amidei,\r {25} K.~Anikeev,\r {24} J.~Antos,\r 1 
G.~Apollinari,\r {11}
% ...
\end{sloppypar}
\vskip .026in
\begin{center}
(CDF Collaboration)
\end{center}

\vskip .026in
\begin{center}
\r 1  {\eightit Institute of Physics, Academia Sinica, Taipei, Taiwan 11529, 
Republic of China} \\
\r 2  {\eightit Argonne National Laboratory, Argonne, Illinois 60439} \\
\r 3  {\eightit Istituto Nazionale di Fisica Nucleare, University of Bologna,
I-40127 Bologna, Italy} \\
\r 4  {\eightit Brandeis University, Waltham, Massachusetts 02254} \\
\r 5  {\eightit University of California at Davis, Davis, California  95616} \\
\r 6  {\eightit University of California at Los Angeles, Los 
Angeles, California  90024} \\  
\r 7  {\eightit Instituto de Fisica de Cantabria, CSIC-University of Cantabria, 
39005 Santander, Spain} \\
\r 8  {\eightit Enrico Fermi Institute, University of Chicago, Chicago, 
Illinois 60637} \\
\r 9  {\eightit Joint Institute for Nuclear Research, RU-141980 Dubna, Russia}
\\
\r {10} {\eightit Duke University, Durham, North Carolina  27708} \\
\r {11} {\eightit Fermi National Accelerator Laboratory, Batavia, Illinois 
60510} \\
\r {12} {\eightit University of Florida, Gainesville, Florida  32611} \\
\r {13} {\eightit Laboratori Nazionali di Frascati, Istituto Nazionale di Fisica
               Nucleare, I-00044 Frascati, Italy} \\
\r {14} {\eightit University of Geneva, CH-1211 Geneva 4, Switzerland} \\
\r {15} {\eightit Glasgow University, Glasgow G12 8QQ, United Kingdom}\\
\r {16} {\eightit Harvard University, Cambridge, Massachusetts 02138} \\
\r {17} {\eightit Hiroshima University, Higashi-Hiroshima 724, Japan} \\
\r {18} {\eightit University of Illinois, Urbana, Illinois 61801} \\
\r {19} {\eightit The Johns Hopkins University, Baltimore, Maryland 21218} \\
\r {20} {\eightit Institut f\"{u}r Experimentelle Kernphysik, 
Universit\"{a}t Karlsruhe, 76128 Karlsruhe, Germany} \\
\r {21} {\eightit Center for High Energy Physics: Kyungpook National
University, Taegu 702-701; Seoul National University, Seoul 151-742; and
SungKyunKwan University, Suwon 440-746; Korea} \\
\r {22} {\eightit High Energy Accelerator Research Organization (KEK), Tsukuba, 
Ibaraki 305, Japan} \\
\r {23} {\eightit Ernest Orlando Lawrence Berkeley National Laboratory, 
Berkeley, California 94720} \\
\r {24} {\eightit Massachusetts Institute of Technology, Cambridge,
Massachusetts  02139} \\   
\r {25} {\eightit University of Michigan, Ann Arbor, Michigan 48109} \\
\r {26} {\eightit Michigan State University, East Lansing, Michigan  48824} \\
\r {27} {\eightit University of New Mexico, Albuquerque, New Mexico 87131} \\
\r {28} {\eightit Osaka City University, Osaka 588, Japan} \\
\r {29} {\eightit University of Oxford, Oxford OX1 3RH, United Kingdom} \\
\r {30} {\eightit Universita di Padova, Istituto Nazionale di Fisica 
          Nucleare, Sezione di Padova, I-35131 Padova, Italy} \\
\r {31} {\eightit University of Pennsylvania, Philadelphia, 
        Pennsylvania 19104} \\   
\r {32} {\eightit Istituto Nazionale di Fisica Nucleare, University and Scuola
               Normale Superiore of Pisa, I-56100 Pisa, Italy} \\
\r {33} {\eightit University of Pittsburgh, Pittsburgh, Pennsylvania 15260} \\
\r {34} {\eightit Purdue University, West Lafayette, Indiana 47907} \\
\r {35} {\eightit University of Rochester, Rochester, New York 14627} \\
\r {36} {\eightit Rockefeller University, New York, New York 10021} \\
\r {37} {\eightit Rutgers University, Piscataway, New Jersey 08855} \\
\r {38} {\eightit Texas A\&M University, College Station, Texas 77843} \\
\r {39} {\eightit Texas Tech University, Lubbock, Texas 79409} \\
\r {40} {\eightit Institute of Particle Physics, University of Toronto, Toronto
M5S 1A7, Canada} \\
\r {41} {\eightit Istituto Nazionale di Fisica Nucleare, University of Trieste/
Udine, Italy} \\
\r {42} {\eightit University of Tsukuba, Tsukuba, Ibaraki 305, Japan} \\
\r {43} {\eightit Tufts University, Medford, Massachusetts 02155} \\
\r {44} {\eightit Waseda University, Tokyo 169, Japan} \\
\r {45} {\eightit University of Wisconsin, Madison, Wisconsin 53706} \\
\r {46} {\eightit Yale University, New Haven, Connecticut 06520} \\
\r {(\ast)} {\eightit Now at Northwestern University, Evanston, Illinois 
60208} \\
\r {(\ast\ast)} {\eightit Now at University of California, Santa Barbara, CA
93106}
\end{center}

\newpage
\begin{abstract}
  We present a detailed examination of the heavy flavor content
  of the  $W $ + jet data sample collected with the CDF detector during the
  1992-1995 collider run at the Fermilab Tevatron. Jets containing heavy 
  flavor quarks are selected via the identification of secondary vertices
  or semileptonic decays of $b$ and $c$ quarks. There is generally good 
  agreement between the rates of secondary vertices and soft leptons
  in the data  and in the standard model simulation including single and pair
  production of top quarks. An exception is the number of events in which a
  single jet has both a soft lepton and a secondary vertex tag. In $W +$ 2,3 
  jet data, we find 13 such events where we expected 4.4 $\pm$ 0.6 events.
  The kinematic properties of this small sample of events are statistically
  difficult to reconcile with the simulation of standard model processes. \\
PACS number(s): 13.85.Qk, 13.38.Be, 13.20.He 
\end{abstract}
\section{ Introduction}
\label{s-intro}
 The production of $\W$ bosons in association with jets in $\ppbar$ collisions
 at the Fermilab Tevatron Collider provides the opportunity to test many 
 standard model (SM)~\cite{SM} predictions. Previous CDF 
 measurements~\cite{cdf-wjet} of the inclusive $\W$ cross section and of the
 yield of $\W$ + jet events as a function of the jet multiplicity and  
 transverse momentum  show agreement between data  and  
 the electroweak and QCD predictions of the standard model. In this study we
 extend the analysis of the jets associated with $\W$ boson production to 
 include the properties of heavy flavor jets identified by the displaced
 vertex or the semileptonic decay of charmed and beauty quarks.

 The present data set consists of 11,076 $\W \rightarrow \ell \nu$ ($\ell = e$ 
 or $\mu$) candidates produced in association with one or more jets selected 
 from 105 $\pm$ 4.0 pb$^{-1}$ of data collected by the CDF experiment at the 
 Fermilab Tevatron~\cite{tot_xsec}.  The $b$ and $c$-quark content of this 
 data set has been evaluated several times as we improved our understanding of
 systematic effects~\cite{cdf-evidence,cdf-discovery,xsec,cdf-tsig}. 
 We use two different methods for identifying (tagging) jets produced by 
 these heavy quarks. The first method uses the CDF silicon microvertex 
 detector (SVX) to locate secondary vertices produced by the decay of $b$ and
 $c$-hadrons in a jet. These vertices (SECVTX tags) are separated from the 
 primary event vertex as a result of the long $b$ and $c$-hadron  lifetimes.
 The second technique is to search a jet for  leptons ($e$ or $\mu$) produced
 by the semileptonic decay of $b$ and  $c$-hadrons. We refer to these as 
 ``soft lepton tags" (SLT's) because these  leptons typically have low 
 momentum compared to leptons from $\W$ decays. Heavy flavors in $\W+$ jet 
 events are mainly contributed by the production and decay of top quarks, 
 by direct $\W c$ production, and by the production of $\W g$ states in which 
 the gluon branches into a heavy-quark pair (gluon splitting). 

 A  recent comparison between measured and predicted rates of $\W$ + jet 
 events with heavy flavor  as a function of the jet multiplicity is presented
 in Ref.~\cite{cdf-tsig}. The focus of that paper, as well as previous CDF 
 publications~\cite{cdf-evidence,cdf-discovery,xsec}, is the measurement of 
 the $\ttbar$ production cross section. By attributing all the excess of 
 $\W + \geq$ 3 jet events with a SECVTX tag over the SM background to  
 $\ttbar$ production, we find $\sigma_{\ttbar}$ = 5.08 $\pm$ 1.54 pb in good
 agreement with the average theoretical prediction which is 5.1 pb with a 15\%
 uncertainty~\cite{tsig-pred}. We derive a numerically larger but not 
 inconsistent value of the cross section, $\sigma_{\ttbar}$ = 9.18 $\pm$ 4.26 
 pb, when using events with one or more SLT tags. The D\O~collaboration has 
 also measured the $t\bar{t}$ production cross section using various 
 techniques~\cite{d0}. D\O~has no measurement based upon displaced secondary 
 vertices, but using $\W+ \geq$ 3 jet events with a muon tag finds  
 $\sigma_{\ttbar}$ = 8.2 $\pm$ 3.5 pb. In the present study, we adopt a 
 different approach to the study of the $\W+$ jet sample and use the 
 theoretical estimate of $\sigma_{\ttbar}$ to test if the SM prediction is
 compatible with the observed yield of different tags as a function of the 
 jet multiplicity. This is of interest for top quark studies and searches for
 new physics, since some mechanisms proposed to explain electroweak symmetry 
 breaking, such as the Higgs mechanism~\cite{higgs} or the dynamics
 of a new interaction~\cite{techni}, predict the existence of new particles
 which can be produced in association with a $\W$ boson and decay into 
 $b\bar{b}$.
 
 Following a description of the CDF detector in Section~\ref{sec:s-det},
 Section~\ref{sec:s-ident} describes the triggers and the reconstruction
 of leptons, jets and the missing transverse energy. The selection of the
 $\W+$ jet sample is described in Section~\ref{sec:ss-dataset}, which also
 contains  a discussion of the algorithms used for the heavy flavor 
 identification followed by a description of the Monte Carlo generators and 
 the detector simulation used to model these events. 
 In Section~\ref{sec:s-secsec} we  summarize the method used in 
 Ref.~\cite{cdf-tsig} to predict the number of  $\W$ + jet events with 
 heavy flavor and then compare the observed yield of different
 tags as a function of the jet multiplicity to the SM prediction
 including single and pair  production of top quarks.
 Following this comparison, in Section~\ref{sec:s-secslt} we study the yield 
 of $\W$ + jet events with a SECVTX and a SLT tag in the same jet 
 (supertag\footnote{The prefix ``super" is used as a generalized term
 of high quality for historical reasons and is not meant  as a reference to
 supersymmetry.});  jets with a supertag will be referred to
 as superjets in the following. Since the semileptonic branching ratios of 
 $b$ and $c$-hadrons are very well measured~\cite{hf-sem}, the measurement of
 the fraction of jets tagged by SECVTX which contain a soft lepton tag 
 provides an additional test of our understanding of the heavy flavor 
 composition of this data sample. The number of these events in the $\W+2$
 and $\W+3$ jet topologies  is larger than the SM prediction.
 In Section~\ref{sec:s-prop} we compare kinematic distributions
 of the events with a superjet to the simulation prediction.
 As a check, we also compare  the simulation to a complementary sample of data.
 We find that the SM simulation models well the kinematics of the 
 complementary sample, but does not describe properly the characteristics of
 the events with a superjet. Some properties of the primary and soft leptons
 are discussed in Section~\ref{sec:s-lepprop}, while 
 Section~\ref{sec:s-sujprop} contains a study of other properties of the 
 superjets. In Section~\ref{sec:s-othersample} we investigate the dependence 
 of this study on the criteria used to select the data.
 Section~\ref{sec:s-concl}  summarizes our conclusions.
\section{ The CDF detector}
\label{sec:s-det}
 CDF is a general purpose detector designed to study $p\bar{p}$ interactions.
 A complete description of CDF can be found in 
 Refs.~\cite{cdf-evidence,cdf-det}. The detector components most relevant to 
 this analysis are summarized below. CDF has azimuthal and forward-backward 
 symmetry. A superconducting solenoid of length 4.8 m and radius 1.5 m 
 generates a 1.4 T magnetic field. Inside the solenoid there are 
 three types of tracking chambers for detecting charged particles and measuring 
 their momenta. A four-layer silicon microstrip vertex detector  surrounds the
 beryllium beam pipe of radius 1.9 cm.  The SVX has an active length of 51 cm;
 the four layers of the SVX are at distances of 2.9, 4.2, 5.5 and 7.9 cm from 
 the beamline. Axial microstrips with 60 $\mu$m pitch provide accurate track 
 reconstruction in the plane transverse to the beam~\cite{cdf-svx}.
 Outside the SVX there is a vertex drift chamber (VTX) which provides track 
 information up to a radius of 22 cm and for pseudo-rapidity $|\eta|\leq$ 3.5.
 The VTX measures the $z$-position (along the beamline) of the primary vertex.
 Both the SVX and VTX are mounted inside the CTC, a 3.2 m long drift chamber
 with an outer radius of 132 cm containing 84 concentric, cylindrical layers 
 of sense wires, which are grouped into alternating axial and stereo 
 superlayers. The solenoid is surrounded by sampling calorimeters used to 
 measure the electromagnetic and hadronic energy of jets and electrons.
 The calorimeters cover the pseudo-rapidity range $|\eta| \leq$ 4.2.
 The calorimeters are segmented into $\eta$-$\phi$ towers  which point to the 
 nominal interaction point. There are three separate $\eta$-regions of 
 calorimeters. Each region has an electromagnetic calorimeter [central (CEM), 
 plug (PEM) and forward (FEM)] and behind it a hadron calorimeter
 [CHA, PHA and FHA, respectively]. Located six radiation lengths inside
 the CEM calorimeter, proportional wire chambers (CES) provide
 shower-position measurements in the $z$ and $r-\phi$ view.
 Proportional chambers (CPR) located between the solenoid and the CEM detect
 early development of electromagnetic showers in the solenoid coil. These 
 chambers provide $r-\phi$ information only.

 The  calorimeter acts as a first hadron absorber for the central muon
 detection system which covers the pseudo-rapidity range $|\eta| \leq$ 1.0.
 The CMU detector consists of four layers of drift
 chambers located outside the CHA calorimeter. This detector covers the 
 pseudo-rapidity range $|\eta| \leq$ 0.6 and can be reached by muons with 
 $p_T \geq 1.4 \; \gevc$. The CMU detector is followed by 0.6 m of steel
 and four additional layers of drift chambers (CMP).
 The CMX system of drift chambers extends the muon 
 detection to $|\eta| \leq $1.0. 
\section{Data collection and identification of jets and leptons}
\label{sec:s-ident}
 The selection of $\W+$ jet events is based upon the identification
 of electrons, muons, missing energy, and jets. Below we discuss
 the criteria used to select these objects.
\subsection{Triggers}
\label{sec:trigger}
 The data acquisition is triggered by a three-level system designed to select
 events that can contain electrons, muons, jets, and missing transverse energy
 ($\MET$).
 
 Central electrons are defined as CEM clusters with $E_T \geq$ 18 GeV and a 
 reconstructed track with $p_T \geq 13\; \gevc$ pointing to it.
 The ratio of hadronic to electromagnetic energy in the cluster
 ($E_{had}/E_{em}$) is required to be less than 0.125. Plug electrons, used 
 for checks, have a higher transverse energy threshold ($E_T \geq$ 20 GeV).
 The inclusive muon trigger requires a match of better than 10 cm
 in $r\Delta \phi$ between a reconstructed track with $p_T \geq 18\; \gevc$,
 extrapolated to the radius of the muon detector, and a track segment
 in the muon chambers. Calorimeter towers are combined
 into  electromagnetic and jet-like clusters by the trigger
 system, which also provides  an estimate of $\MET$.
 Trigger efficiencies have been measured using the data and are included in 
 the detector simulation.
\subsection{Electron  selection}
 We use electrons in the central pseudo-rapidity region ($|\eta| \leq$ 1.0).
 Stricter selection cuts are applied to central electrons which passed the 
 trigger prerequisites. The following variables are used to discriminate 
 against charged hadrons: (1) the ratio  of hadronic to electromagnetic energy
 of the cluster, $E_{had}/E_{em}$; (2) the ratio of cluster energy to track
 momentum, $E/P$; (3) a comparison of the lateral shower profile in the 
 calorimeter cluster with that of test-beam electrons, $L_{shr}$;
 (4) the distance between the extrapolated track-position and the 
 CES measurement in the $r-\phi$ and $z$ views, $\Delta x$ and
 $\Delta z$, respectively; (5) a $\chi^{2}$ comparison of the CES shower 
 profile with that of test-beam electrons, $\chi^{2}_{strip}$;
 (6) the distance between the interaction vertex and the reconstructed track
 in the $z$-direction, $z$-vertex match; and (7) the isolation, $I$,
 defined as the ratio of additional transverse energy
 in a cone of radius $R=\sqrt{(\Delta \phi)^{2}+(\Delta \eta)^{2}}$ = 0.4
 around the electron direction to the electron transverse energy.
 Fiducial cuts on the shower position measured by the CES are applied to 
 ensure that the electron candidate is away from calorimeter boundaries and 
 therefore provide a reliable energy measurement. Electrons from photon 
 conversions are removed with high efficiency using the tracking information 
 in the event. A more detailed description  of the primary electron 
 selection can be found in  Refs.~\cite{cdf-evidence,cdf-tsig}.

 The $\eta$ coverage for electron detection is extended by using the plug 
 calorimeter. When selecting plug electrons we replace the variables $L_{shr}$,
 $\chi^{2}_{strip}$, $\Delta x$, and $\Delta z$ used for central electrons 
 with the $\chi^{2}$ comparison of the longitudinal and transverse shower
 profiles, $\chi^{2}_{depth}$ and $\chi^{2}_{transv}$, respectively. We 
 require $\chi^{2}_{depth} \leq$ 15 and $\chi^{2}_{transv} \leq$ 3. We do not
 use the $E/P$ cut, as the momentum measurement is not accurate at large  
 rapidities. However, we require that a track pointing to the electromagnetic
 cluster has hits in at least three CTC axial layers. We also require that the
 ratio of the number of VTX hits found along the electron path to the 
 predicted number be larger than 50\%. Because of the CTC geometrical 
 acceptance and of fiducial cuts to ensure a reliable energy measurement, the
 effective  coverage for plug electrons is 1.2 $\leq |\eta| \leq$ 1.5.
\subsection{Muon selection} 
 Muons are identified in the pseudo-rapidity region $|\eta | \leq$ 1.0 by 
 requiring a match between a CTC track and a track segment measured by the 
 CMU, CMP or CMX muon chambers. 
 
 The following variables are used to separate muons from hadrons interacting
 in the calorimeter and cosmic rays: (1) an energy deposition in the 
 electromagnetic and hadronic calorimeters characteristic of minimum ionizing
 particles, $E_{em}$ and $E_{had}$, respectively; (2) the distance of closest
 approach of the reconstructed track to the beam line (impact parameter), $d$;
 (3) the $z$-vertex match; (4) the distance between the extrapolated track and 
 the track segment in the muon chamber, $\Delta x = r \Delta \phi$;
 and (5) the isolation $I$. A more detailed description of the primary muon 
 selection can  be found in Refs.~\cite{cdf-evidence,cdf-tsig}.
 Selection efficiencies for electrons and muons in the simulation are adjusted
 to those  of $Z \rightarrow \ell \ell $ events in the data.  
\subsection{Loose leptons}
\label{sec:s-loose}
 In order to be more efficient in rejecting events containing two leptons 
 from $Z$ decays, $\ttbar$ decays and other sources we use looser selection 
 criteria to search for additional isolated leptons. These selection criteria
 are described in detail in Ref.~\cite{cdf-tsig}.
\subsection{Jet identification and corrections}
\label{sec:s-jet}
 Jets are reconstructed from the energy deposited in the calorimeter using a 
 clustering algorithm with a fixed cone of radius $R=0.4$ in the $\eta-\phi$ 
 space. A detailed description of the algorithm can be found in 
 Ref.~\cite{jet_clus}. Jet energies can be mismeasured for a variety of 
 reasons (calorimeter non-linearity, loss of low momentum particles because of 
 bending in the magnetic field, contributions from the underlying event, 
 out-of-cone losses, undetected energy carried by muons and neutrinos). 
 Corrections, which depend on the jet $E_T$ and $\eta$, are applied to jet 
 energies; they compensate for these mismeasurements on  average  
 but do not improve the jet energy resolution. We estimate a 10\% 
 uncertainty on the corrected jet energy~\cite{cdf-evidence,clus_err}.
 Where appropriate, we apply additional corrections to jet energies in order 
 to extrapolate on average to the energy of the parton producing the  
 jet~\cite{cdf-evidence,top_mass,blusk}.
\subsection{$\MET$ Measurement} 
 The missing transverse energy ($\MET$) is defined as the negative
 of the vector sum of the transverse energy in all calorimeter towers
 with $|\eta| \leq$ 3.5. For events with muon candidates 
 the vector sum of the calorimeter transverse energy is corrected by 
 vectorially subtracting the energy deposited by the muon and then adding 
 the $p_T$ of the muon as measured by the tracking detectors. This is done
 for all muon candidates with $p_T \geq 5\; \gevc$ and $I \leq$ 0.1.
 When jet energy corrections are used,  the $\met$ calculation
 accounts for them  as detailed in Ref.~\cite{top_mass}.
\section{The $\W$ + jet sample}
\label{sec:ss-dataset}
 The $\W$ selection requires an isolated, $I \leq$ 0.1, electron (muon) 
 to pass the trigger and offline requisites outlined in 
 Section~\ref{sec:s-ident}, and also to have $E_T \geq 20\; \gev$ 
 ($p_T \geq 20\; \gevc$). We require the  $z$-position of the event vertex
 ($Z_{\rm vrtx}$) to be within 60 cm of the center of the CDF detector. 
 We additionally require $\MET \geq$ 20 GeV to reduce the background from
 misidentified leptons and semileptonic $b$-hadron decays. Events containing
 additional loose lepton candidates with isolation $I \leq$ 0.15  and
 $p_T \geq 10\; \gevc$ are removed from the sample. We bin the $\W$ 
 candidate events according to the observed jet multiplicity (a jet is a
 $R = 0.4$ cluster with uncorrected $E_T \geq 15\; \gev$ and $|\eta| \leq$ 2.0).

 The heavy flavor content of the $\W+$ jet sample is enhanced by selecting
 events with jets containing a displaced secondary vertex or a soft lepton.
\subsection{Description of the tagging algorithms}
\label{sec:tag-alg}
 The secondary vertex tagging algorithm (SECVTX) is described in detail in
 Refs.~\cite{cdf-evidence,cdf-tsig}. SECVTX is based on the determination of 
 the primary event vertex and the reconstruction of additional secondary 
 vertices using displaced tracks contained inside jets. The search for a 
 secondary vertex in a jet is a two-stage process. In both stages, tracks 
 in the jet are selected for reconstruction of a secondary vertex based on 
 the significance of their impact parameter $d$ with respect to the primary
 vertex, $d/\sigma_{d}$, where $\sigma_{d}$ is the estimated uncertainty on 
 $d$. The first stage requires at least three candidate tracks for the 
 reconstruction of the secondary vertex. Tracks consistent with coming from 
 the decay $K_s \rightarrow \pi^+ \pi^-$ or $\Lambda \rightarrow \pi^- p$ are
 not used as candidate tracks. Two candidate tracks are constrained to pass 
 through the same space point to form a seed vertex. If at least one 
 additional candidate track is consistent with intersecting this seed vertex,
 then the seed vertex is used as the secondary vertex. If the first stage is
 not successful in finding a secondary vertex, a second pass is attempted. 
 More stringent track requirements (such as $d/\sigma_{d}$ and $p_T$) 
 are imposed on the candidate tracks. All candidate tracks satisfying 
 these stricter criteria are constrained to pass through the same space 
 point to form a seed vertex. This vertex has an associated $\chi^2$. 
 Candidate tracks that contribute too much to the $\chi^2$ are removed 
 and a new seed vertex is formed. This procedure is iterated until a seed
 vertex remains that has at least two associated tracks and an acceptable
 value of $\chi^2$.

 The decay length of the secondary vertex $L_{xy}$ is the projection in the 
 plane transverse to the beam line of the vector pointing from the primary 
 vertex to the secondary vertex onto the jet axis. If the cosine of the angle
 between these two vectors is positive (negative), then $L_{xy}$ is positive
 (negative). Most of the secondary vertices from the decay of $b$ and 
 $c$-hadrons are expected to have positive $L_{xy}$; conversely, secondary 
 vertices constructed  from a random combination of mismeasured tracks 
 (mistags) have a symmetric distribution around $L_{xy}$=0. To reduce the 
 background, a jet is considered tagged by SECVTX if it contains a secondary
 vertex with ${\displaystyle \frac{L_{xy}}{\sigma_{L_{xy}}} } \geq 3.0$, where 
 $\sigma_{L_{xy}}$ is the estimated uncertainty on $L_{xy}$ (typically about
 130 $\mu$m). The mistag contribution to positive SECVTX tags is evaluated 
 using a parameterization derived from negative tags in generic-jet 
 data~\cite{cdf-tsig}.

 A second $b$-tagging method is represented by the jet-probability (JPB)
 algorithm described in detail in Ref.~\cite{cdf-tsig}. This tagging 
 method compares track impact parameters to measured resolution functions
 in order to calculate for each jet a probability that there are no 
 long-lived particles in the jet cone. The sign of the impact parameter
 is defined to be positive if the point of closest approach to the  primary
 vertex lies in the same hemisphere as the jet direction, and negative 
 otherwise. Jet-probability is defined using tracks with positive impact
 parameter; we also define a negative jet-probability where we select only
 tracks with negative impact parameter in the calculation.
 Jet-probability is uniformly distributed for light quark or gluon jets, 
 but is very small for jets containing displaced vertices from heavy 
 flavor decays. A jet has a positive (negative) JPB tag if a 
 jet-probability value smaller than 0.05 is derived using at least two 
 tracks with positive (negative) impact parameter.
  
 An alternative way to tag $b$ quarks is to search a jet for soft leptons 
 produced by $b \rightarrow l \nu c$ or $b \rightarrow c \rightarrow l \nu s$
 decays. The soft lepton tagging algorithm is applied to sets of CTC tracks 
 associated with jets with $E_T \geq 15\; \gev$  and $|\eta| \leq $2.0. CTC
 tracks are associated with a jet if they are inside a cone of radius 0.4
 centered around the jet axis. In order to maintain high efficiency, 
 the lepton $p_T$ threshold is set low at 2 $\gevc$.

 To search for soft electrons the algorithm extrapolates each track to the 
 calorimeter and attempts to match it to a CES cluster. The matched CES 
 cluster is required to be consistent in shape and position with the 
 expectation for electron showers. In addition, it is required that 
 0.7 $\leq E/P \leq$ 1.5 and $E_{had}/E_{em} \leq$ 0.1. The track specific
 ionization ($dE/dx$), measured in the CTC, is required to be consistent 
 with the electron hypothesis. Electron candidates must also have an energy
 deposition in the CPR corresponding to that left by at least four 
 minimum-ionizing particles. The efficiency of the selection criteria has 
 been determined using a sample of electrons produced by photon 
 conversions~\cite{cdf-evidence}.

 To identify soft muons, track segments reconstructed in the CMU, CMP or CMX 
 systems are matched to  CTC tracks. Only the CMU or CMX systems are used
 to identify muons with $2 \leq p_T \leq 3\; \gevc$. Muon candidate tracks 
 with $ p_T \geq 3\; \gevc$  within the CMU and CMP fiducial volume are 
 required to match to track segments in both systems. The reconstruction 
 efficiency has been measured using samples of muons from 
 $J/\psi \rightarrow \mu^{+}\mu^{-}$ and $Z \rightarrow \mu^{+}\mu^{-}$
 decays~\cite{cdf-evidence}.

 In the data, the rate of fake soft lepton tags which are not due to heavy 
 flavor semileptonic decays is evaluated using a parameterization of the SLT
 fake probability per track as a function of the track isolation and $p_T$. 
 This parameterization has been derived in a large sample of generic-jet data
 ~\cite{cdf-evidence} after removing the fraction of soft lepton tags 
 contributed by heavy flavor (about 26\%)~\cite{cdf-tsig}. In the simulation,
 a SLT track is required to match at generator level a lepton coming from a 
 $b$ or $c$-hadron decay~\cite{cdf-tsig}.
\subsection{Monte Carlo generators and detector simulation}
\label{sec:sec-mc}
 We use three different Monte Carlo generators to estimate the contribution
 of SM processes to the $\W+$ jet sample. The settings and the calibration 
 of these Monte Carlo generators are described in Ref.~\cite{cdf-tsig}.

 A few processes, including $t\bar{t}$ production, are evaluated using
 version 5.7 of {\sc pythia}~\cite{pythia}. These processes are detailed 
 in the next section.

 The fraction of $\W+$ jet direct production with heavy flavor, namely
 $p\bar{p} \rightarrow Wg$ with $g \rightarrow b\bar{b}$, $c\bar{c}$
 (gluon splitting) and $p\bar{p}\rightarrow \W c$, is calculated using version 
 5.6 of the {\sc herwig} generator~\cite{herwig}. The part of the phase space
 region of these hard scattering processes that is not correctly mapped by 
 {\sc herwig}  (namely $\W b\bar{b}$ and $\W c\bar{c}$ events in which the two 
 heavy flavor partons produce two well separated jets) is evaluated using the
 {\sc vecbos} generator~\cite{vecbos}. {\sc vecbos} is a parton-level Monte 
 Carlo generator and we transform the partons produced by {\sc vecbos} into
 hadrons and jets using {\sc herwig} adapted to perform the coherent shower 
 evolution of both initial and final state partons from an arbitrary
 hard-scattering subprocess~\cite{heprt}. In summary, we use {\sc herwig} to 
 predict the fraction of $\W+\geq$ 1 jet events where only one jet contains $b$
 or $c$-hadrons while we rely on {\sc vecbos} to extend the prediction to the
 cases where two different jets contain heavy-flavored hadrons.
 The MRS~D$_0^{\prime}$ set of structure functions~\cite{mrsd0} is used with
 these generators. We set the $b$-mass value to 4.75 $\gevcc$ and 
 the $c$-mass value to 1.5 $\gevcc$.

 The fraction of jets containing heavy flavor hadrons from gluon splitting 
 predicted by the Monte Carlo generators has been tuned using generic-jet data.
 As a result,  the fraction of $\gbb$ calculated by the  generators is
 increased by the factor 1.40 $\pm$ 0.19 and the fraction of $\gcc$ by the 
 factor 1.35 $\pm$ 0.36. These factors are of the same size as those measured
 by the SLC and LEP experiments for the rate of $g \rightarrow b\bar{b}$
 and $g \rightarrow c\bar{c}$ in $Z$ decays~\cite{lep}, and are within
 the estimated theoretical uncertainties~\cite{gbb}.

 We use the CLEO Monte Carlo generator, {\sc qq}, to model the decay of $b$ and
 $c$-hadrons~\cite{cleo}. All particles produced in the final state by the 
 {\sc herwig} (or {\sc pythia}) + {\sc qq} generator package are decayed and 
 interacted with the CDF-detector simulation (called QFL). The detector 
 response is based upon parameterizations and simple models which depend on 
 the particle kinematics. After the simulation of the CDF detector, the Monte
 Carlo events are treated as if they were real data.
 Ref.~\cite{cdf-tsig} describes the calibration of the detector simulation,
 including tagging efficiencies, using several independent data samples.
\section{Comparison of measured and predicted rates of
         $\W+ \geq$ 1 jet events with heavy flavor tags}
\label{sec:s-secsec}
 In this study, we compare the observed numbers of tagged $\W+$ jet events 
 as a function of the jet multiplicity to the SM prediction which uses the
 NLO calculation of the $t\bar{t}$ cross section. The various contributions
 to $\W+$ jet events are discussed in subsection A, and the results of the
 comparisons are summarized in subsection B.
\subsection{Predicted contributions to the $\W+$ jet event sample}
 A detailed study of the non-$t\bar{t}$ contributions to the W + jet events 
 was made in Ref.~\cite{cdf-tsig}. These studies are reviewed here,  
 along with the $t\bar{t}$ contribution derived using the theoretical 
 prediction.

 The small number of  events contributed by non-$\W$ sources, including 
 $b\bar{b}$ production, is estimated using the data. The number of non-$\W$ 
 events in the signal region (lepton $I \leq$ 0.1 and $\MET \geq$ 20 GeV) is
 predicted by multiplying the number of events with $I \leq$ 0.1 and 
 $\MET \leq$ 10 GeV by the ratio $R$ of events with $I \geq$ 0.2 and 
 $\MET \geq$ 20 GeV to events with $I \geq$ 0.2 and $\MET \leq$ 10 GeV. The 
 number of tagged non-$\W$ events is predicted by multiplying the number of 
 tagged events with $I \leq$ 0.1 and $\MET \leq$ 10 GeV by the same ratio $R$.

 The number of $Z$ + jet events in which one lepton from the $Z$ decay is not
 identified (unidentified-$Z$) is calculated using the {\sc pythia} generator.
 The simulated sample is normalized to the number of $Z  \rightarrow \ell \ell$
 decays observed in the data for each jet bin. Unidentified-$Z$ + jet events 
 can be tagged either because a jet is produced by a $\tau$ originating from  
 $Z\rightarrow \tau \bar{\tau} $ decays or because a jet contains heavy flavor.
 The number of tagged $Z \rightarrow \tau \bar{\tau} $ events is estimated 
 using the  {\sc pythia} simulation. The number of tags contributed by
 unidentified-$Z$ + jet events with heavy flavor is estimated with a 
 combination of the {\sc pythia}, {\sc herwig} and {\sc vecbos} generators.

 The contribution of diboson  production before and after tagging is 
 calculated using the {\sc pythia} generator. The values of the diboson 
 production cross sections
 [$\sigma_{\W\W}$ = 9.5 $\pm$ 0.7 pb,
 $\sigma_{\W Z}$ = 2.60 $\pm$ 0.34 pb
 and $\sigma_{Z Z}$= 1.0 $\pm$ 0.2 pb] are taken from Ref.~\cite{dibos}.

 The contribution from  single top production before and after tagging is
 estimated using {\sc pythia} to model the process 
 $p\bar{p} \rightarrow t\bar{b}$ via a virtual $s$-channel 
 $\W$ and {\sc herwig} to model the process
 $p \bar{p} \rightarrow t \bar{b}$ via a virtual $t$-channel $\W$.
 The production cross sections [0.74 $\pm$ 0.05 pb and
 1.5 $\pm$ 0.4 pb for the $s$ and $t$-channel, respectively]
 are derived using the NLO calculation of Ref.~\cite{single_top}.

 The $t\bar{t}$ contribution is calculated using the {\sc pythia}
 generator. We use $\sigma_{t\bar{t}}$ = 5.1 pb with a 15\% uncertainty.
 This number is the average of several NLO calculations of the $t\bar{t}$
 production cross section~\cite{tsig-pred}.

 The direct production of $\W+$ jets with heavy flavor is estimated using
 a combination of data and simulation. Since the leading-order matrix
 element calculation has a 40\% uncertainty~\cite{mlm}, we first evaluate
 in each jet bin the number of events due to $\W+$ jet direct production as 
 the difference between the data and the sum of all processes listed above,
 including $t\bar{t}$ production, before tagging. We then use the {\sc herwig}
 and {\sc vecbos} generators, calibrated with generic jet data as discussed in
 Section IV B, to estimate  the fraction of $\W+$ jet events which contain 
 $c\bar{c}$ or $b\bar{b}$ pairs and their tag contribution. The fraction of 
 $\W c$ events and their tag contribution is determined using {\sc herwig}.

 The number of events in which a jet without heavy flavor (h.f.) is tagged 
 because of detector effects (mistags) is estimated using a parametrization 
 of the mistag probability (as a function of the jet transverse energy and
 track multiplicity), which has been derived from generic jet data.
 \subsection{Comparison with a SM prediction using
             the theoretical estimate of $\sigma_{\ttbar}$ }
 The composition of the $\W+$ jet event candidates before heavy flavor 
 tagging is summarized in  Table~\ref{tab:tab_3.0}.  
 As previously discussed in Section IV A, the heavy flavor content of the
 $\W+$ jet sample is enriched by searching jets for a displaced secondary 
 vertex (SECVTX tag) or an identified lepton (SLT tag). 

 The composition of the $\W+$ jet events with SECVTX tags is shown in 
 Table~\ref{tab:tab_3.1} and those with SLT tags in Table~\ref{tab:tab_3.3}. 
 The numbers of observed events with one  (ST) or two (DT) jets tagged by
 the SECVTX or SLT algorithms are compared to predictions for each value of 
 the jet multiplicity.

 There is good agreement between the  observed and predicted numbers of
 tagged events for the four jet multiplicity bins. The probability~\cite{prob}
 that the observed numbers of events with at least one SECVTX tag are 
 consistent with the predictions in all four jet bins is 80\%. The 
 probability~\cite{prob} that the observed number of events with at least 
 one SLT tag are consistent with the predictions in all four jet bins is 56\%.

 In the next section we perform a more detailed study of heavy flavor content 
 of the $\W+$ jet sample by selecting events with jets containing both a 
 displaced vertex and a soft lepton.
\newpage
\begin{table}[p]
\begin{center}
\def\arraystretch{0.9}
\caption[]{Estimated composition of the $\W+\geq $ 1 jet sample before tagging.}
 \begin{tabular}{lcccc}
 Source         &  $\W+1 \,{\rm jet}$        &  $\W+2 \,{\rm jet}$        &  $\W+3 \,{\rm jet}$        &  
$\W+\geq4 \,{\rm jet}$     \\
 \hline
 Data                 &   9454 &   1370 &    198 &     54 \\
 Non-$\W$                 &  ~560.1 $\pm$ 14.9 & ~~71.2 $\pm$ 2.7 & ~12.4 $\pm$  2.0 & ~5.1 $\pm$  1.7 \\
 $\W\W$                   & ~~31.2 $\pm$  5.4 & ~~31.1 $\pm$ 5.4 & ~~5.2 $\pm$ 1.0 & ~0.8 $\pm$  0.2 \\
 $\W Z$                   & ~~~4.4 $\pm$  0.9 & ~~~4.8 $\pm$ 1.0 & ~~0.9 $\pm$  0.2 & ~0.1 $\pm$  0.0 \\
 $ZZ$                     & ~~~0.3 $\pm$  0.1 & ~~~0.4 $\pm$ 0.1 & ~~0.1 $\pm$  0.0 & ~0.0 $\pm$  0.0 \\
 Unidentified-$Z$ + jets  & ~234.8 $\pm$ 14.5 & ~~38.5 $\pm$ 5.9 & ~~7.9 $\pm$  2.4 & ~0.7 $\pm$  0.7 \\
 Single top               & ~~14.1 $\pm$  2.1 & ~~~7.9 $\pm$ 1.7 & ~~1.7 $\pm$  0.4 & ~0.3 $\pm$  0.1 \\
 $t\bar{t}$               & ~~~1.8 $\pm$  0.5 & ~~10.1 $\pm$ 2.8 & ~20.3 $\pm$  5.7 & 21.3 $\pm$  5.9 \\
 $\W +$  jets without h.f. &7952.0 $\pm$ 133.6 & 1027.7 $\pm$ 31.1 & 121.1 $\pm$ 7.7 & 19.9 $\pm$ 6.1 \\
 $\W c$                   & ~413.1 $\pm$ 123.9 & ~~86.8 $\pm$ 26.1 &  ~11.2 $\pm$ 3.4 & ~1.9 $\pm$ 0.7 \\
 $\W c\bar{c}$            & ~173.1 $\pm$ 46.2 &  ~~61.9 $\pm$ 13.6 &  ~11.4 $\pm$ 2.6 & ~2.3 $\pm$ 0.9 \\
$\W b\bar{b}$             & ~~69.0 $\pm$  9.5 &  ~~29.7 $\pm$  5.1 &  ~~5.7  $\pm$ 1.1 & ~1.5 $\pm$ 0.5 \\
 \end{tabular}
\label{tab:tab_3.0}
\end{center}
\end{table}

\newpage
\begin{table}[p]
\begin{center}
\caption[]{Summary of observed and predicted number of $\W$ events with one
           (ST) and two (DT) SECVTX tags.}
 \begin{tabular}{lcccc}
  Source         &  $\W+1 \,{\rm jet}$     &  $\W+2 \,{\rm jet}$     &  $\W+3 \,{\rm jet}$        &  
$\W+\geq4 \,{\rm jet}$     \\
 \hline
 Mistags                     &   10.82 $\pm$ 1.08 &~3.80 $\pm$ 0.38 &~0.99 $\pm$ 0.10 &  0.35$ \pm$ 0.04 \\
 Non-$\W$                    &   ~8.18 $\pm$ 0.78 &~1.49 $\pm$ 0.47 &~0.76 $\pm$ 0.38 &  0.31 $\pm$ 0.16 \\
 $\W\W,\W Z,ZZ$              &   ~0.52 $\pm$ 0.14 &~1.38 $\pm$ 0.28 &~0.40 $\pm$ 0.13 &  0.00 $\pm$ 0.00 \\
 Single top                  &   ~1.36 $\pm$ 0.35 &~2.38 $\pm$ 0.54 &~0.63 $\pm$0.14 &  0.14 $\pm$ 0.03 \\
 $\W c$                      &   16.89 $\pm$ 5.38 &~3.94 $\pm$ 1.30 &~0.51 $\pm$ 0.17 &  0.09 $\pm$ 0.04 \\
 $\W c\bar{c}$ (ST)          &   ~7.89 $\pm$ 2.17 &~3.54 $\pm$ 0.88 &~0.77 $\pm$ 0.25 & 0.16 $\pm$ 0.07 \\
 $\W c\bar{c}$ (DT)          &                   & ~0.06 $\pm$ 0.04 &~0.00 $\pm$ 0.00 & 0.00 $\pm$ 0.00 \\
 $\W b\bar{b}$ (ST)          &   17.00 $\pm$ 2.41 &~8.35 $\pm$ 1.74 &~1.62 $\pm$ 0.40 & 0.41 $\pm$ 0.14 \\
 $\W b\bar{b}$ (DT)          &                   & ~1.51 $\pm$ 0.52 &~0.31 $\pm$ 0.13 & 0.07 $\pm$ 0.03 \\
 $Z\rightarrow \tau\tau$     &   ~0.96 $\pm$ 0.30 &~0.70 $\pm$ 0.25 &~0.17 $\pm$ 0.12 & 0.00 $\pm$ 0.00 \\
 $Zc$                        &   ~0.14 $\pm$ 0.04 &~0.03 $\pm$ 0.01 &~0.01 $\pm$ 0.00 & 0.00 $\pm$ 0.00 \\
 $Zc\bar{c}$ (ST)            &   ~0.22 $\pm$ 0.06 &~0.10 $\pm$ 0.03 &~0.04 $\pm$ 0.02 & 0.00 $\pm$ 0.00 \\
 $Zc\bar{c}$ (DT)            &                   & ~0.00 $\pm$ 0.00 &~0.00 $\pm$ 0.00 & 0.00 $\pm$ 0.00 \\
 $Zb\bar{b}$ (ST)            &   ~0.93 $\pm$ 0.14 &~0.46 $\pm$ 0.12 &~0.17 $\pm$ 0.06 & 0.02 $\pm$ 0.02 \\
 $Zb\bar{b}$ (DT)            &                   & ~0.08 $\pm$ 0.03 &~0.03 $\pm$ 0.02 & 0.00 $\pm$ 0.00 \\
 $t\bar{t}$  (ST)            &   ~0.54 $\pm$ 0.14 & ~3.34 $\pm$ 0.87 &~6.76 $\pm$ 1.76 & 7.42 $\pm$ 1.93 \\
 $t\bar{t}$  (DT)            &                   & ~0.76 $\pm$ 0.20 &~2.88 $\pm$ 0.75 & 3.96 $\pm$ 1.03 \\
 \hline
 SM prediction (ST)       &   65.44 $\pm$ 6.45 & 29.61 $\pm$ 2.66 & 12.87 $\pm$ 1.89 & 8.92 $\pm$ 1.95 \\
 SM prediction (DT)       &                   &  ~2.41 $\pm$ 0.56 & ~3.23 $\pm$ 0.76 & 4.03 $\pm$ 1.03 \\
 \hline
 Data  (ST)                  &     66 &     35 &     10 &     11 \\
 Data  (DT)                  &                   &      5 &      6 &      2 \\
 \end{tabular}
\label{tab:tab_3.1}
\end{center}
\end{table}

\newpage
\begin{table}[p]
\begin{center}
\caption[]{Summary of observed and predicted number of $\W$ events with
           one (ST) and two (DT) SLT tags.}
 \begin{tabular}{lcccc}
 Source         &  $\W+1 \,{\rm jet}$        &  $\W+2 \,{\rm jet}$    &  $\W+3 \,{\rm jet}$        &  
$\W+\geq4 \,{\rm jet}$     \\

 Mistags                 & 101.92 $\pm$ 10.19 & 30.90 $\pm$ 3.09 & ~7.34 $\pm$ 0.73 &  3.01 $\pm$ 0.30 \\
 Non-$\W$                & ~~8.96 $\pm$ 0.84 &  ~2.09 $\pm$ 0.56 & ~0.38 $\pm$ 0.27 & 0.16 $\pm$ 0.11 \\
 $\W\W,\W Z,ZZ$          & ~~0.50 $\pm$ 0.16 &  ~0.88 $\pm$ 0.22 & ~0.10 $\pm$ 0.05 &  0.00 $\pm$ 0.00 \\
 Single top              & ~~0.38 $\pm$ 0.10 &  ~0.67 $\pm$ 0.15 & ~0.18 $\pm$ 0.05 & 0.05 $\pm$ 0.01 \\
 $\W c$                  & ~13.12 $\pm$ 4.27 &  ~4.29 $\pm$ 1.46 & ~0.73 $\pm$ 0.32 & 0.13 $\pm$ 0.06 \\
 $\W c\bar{c}$ (ST)      & ~~6.41 $\pm$ 1.89 &  ~2.70 $\pm$ 0.67 & ~0.69 $\pm$ 0.22 & 0.14 $\pm$ 0.06 \\
 $\W c\bar{c}$ (DT)      &                   &  ~0.02 $\pm$ 0.02 & ~0.00 $\pm$ 0.00 & 0.00 $\pm$ 0.00 \\
 $\W b\bar{b}$ (ST)      & ~~5.31 $\pm$ 0.96 &  ~2.86 $\pm$ 0.67 & ~0.47 $\pm$ 0.14 & 0.12 $\pm$ 0.05 \\
 $\W b\bar{b}$  (DT)     &                   &  ~0.09 $\pm$ 0.05 & ~0.01 $\pm$ 0.01 &  0.00 $\pm$ 0.00 \\
 $Z\rightarrow \tau\tau$ & ~~0.43 $\pm$ 0.20 &  ~0.09 $\pm$ 0.09 & ~0.09 $\pm$ 0.09 &  0.00 $\pm$ 0.00 \\
 $Zc$                    & ~~0.11 $\pm$ 0.04 &  ~0.04 $\pm$ 0.01 & ~0.01 $\pm$ 0.01 &  0.00 $\pm$ 0.00 \\
 $Zc\bar{c}$ (ST)        & ~~0.17 $\pm$ 0.05 &  ~0.08 $\pm$ 0.02 & ~0.03 $\pm$ 0.01 & 0.00 $\pm$ 0.00 \\
 $Zc\bar{c}$ (DT)        &                   &  ~0.00 $\pm$ 0.00 & ~0.00 $\pm$ 0.00 & 0.00 $\pm$ 0.00 \\
 $Zb\bar{b}$ (ST)        & ~~0.29 $\pm$ 0.06 &  ~0.16 $\pm$ 0.04 & ~0.05 $\pm$ 0.02 & 0.01 $\pm$ 0.01 \\
 $Zb\bar{b}$ (DT)        &                   &  ~0.00 $\pm$ 0.00 & ~0.00 $\pm$ 0.00 &  0.00 $\pm$ 0.00 \\
 $t\bar{t}$  (ST)        & ~~0.14 $\pm$ 0.06 &  ~1.35 $\pm$ 0.61 & ~2.85 $\pm$ 1.30 & 3.36 $\pm$ 1.53 \\
 $t\bar{t}$  (DT)        &                   &  ~0.04 $\pm$ 0.02 & ~0.13 $\pm$ 0.06 & 0.18 $\pm$ 0.08 \\
 \hline
  SM prediction (ST)  & 137.75 $\pm$ 11.29 & 46.08 $\pm$ 3.65 & 12.91 $\pm$ 1.57 & 6.98 $\pm$ 1.57 \\
  SM prediction (DT)  &                   &  ~0.14 $\pm$ 0.06 & ~0.14 $\pm$ 0.06 &  0.18 $\pm$ 0.08 \\
 \hline
 Data   (ST)              &    146 &     56 &     17 &      8 \\
 Data   (DT)              &        &      0 &      0 &      0 \\
 \end{tabular}
\label{tab:tab_3.3}
\end{center}
\end{table}

\clearpage
\section{Comparison of measured and predicted rates of $\W+$ jet events
         with both a SECVTX and SLT heavy flavor tag}
\label{sec:s-secslt}
 We begin this study by selecting $\W+$ jet events with both SECVTX and SLT 
 tags. In Table~\ref{tab:tab_3.4} the predicted and observed $\W+$ jet events
 with a SLT tag are split into samples without (top part of 
 Table~\ref{tab:tab_3.4}) and with (bottom part of Table~\ref{tab:tab_3.4})
 SECVTX tags. There is good agreement between data and predictions for the 
 $\W+$ jet events with a SLT tag and no SECVTX tag, where a large fraction of
 the events have fake SLT tags in jets without heavy flavor. In contrast, the 
 numbers of events with both SECVTX and SLT tags, which are mostly contributed
 by real heavy flavor, are not well predicted by the simulation. Therefore, we
 check if the rate of SLT tags in jets tagged by SECVTX (superjets) is 
 consistent with the expected production and decay of hadrons with heavy flavor.

 After tagging with SECVTX, we estimate that approximately 70\% of the $\W+$ 
 jet sample contains $b$-jets and 20\% contains $c$-jets  (see 
 Table~\ref{tab:tab_3.1}).  On average, 20\% of the $b$ and $c$-hadron decays
 produce a lepton ($e$ or $\mu$).  Only 50\% of the leptons resulting from a
 $b$-hadron satisfy the 2 $\gevc$ transverse momentum requirement
 of the soft lepton tag (this fraction is slightly smaller for $c$-hadron 
 decays). In addition, the SLT tagger  is approximately 
 90\% efficient in identifying muons and  50\% efficient in identifying 
 electrons. Altogether, we then expect that about 7\% of the jets tagged by 
 SECVTX will contain an additional SLT tag if the heavy flavor composition of 
 $\W+$ jet events is correctly understood.

 The observed numbers of events with a superjet are compared to the SM 
 prediction in Table~\ref{tab:tab_4.0}. The information in 
 Table~\ref{tab:tab_4.0} is similar to that presented in 
 Table~\ref{tab:tab_3.4}, except that two events listed in 
 Table~\ref{tab:tab_3.4} have the SLT and SECVTX tags in
 different jets.  The probability~\cite{prob} that the observed 
 numbers of events with at least one superjet 
 are consistent with the prediction in all four jet bins is 0.4\%.
 This low probability value is mostly driven by
 an excess in the $\W+$ 2,3 jet bins where 13 events are
 observed\footnote{The 13 events include $t\bar{t}$ candidates and four
 of these events are included in the sample used to
 measure the top quark mass~\cite{blusk}(see also Appendix~B).}
 and 4.4 $\pm$ 0.6 are expected from SM sources. 
 The {\em a posteriori} probability of observing no less than 13 events 
 is 0.1\%. The  probability for observing this excess of $\W+$ 2,3 jet 
 events with a superjet does not take into account
 the number of comparisons made in our studies in various jet-multiplicity
 bins and using different tagging algorithms.
 It is not possible to quantify precisely the effect of this ``trial factor''. 
 We have carried out several statistical tests using different combinations
 of the observed and predicted numbers of single and double tags reported in
 Tables~\ref{tab:tab_3.1} through~\ref{tab:tab_4.0}. These combinations always 
 include the observed numbers of supertags. We have used both a likelihood 
 method~\cite{prob} and other statistical techniques,
 which combine the probabilities of observing a number of tagged events 
 at least as large as the data. These studies yield probabilities in the
 range of one to several percent.

 The cause of the excess of $\W+$ 2,3 jet events with supertags could be 
 a discrepancy in the correlation between the SLT and SECVTX
 efficiencies in the data and simulation. These simulated efficiencies  
 have been tuned separately using the data and, in principle, the SLT tagging 
 efficiency in jets already tagged by SECVTX could be higher in the data than
 in the simulation. We have checked this using generic-jet data (see Appendix~A)
 and we conclude that the excess of $\W+$ 2,3 jet events with a supertag
 cannot be explained by this type of simulation deficiency.
\newpage
\begin{table}[p]
\begin{center}
\caption[]{Summary of observed and predicted number of $\W$ events with a 
           soft lepton tag. The data sample is split in events with and 
           without SECVTX tags.}
\def\arraystretch{0.7}
 \begin{tabular}{lcccc}
 Source            &  $\W+1 \,{\rm jet}$        &  $\W+2 \,{\rm jet}$    &  $\W+3 \,{\rm
jet}$ &   $\W+\geq4 \,{\rm jet}$     \\
 \hline
 \multicolumn{5}{c}{ Events without SECVTX tags }  \\
Data                 &   9388 &   1330 &    182 &     41 \\
\hline
 SLT mistags in   &        &        &        &  \\
 $\W+$ jet without h.f.    & 93.31 $\pm$ 9.33 &  24.81 $\pm$ 2.48 & ~4.74 $\pm$ 0.47 & 1.26 $\pm$ 0.13 \\
 Non-$\W$                  & ~~8.39 $\pm$ 0.67 &  ~1.67 $\pm$ 0.44 & ~0.31 $\pm$ 0.22 & 0.13 $\pm$ 0.09 \\ 
 $\W\W,\W Z,ZZ$            & ~~0.83 $\pm$ 0.15 &  ~1.58 $\pm$ 0.21 & ~0.31 $\pm$ 0.04 & 0.05 $\pm$ 0.00 \\
 Single top                & ~~0.27 $\pm$ 0.06 &  ~0.46 $\pm$ 0.09 & ~0.13 $\pm$ 0.03 & 0.03 $\pm$ 0.01 \\
 $\W c$                    & ~16.97 $\pm$ 4.08 &  ~5.99 $\pm$ 1.40 & ~1.10 $\pm$ 0.30 & 0.22 $\pm$ 0.06 \\
 $\W c\bar{c}$             & ~~7.99 $\pm$ 1.81 &  ~3.78 $\pm$ 0.51 & ~1.02 $\pm$ 0.39 & 0.25 $\pm$ 0.12 \\
 $\W b\bar{b}$             & ~~4.47 $\pm$ 0.68 &  ~2.26 $\pm$ 0.43 & ~0.31 $\pm$ 0.07 & 0.10 $\pm$ 0.03 \\
 $Z\rightarrow \tau\tau$   & ~~0.83 $\pm$ 0.20 &  ~0.40 $\pm$ 0.09 & ~0.15 $\pm$ 0.09 & 0.02 $\pm$ 0.00 \\
 $Zc$                      & ~~0.14 $\pm$ 0.03 &  ~0.05 $\pm$ 0.01 & ~0.02 $\pm$ 0.01 & 0.00 $\pm$ 0.00 \\
 $Zc\bar{c}$               & ~~0.22 $\pm$ 0.05 &  ~0.11 $\pm$ 0.03 & ~0.05 $\pm$ 0.02 & 0.01 $\pm$ 0.00 \\
 $Zb\bar{b}$               & ~~0.23 $\pm$ 0.04 &  ~0.11 $\pm$ 0.03 & ~0.03 $\pm$ 0.01 & 0.00 $\pm$ 0.00 \\
 $t\bar{t}$                & ~~0.11 $\pm$ 0.05 &  ~0.85 $\pm$ 0.31 & ~1.90 $\pm$ 0.65 & 2.15 $\pm$ 0.69 \\
 \hline
   SM prediction       & 133.75 $\pm$ 10.38 & 42.06 $\pm$ 2.99 & 10.06 $\pm$ 0.98 & 4.22 $\pm$ 0.72 \\
 Data with SLT tags       &    145 &     47 &     12 &      5 \\
 \hline
\hline
\multicolumn{5}{c}{ Events with SECVTX tags }   \\
 Data                     &     66 &     40 &     16 &     13 \\
\hline
 SECVTX mistags in  &     &        &        &  \\
 events with SLT tags     &  0.28 $\pm$ 0.03 & 0.20 $\pm$ 0.02 & 0.16 $\pm$ 0.02 & 0.05 $\pm$ 0.01 \\
 Non-$\W$                 &  0.57 $\pm$ 0.05 & 0.42 $\pm$ 0.11 & 0.08 $\pm$ 0.05 & 0.03 $\pm$ 0.02 \\
 $\W\W,\W Z,ZZ$           &  0.02 $\pm$ 0.02 & 0.16 $\pm$ 0.03 & 0.03 $\pm$ 0.01 & 0.00 $\pm$ 0.00 \\
 Single top               &  0.12 $\pm$ 0.04 & 0.32 $\pm$ 0.06 & 0.09 $\pm$ 0.02 & 0.02 $\pm$ 0.01 \\
 $\W c$                   &  0.88 $\pm$ 0.29 & 0.38 $\pm$ 0.12 & 0.17 $\pm$ 0.02 & 0.02 $\pm$ 0.00 \\
 $\W c\bar{c}$            &  0.41 $\pm$ 0.13 & 0.41 $\pm$ 0.13 & 0.14 $\pm$ 0.05 & 0.03 $\pm$ 0.01 \\
 $\W b\bar{b}$            &  1.58 $\pm$ 0.33 & 1.40 $\pm$ 0.30 & 0.40 $\pm$ 0.08 & 0.11 $\pm$ 0.02 \\
 $Z\rightarrow \tau\tau$  &  0.00 $\pm$ 0.00 & 0.00 $\pm$ 0.00 & 0.00 $\pm$ 0.00 & 0.00 $\pm$ 0.00 \\
 $Zc$                     &  0.01 $\pm$ 0.00 & 0.00 $\pm$ 0.00 & 0.00 $\pm$ 0.00 & 0.00 $\pm$ 0.00 \\
 $Zc\bar{c}$              &  0.01 $\pm$ 0.00 & 0.01 $\pm$ 0.00 & 0.01 $\pm$ 0.00 & 0.00 $\pm$ 0.00 \\
 $Zb\bar{b}$              &  0.08 $\pm$ 0.02 & 0.06 $\pm$ 0.02 & 0.03 $\pm$ 0.01 & 0.01 $\pm$ 0.00 \\
 $t\bar{t}$               &  0.04 $\pm$ 0.02 & 0.78 $\pm$ 0.30 & 1.88 $\pm$ 0.65 & 2.65 $\pm$ 0.85 \\
 \hline
  SM prediction        &  4.00 $\pm$ 0.47 & 4.15 $\pm$ 0.50 & 2.99 $\pm$ 0.66 & 2.93 $\pm$ 0.85 \\
 Data with SECVTX and SLT tags   &      1 &      9 &      5 &      3 \\
 \end{tabular}
\label{tab:tab_3.4}
\end{center}
\end{table}

\newpage
\begin{table}[p]
\begin{center}
\caption[]{Observed and predicted number of $\W+$ jet events with a supertag.
           The subsample of events with an additional SECVTX tag (DT) is also
           listed.}
 \begin{tabular}{lcccc}
 Source         &  $\W+1 \,{\rm jet}$        &  $\W+2 \,{\rm jet}$      &  $\W+3
\,{\rm jet}$        &
$\W+\geq4 \,{\rm jet}$     \\
SECVTX mistags  in & & & & \\
events with SLT tags  &  0.28 $\pm$ 0.03 &    0.09 $\pm$ 0.01 
&    0.07 $\pm$ 0.01 &    0.02 $\pm$ 0.00\\
 Non-$\W$                                  &    0.57 $\pm$ 0.05 &  0.13 $\pm$ 0.03
&    0.00 $\pm$ 0.00 &    0.00 $\pm$ 0.00\\
 $\W\W,\W Z,ZZ$                               &    0.02 $\pm$ 0.02 &  0.13 $\pm$ 0.06
&    0.01 $\pm$ 0.01 &    0.00 $\pm$ 0.00\\
 Single top                               &    0.12 $\pm$ 0.04 &  0.24 $\pm$ 0.05
&    0.07 $\pm$ 0.02 &    0.02 $\pm$ 0.00\\
 $\W c$                                     &    0.88 $\pm$ 0.29 &  0.24 $\pm$ 0.14
&    0.14 $\pm$ 0.10 &    0.00 $\pm$ 0.00\\
 $\W c\bar{c}$                          &    0.41 $\pm$ 0.13 &    0.25 $\pm$ 0.09
&    0.13 $\pm$ 0.06 &    0.00 $\pm$ 0.00\\
 $\W b\bar{b}$                          &    1.58 $\pm$ 0.33 &    1.07 $\pm$ 0.26
&    0.19 $\pm$ 0.09 &    0.01 $\pm$ 0.00\\
 $Z\rightarrow \tau\tau$                  &    0.00 $\pm$ 0.00 &  0.00 $\pm$ 0.00
&    0.00 $\pm$ 0.00 &    0.00 $\pm$ 0.00\\
 $Zc$                                 &    0.01 $\pm$ 0.00 &    0.00 $\pm$ 0.00
&    0.00 $\pm$ 0.00 &    0.00 $\pm$ 0.00\\
 $Zc\bar{c}$                         &    0.01 $\pm$ 0.00 &    0.01 $\pm$ 0.00
&    0.01 $\pm$ 0.00 &    0.00 $\pm$ 0.00\\
 $Zb\bar{b}$                          &    0.08 $\pm$ 0.02 &    0.05 $\pm$ 0.02
&    0.02 $\pm$ 0.01 &    0.00 $\pm$ 0.00\\
 $t\bar{t}$                         &    0.04 $\pm$ 0.02 &    0.48 $\pm$ 0.19
&    1.08 $\pm$0.40  &    1.42 $\pm$ 0.49 \\
 \hline
SM prediction (supertags) & 4.00 $\pm$ 0.50 & 2.69 $\pm$ 0.41 &1.71 $\pm$ 0.40 &1.47 $\pm$ 0.51 \\
SM prediction (DT)        &   & 0.26 $\pm$ 0.06 & 0.36 $\pm$ 0.08 & 0.50 $\pm$ 0.13 \\
\hline
Data (supertags)   &  1           & 8          & 5          &      2  \\
Data (DT)          &   & 2 & 3 &  0   \\
\end{tabular}
\label{tab:tab_4.0}
\end{center}
\end{table}    

\clearpage
\section{Properties of the events with a superjet}
\label{sec:s-prop}
 Having observed an excess of $\W+$ 2,3 jet events with a supertag, we
 next compare the kinematics of these events with the SM simulation.
 We check the simulation using a complementary $\W+$ 2,3 jet sample of data.
 This sample is described in subsection~A. In subsection~B we compare the 
 heavy flavor content of the additional jets in events with a superjet and in
 the complementary sample. In subsections~C and D we compare several
 kinematical distributions of these events to the simulation.
\subsection{Complementary data sample}
\label{sec:ss-cssa}
 We check our simulation by studying a larger data sample consisting of
 $\W+$ 2,3 jet events with a SECVTX tag, but no supertags. The number of 
 observed and predicted events are compared in Table~\ref{tab:tab_5.0}
 (43 $\W+$ 2,3 jet events are observed, in agreement with the 
 SM prediction of 43.6 $\pm$ 3.3). We have chosen this sample because, 
 as shown by the comparison of Table~\ref{tab:tab_5.0} with  
 Table~\ref{tab:tab_4.0}, its composition is quite 
 similar to  $\W+$ jet events with a supertag\footnote{$\W+$ 2,3 jet
 events with a SLT tag and no supertags are another larger statistics data 
 set, however the heavy flavor composition is quite different from that
 expected for events with a superjet.}.
 In order to have  a complementary sample of data  with the same kinematical 
 acceptance  of the events with a supertag, we also require that at least one
 of the jets tagged by SECVTX contains a soft lepton candidate track. After 
 this additional requirement this  sample of  $\W+$ 2,3 jet events 
 consists of 42 events (the SM prediction is 41.2 $\pm$ 3.1 events).
 We note that, while closely related, this event sample has still a few 
 features which are different from the superjet sample. For instance, 
 most of the superjets are expected to be produced by heavy flavor semileptonic
 decays, in which the corresponding neutrino escapes detection, while
 in the complementary sample SECVTX tagged jets are predominantly produced
 by purely hadronic decays of heavy flavors. However, according to the 
 simulation, a large fraction of heavy flavor semileptonic decays is not 
 identified by the SLT algorithm and is also included in the complementary 
 sample. All such effects are in principle described by the simulation.
\subsection{Heavy flavor content of additional jets} 
\label{sec:ss-dt}
 The heavy flavor content of the second and third jet in the events 
 can be inferred from the rate of additional SECVTX tags.
 Tables~\ref{tab:tab_4.0} and~\ref{tab:tab_5.0} show the number of observed
 and predicted events with an additional jet tagged by SECVTX in superjet 
 events and in the complementary sample. In the latter data sample, in which
 according to the simulation in Table~\ref{tab:tab_5.0} most of the events 
 contain a second jet with $b$ flavor, there are 6 $\W+$ 2,3 jet events with
 a double SECVTX tag, in agreement with the expectation of 
 5.02 $\pm$ 0.84 events.

 Of the 13 $\W+$ 2,3 jet events with a superjet 5 contain an additional SECVTX
 tag. If the 13 events are a fluctuation of SM processes, we  expect to find
 1.8 $\pm$ 0.3 events with a double tag\footnote{The prediction is 
 0.62 $\pm$ 0.10 events with a double tag in 4.4 events with a superjet.}. 
 The probability of observing 5 or more  $\W+$ 2,3 jet events with double tags
 is 4.1\%. Given the high probability of finding an additional SECVTX tag,
 we apply $b$-jet specific energy corrections to the additional jets in the 
 event. These jets are later referred to as ``$b$-jets''.
\newpage
\begin{table}[p]
\begin{center}
\caption[]{Observed and predicted number of $W+$ jet events tagged by SECVTX 
           after removing events with a supertag. The subsample of events
           with an additional SECVTX tag (DT) is also listed.}
 \begin{tabular}{lcccc}
 Source         &  $\W+1 \,{\rm jet}$        &  $\W+2 \,{\rm jet}$     &  $\W+3
\,{\rm jet}$        &
$\W+\geq4 \,{\rm jet}$     \\
 Mistags  &  ~10.52 $\pm$ 1.00 & ~3.72 $\pm$ 0.34& ~0.93 $\pm$ 0.09 & ~0.34 $\pm$ 0.04\\
 Non-$\W$                               &   ~7.61 $\pm$ 0.06 &  ~1.36 $\pm$ 0.04
&   ~0.76 $\pm$ 0.03 &   ~0.31 $\pm$ 0.03\\
 $\W\W,\W Z,ZZ$                           &   ~0.50 $\pm$ 0.14  &  ~1.25 $\pm$ 0.25
&   ~0.40 $\pm$ 0.13 &   ~0.00 $\pm$ 0.00\\
 Single top                           &   ~1.24 $\pm$ 0.31  &  ~2.15 $\pm$ 0.49
&   ~0.56 $\pm$ 0.13 &   ~0.12 $\pm$ 0.03\\
 $\W c$                                 &    16.02 $\pm$ 5.13 &  ~3.70 $\pm$ 1.29
&   ~0.37 $\pm$ 0.13 &   ~0.09 $\pm$ 0.03\\
 $\W c\bar{c}$                          &   ~7.48 $\pm$ 2.08  &  ~3.35 $\pm$ 0.86
&   ~0.64 $\pm$ 0.22 &   ~0.16 $\pm$ 0.06\\
 $\W b\bar{b}$                          &    15.42 $\pm$ 2.21 &  ~8.80 $\pm$ 1.63
&   ~1.74 $\pm$ 0.40 &   ~0.47 $\pm$ 0.13\\
 $Z\rightarrow \tau\tau$              &    ~0.96 $\pm$ 0.30 &  ~0.70 $\pm$ 0.25
&   ~0.17 $\pm$ 0.12 &   ~0.00 $\pm$ 0.00\\
 $Zc$                                 &    ~0.13 $\pm$ 0.04 &  ~0.03 $\pm$ 0.01
&   ~0.01 $\pm$ 0.00 &   ~0.00 $\pm$ 0.00\\
 $Zc\bar{c}$                          &    ~0.21 $\pm$ 0.06 &  ~0.10 $\pm$ 0.03
&   ~0.03 $\pm$ 0.02 &   ~0.00 $\pm$ 0.00\\
 $Zb\bar{b}$                          &    ~0.85 $\pm$ 0.13 &  ~0.48 $\pm$ 0.11
&   ~0.19 $\pm$ 0.06 &   ~0.02 $\pm$ 0.02\\
 $t\bar{t}$                           &    ~0.50 $\pm$ 0.16 &  ~3.62 $\pm$ 1.00
&   ~8.56 $\pm$ 2.38 &   ~9.96 $\pm$ 2.40 \\
 \hline
SM prediction & 61.44 $\pm$ 6.09 & 29.26 $\pm$ 2.58 &14.39 $\pm$ 2.34 &11.48 $\pm$ 2.37 \\
SM prediction (DT) &  & 2.15 $\pm$ 0.50 & 2.87 $\pm$ 0.67 & 3.53 $\pm$  0.90\\
\hline
Data    &  65           & 32         & 11          &      11  \\
Data (DT) & & 3 & 3 & 2 \\
\end{tabular}
\label{tab:tab_5.0}
\end{center}
\end{table}    

\clearpage
\subsection{Method for testing if the data are consistent
            with the SM simulation}
\label{sec:ss-kolmo}
 In the next subsection we study distributions  of several simple kinematic
 variables $x_i$ for the 13 events with a superjet and the complementary 
 sample of 42 events. Each data distribution is compared with the sum of the
 12 SM~contributions,  $SM_{j}(x_i)$, listed in Tables~\ref{tab:tab_4.0} 
 and~\ref{tab:tab_5.0} using a Kolmogorov-Smirnov (K-S) 
 test~\cite{kuiper,sadoulet}. Using the cumulative distribution functions 
 $F(x_i)$ and $H(x_i)$ of the two distributions to be compared, the K-S  
 distance is defined as 
 $\delta$ = max $(F(x_i)-H(x_i))$ + max $(H(x_i)-F(x_i))$.
 This is the Kuiper's definition of the K-S distance~\cite{numrep}.
 
 For each variable $x_i$, the probability distribution of the K-S distance,
 $W_{i}(\delta)$, is determined with Monte Carlo pseudo-experiments. In each 
 experiment, we randomly generate parent distributions
 ${\displaystyle \sum_{j=1}^{12}} \frac {{\cal I}_j^{r}}
 {{\cal I}_j} SM_{j}(x_i)$ 
 for two and three jet events independently.
 The integral ${\cal I}_j= {\displaystyle \int} SM_{j}(x_i) dx_i $ corresponds
 to the average number of events contributed by the process $j$ and, 
 in each pseudo-experiment, the value ${\cal I}_j^r$ accounts for
 Poisson fluctuations and Gaussian uncertainties in ${\cal I}_j$.
 We use these parent distributions to randomly generate
 the same  number of $x_i$-values as  in the data,
 but we evaluate the K-S distance of the $x_i$ distribution in each 
 pseudo-experiment with respect to the parent distribution 
 ${\displaystyle \sum_{j=1}^{12}} SM_{j}(x)$.
 Using the so derived $W_i(\delta)$ distribution, we define the probability 
 $P_{i}$ that the $x_i$ distribution of the data is consistent with the SM
 simulation as $P_{i}= {\displaystyle \int_{\delta_{i}^{0}}^{\infty} }
 W_i(\delta) d\delta $, where $\delta^{0}$ is the K-S distance of
 the data.
\subsection{Comparison of  kinematical distributions in the data with the
            SM simulation}
 We test if the events with a superjet are consistent with the SM
 prediction by comparing the production cross sections
 ${\displaystyle \frac {d^2 \sigma} {dp_T d\eta} }$ of each object in the 
 final state. In all SM processes contributing to these events,
 these differential  cross sections approximately factorize, and
 ${\displaystyle \frac {d^2 \sigma} {dp_T d\eta}} \simeq f(p_T) \cdot g(\eta)$.
 Therefore we compare data and SM simulation in the following kinematical 
 variables: the transverse energy and pseudo-rapidity distributions of the
 primary leptons, the superjets, the additional jets in the event (referred 
 to as $b$-jets), and the neutral object producing the missing energy in 
 the event\footnote{Jet energies are corrected using the full set of
 correction functions developed to measure the top 
 mass~\cite{cdf-evidence,top_mass,blusk}.}.
 The kinematics of the neutral object producing the missing energy cannot be
 measured directly. However, correlated quantities are
 the transverse energy and the rapidity of the recoiling system $l+b+suj$
 composed of the primary lepton ($l$), the superjet ($suj$) and each 
 additional jet ($b$) in the event. Since the total transverse momentum 
 of the events is conserved, in $\W+$ 2 jet events the transverse energy 
 $E_{T}^{l+b+suj}$ of the system $l+b+suj$ is a measure of the missing 
 transverse energy. In the rest frame of the initial state partons producing
 $\W+$ 2 jet events, the rapidities of the system $l+b+suj$ and of the
 object producing the missing energy are also correlated. This correlation
 is however smeared by the unknown Lorentz boost of the initial parton system.
 For uniformity, in $\W+$ 3 jet events we use the same variables with two
 entries per event (corresponding to the two possible choices for the $b$-jet). 

 We finally test the distribution of the azimuthal angle 
 $\delta \phi^{l,b+suj}$ between the primary lepton and the system
 $b+suj$ composed by the superjet and each additional $b$-jet with 
 the purpose of checking if the events are consistent with the simulated
 production and decay of $\W$ bosons. The $\W$ transverse mass can be 
 described with the variables $E_T^{l}$ and $\MET$, which are already used,
 and the azimuthal angle between the primary lepton and the $\W$ direction.
 Since the total transverse momentum of the events is conserved, in  
 $\W+$ 2 jet events this azimuthal angle can be inferred from the 
 supplementary angle $\delta \phi^{l,b+suj}$. For uniformity, in $\W+$ 3 jet
 events we use the same variable with two entries per event.

\begin{figure}[htb]
\begin{center}
\leavevmode
\epsfxsize \textwidth
\epsffile{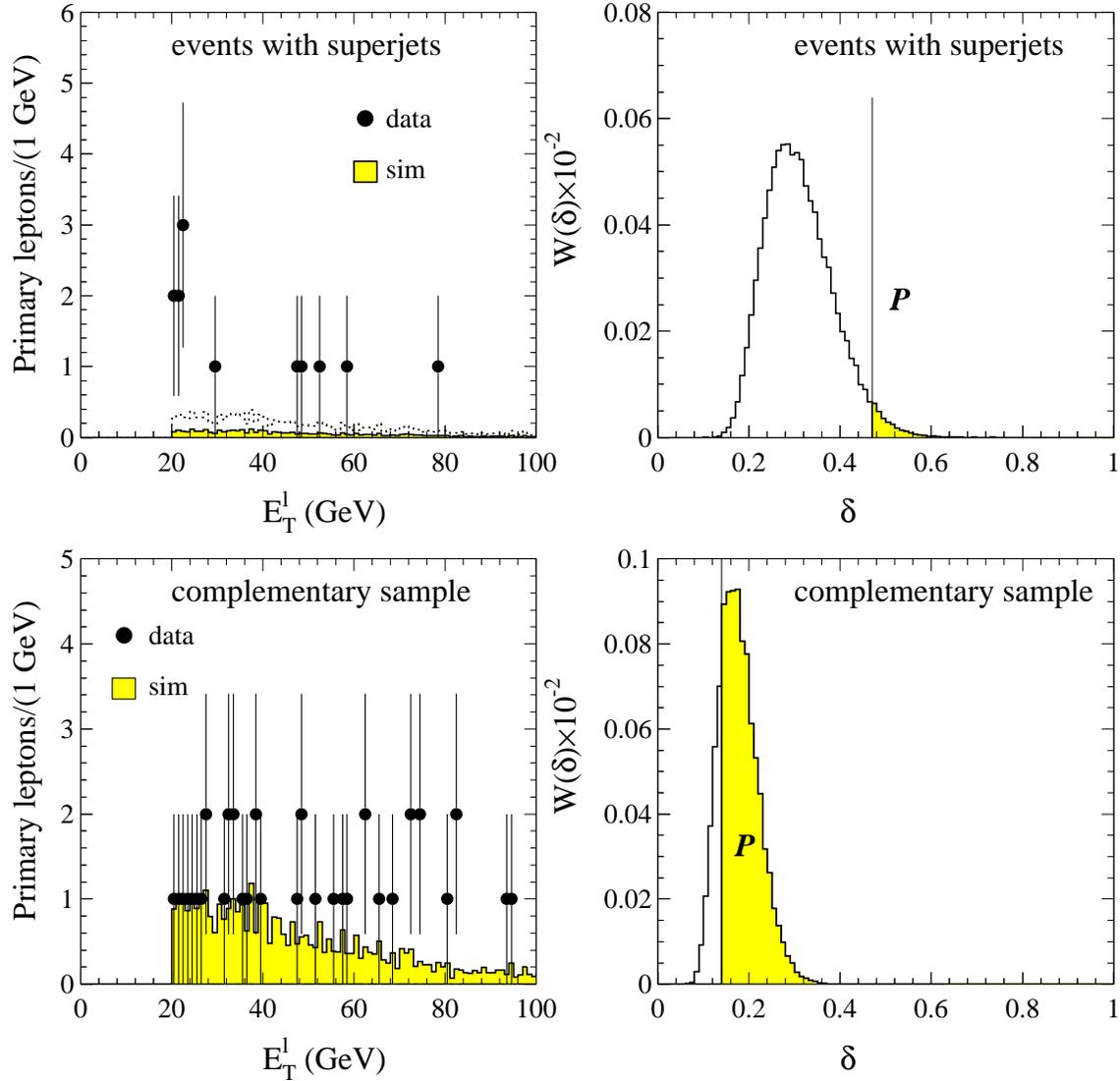}
\caption[]{Distributions of the transverse energy of the primary lepton 
           for the data ($\bullet$) are compared to the SM prediction 
           (shaded histograms). The dotted histograms show the SM simulation 
           normalized to the data. The probability distribution of the K-S 
           distance $\delta$ is calculated with Monte Carlo 
           pseudo-experiments (see text). The vertical line indicates the 
           observed distance $\delta^{0}$ between the cumulative distributions
           of the data and the simulation. The integral of the shaded area 
           represents the probability $P$ of measuring a K-S distance 
           no smaller than $\delta^{0}$.}
\label{fig:fig_5.2}
\end{center}
\end{figure}

\begin{figure}[htb]
\begin{center}
\leavevmode
\epsfxsize \textwidth
\epsffile{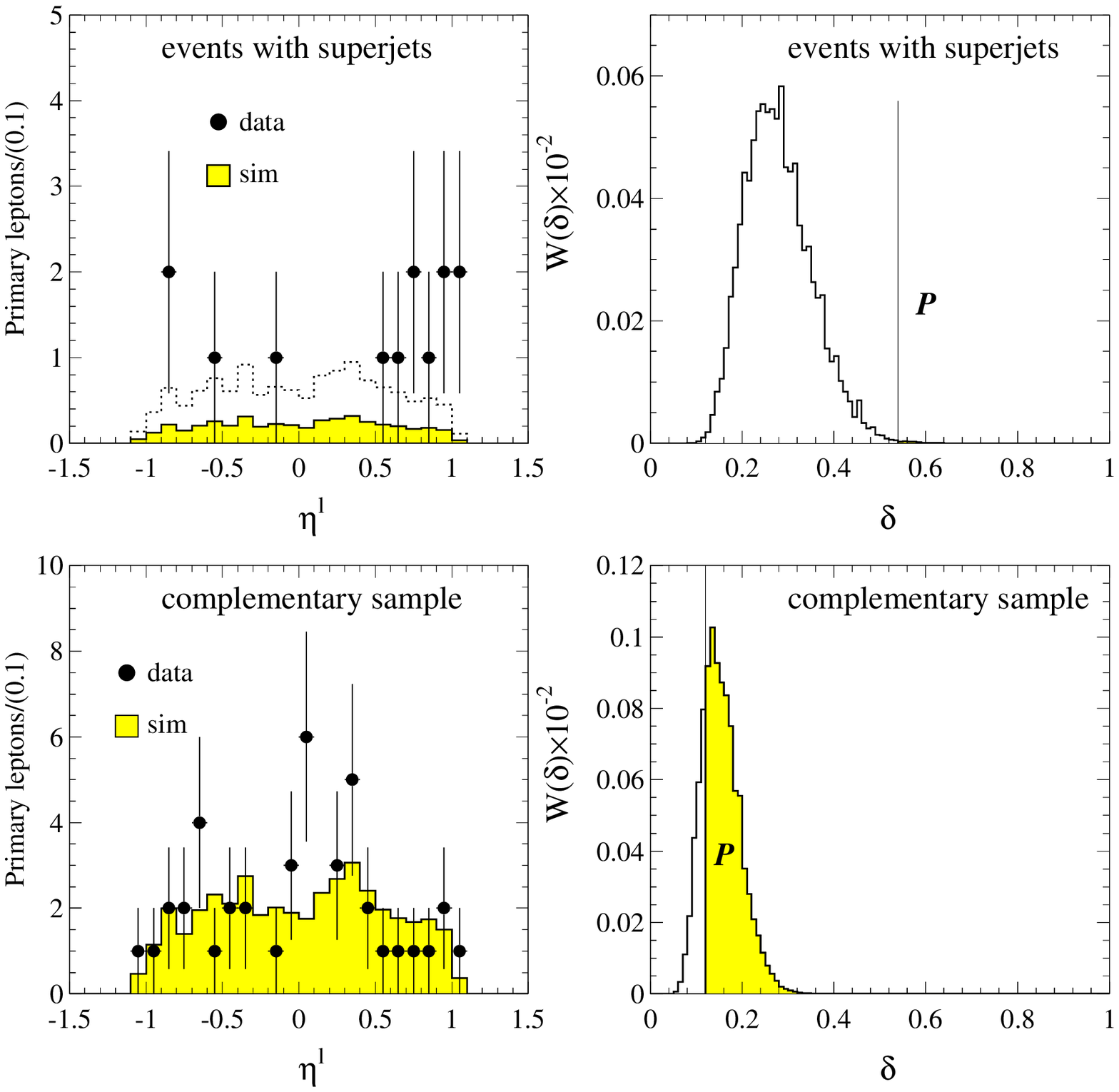}
\caption[]{Distribution of the pseudo-rapidity of the primary lepton in 
           events with a superjet and in the complementary sample.}
\label{fig:fig_5.3}
\end{center}
\end{figure}

\begin{figure}[htb]
\begin{center}
\leavevmode
\epsfxsize \textwidth
\epsffile{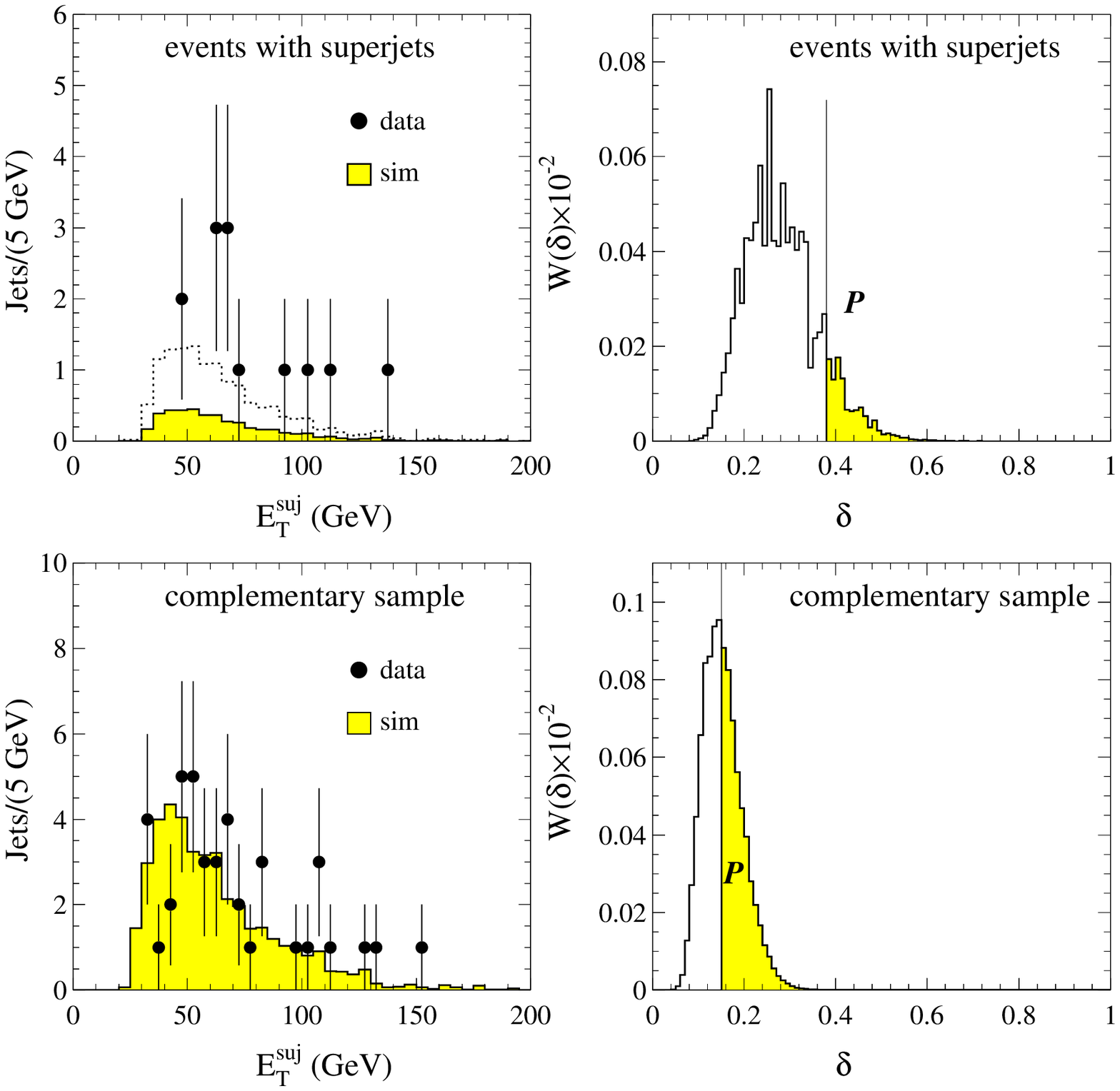}
\caption[]{Distribution of the transverse energy of the superjet in 
           events with a superjet and in the complementary sample.}
\label{fig:fig_5.4}
\end{center}
\end{figure}

\begin{figure}[htb]
\begin{center}
\leavevmode
\epsfxsize \textwidth
\epsffile{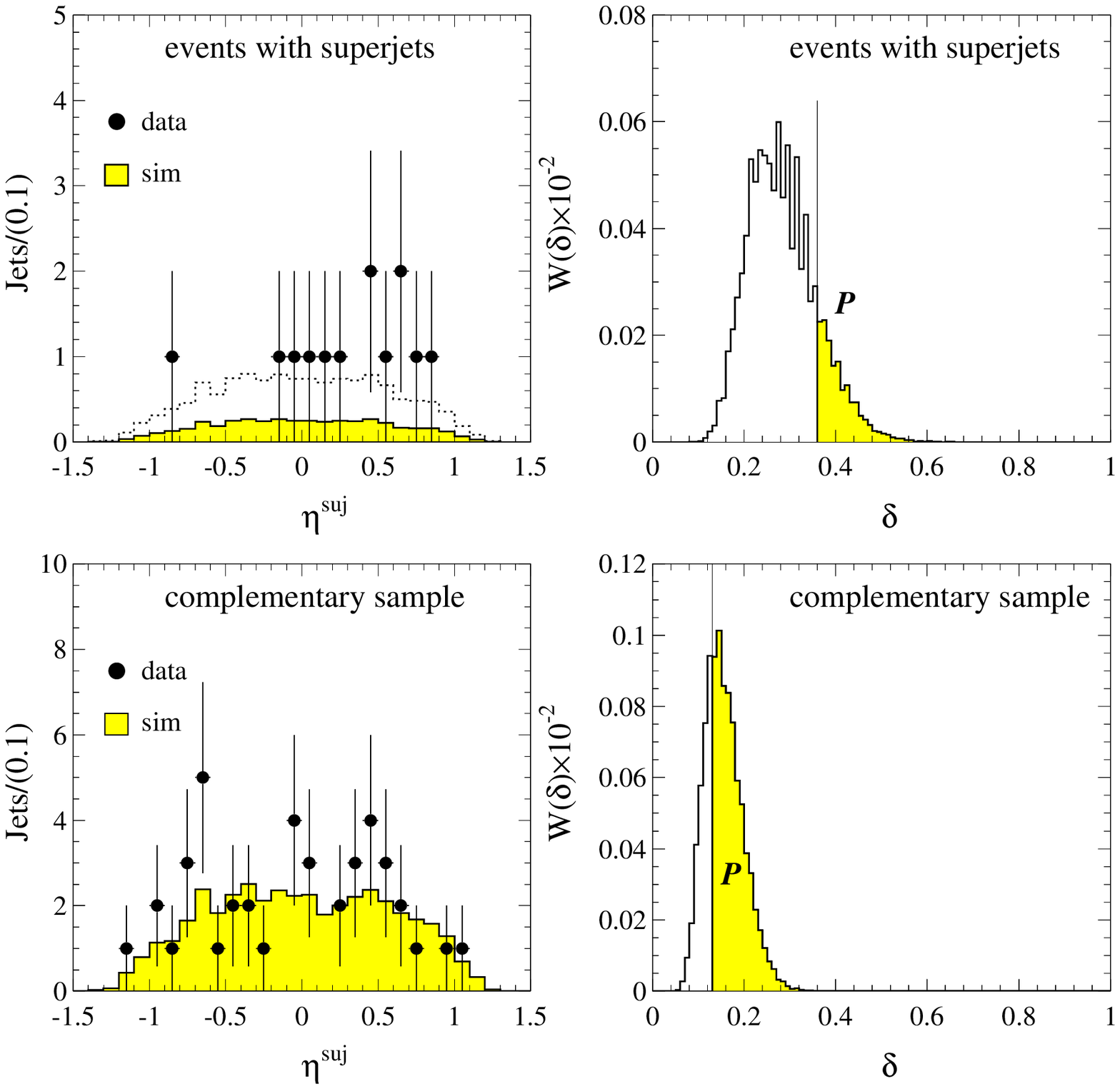}
\caption[]{Distribution of the pseudo-rapidity of the superjet in 
           events with a superjet and in the complementary sample.}
\label{fig:fig_5.5}
\end{center}
\end{figure}

\begin{figure}[htb]
\begin{center}
\leavevmode
\epsfxsize \textwidth
\epsffile{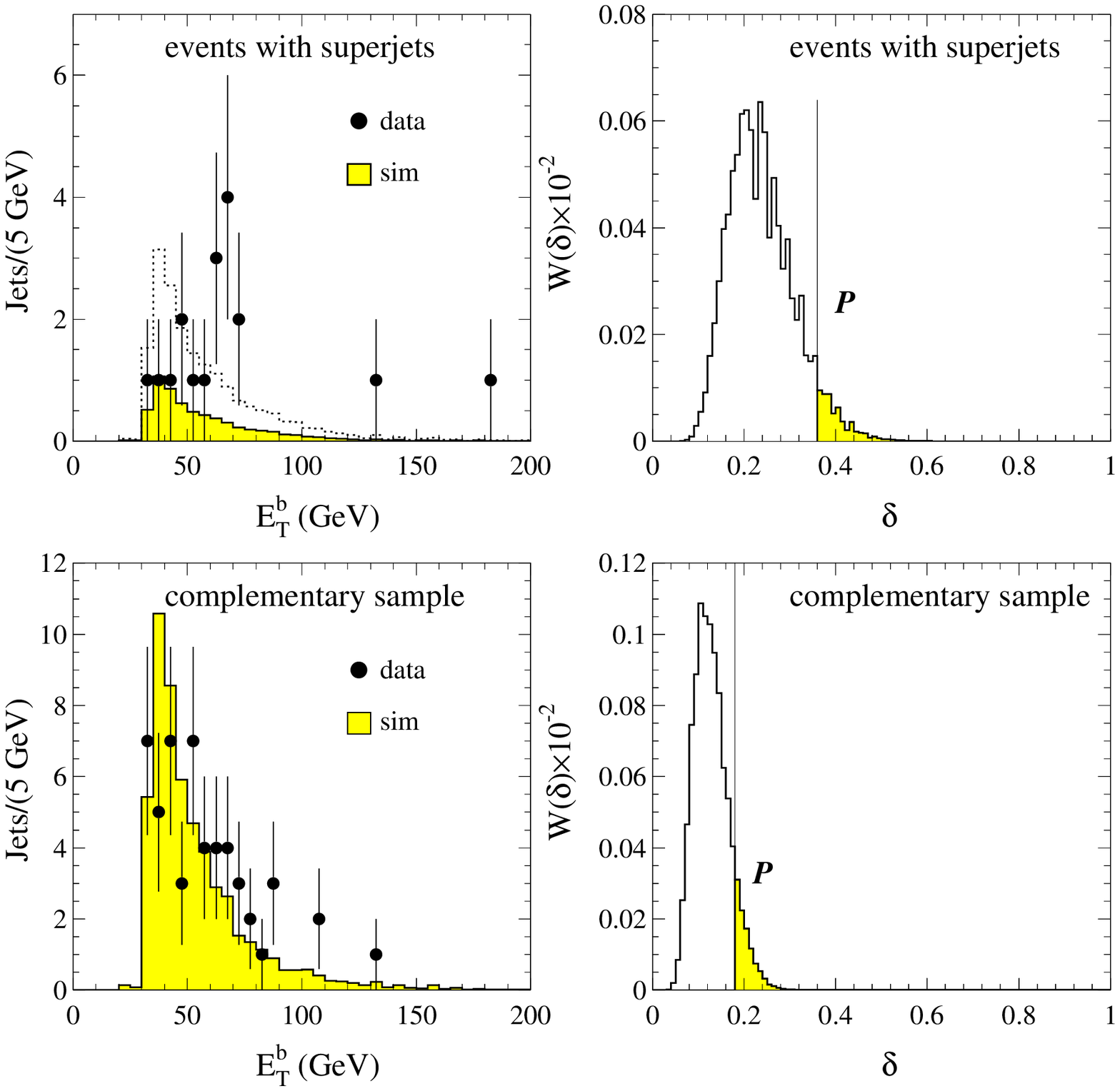}
\caption[]{Distribution of the transverse energy of all $b$-jets 
           in events with a superjet and in the complementary sample.}
\label{fig:fig_5.6}
\end{center}
\end{figure}

\begin{figure}[htb]
\begin{center}
\leavevmode
\epsfxsize \textwidth
\epsffile{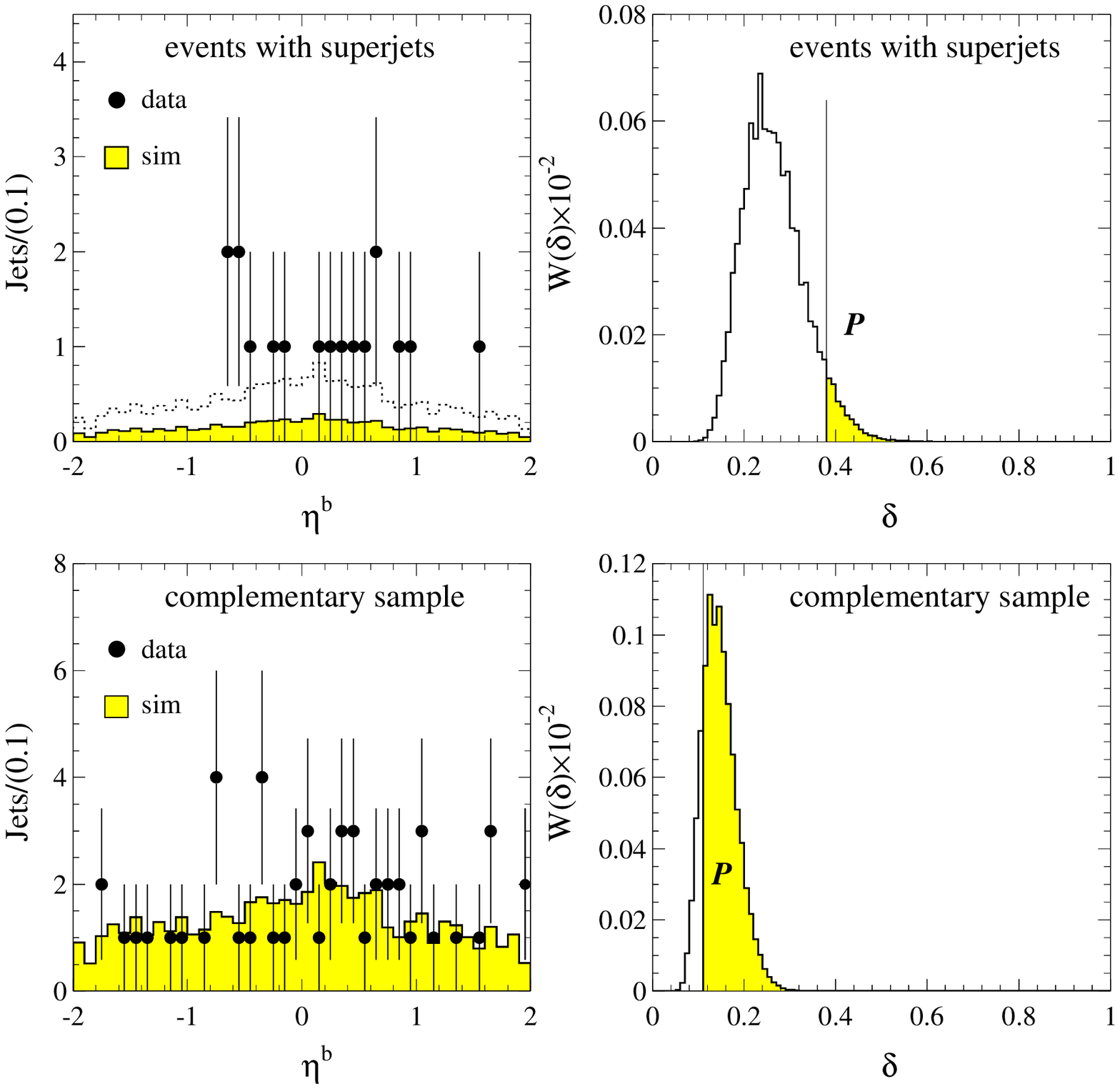}
\caption[]{Distribution of the pseudo-rapidity of all $b$-jets in 
           events with a superjet and in the complementary sample.}
\label{fig:fig_5.7}
\end{center}
\end{figure}

\begin{figure}[htb]
\begin{center}
\leavevmode
\epsfxsize \textwidth
\epsffile{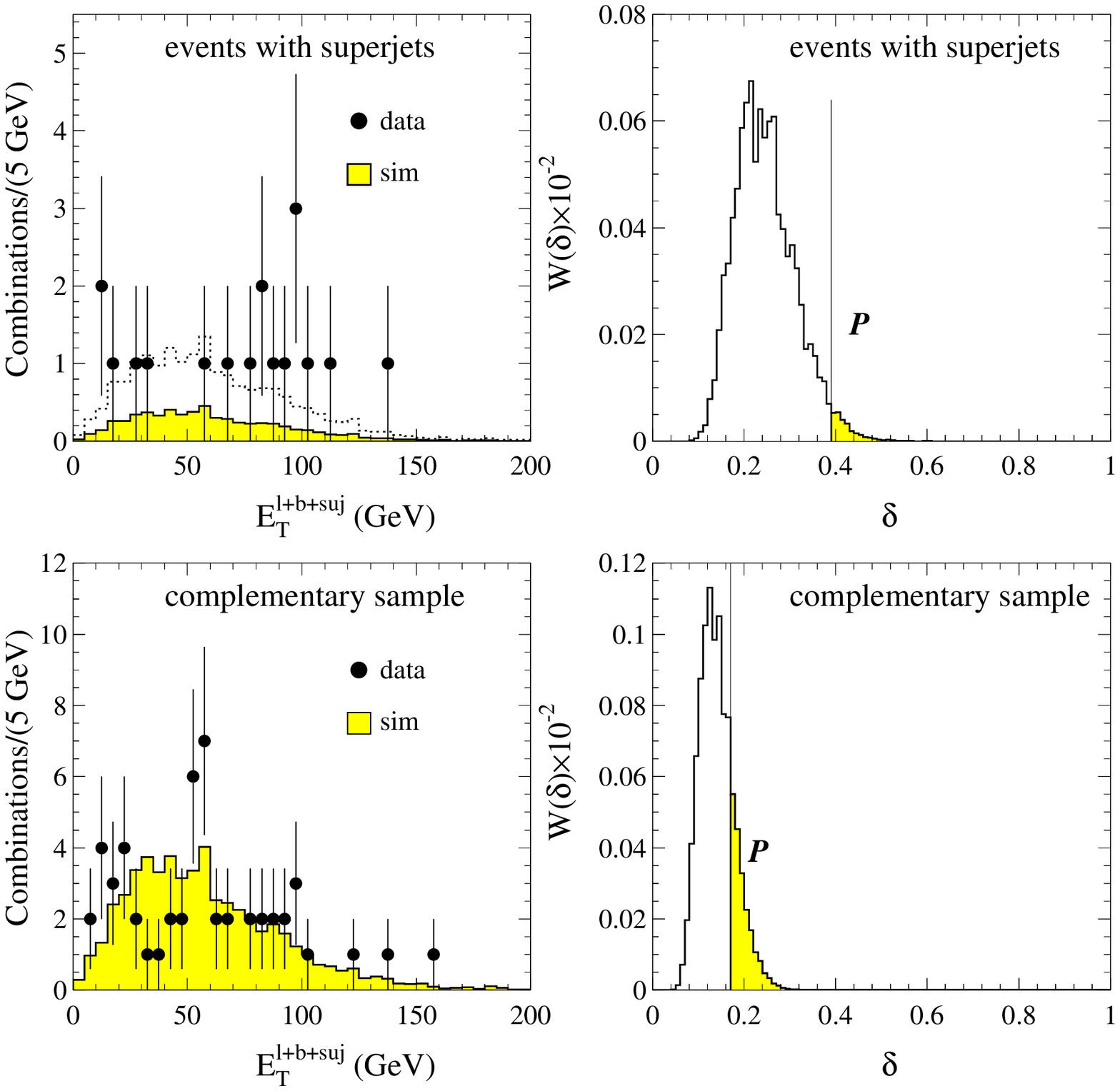}
\caption[]{Distribution of the transverse energy of the system
           $l+$superjet$+b$-jet in events with a superjet and in the 
           complementary sample.}
\label{fig:fig_5.8}
\end{center}
\end{figure}

\begin{figure}[htb]
\begin{center}
\leavevmode
\epsfxsize \textwidth
\epsffile{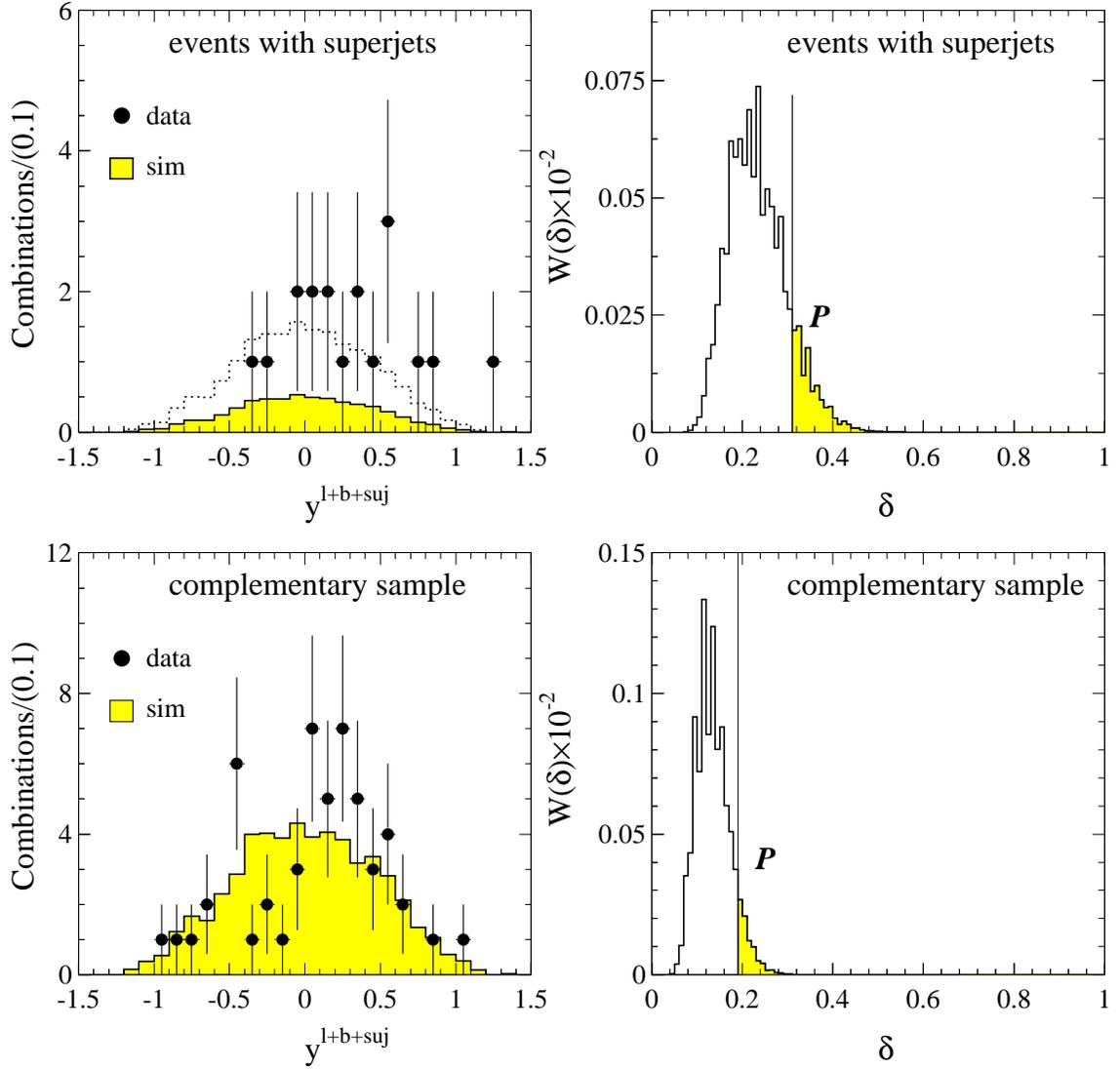}
\caption[]{Distribution of the rapidity of the system $l+$superjet$+b$-jet
          in events with a superjet and in the complementary sample.}
\label{fig:fig_5.9}
\end{center}
\end{figure}

\begin{figure}[htb]
\begin{center}
\leavevmode
\epsfxsize \textwidth
\epsffile{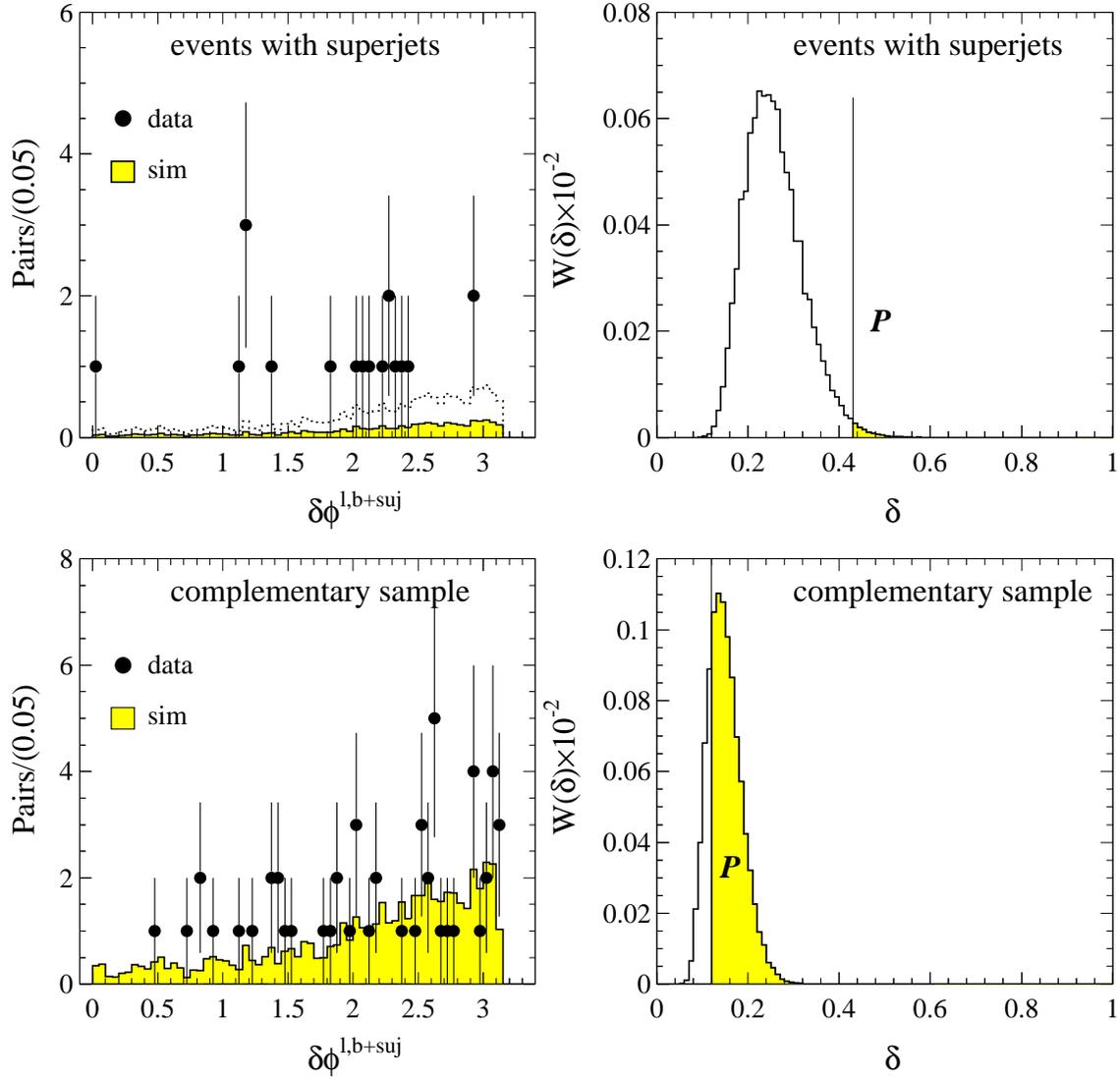}
\caption[]{Distribution of the azimuthal angle between the primary lepton and
           the superjet$+b$-jet system in events with a superjet and in the
           complementary sample.}
\label{fig:fig_5.10}
\end{center}
\end{figure}

 This minimal set of 9 variables is sufficient to describe the kinematics of
 the final state with relatively modest correlations. The observed and 
 predicted distributions of these kinematical variables are compared in 
 Figures~\ref{fig:fig_5.2}  to~\ref{fig:fig_5.10}. For each comparison, we
 show the probability $P$ that the data are consistent with the simulation.
 Table~\ref{tab:tab_5.2} summarizes the probabilities of these comparisons.
 The SM simulation models correctly the complementary sample of data,
 but has a systematically low probability of being consistent with the 
 kinematical distributions of the events with a superjet.

\begin{figure}[htb]
\begin{center}
\leavevmode
\epsfxsize \textwidth
\epsffile{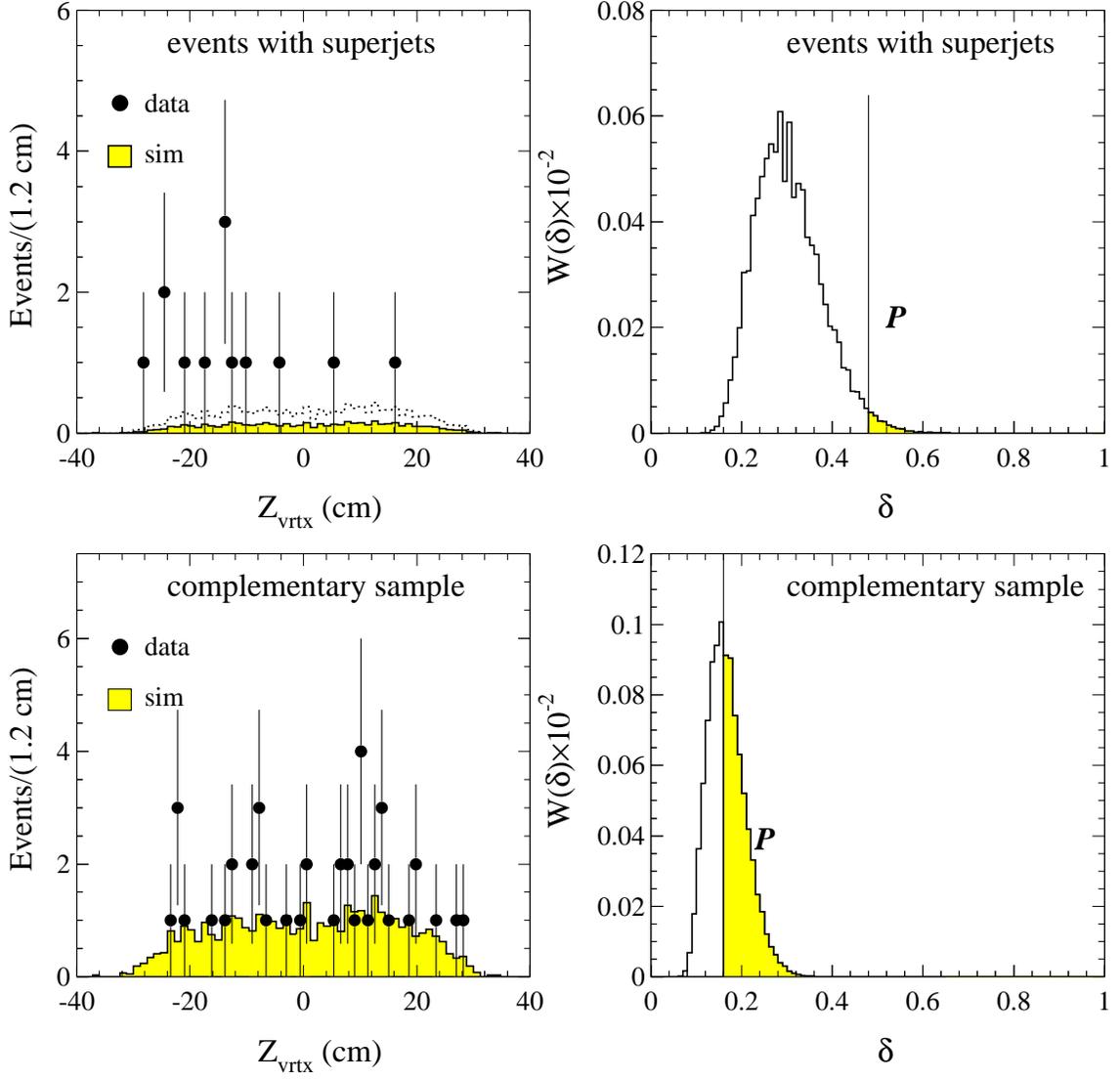}
\caption[]{Distribution of the event-vertex position along the beam line 
          ($z$-axis) in events with a superjet and in the complementary sample.}
\label{fig:fig_5.11}
\end{center}
\end{figure}

 In addition, one notices that the rapidity distributions of the primary 
 lepton and the jets in the 13 events (Figures~\ref{fig:fig_5.3},
 ~\ref{fig:fig_5.5},~\ref{fig:fig_5.7}, and~\ref{fig:fig_5.9})
 are not symmetric around $\eta=0$ and are more populated at positive 
 rapidities. These observations led to additional investigations of the 
 characteristics of the 13 events exploring the possibility that some 
 detector effects were not properly modeled by the simulation.
 These studies have not revealed any anomaly which could be taken as an
 indication of detector problems. In particular, asymmetries due to detector
 problems are not visible in the complementary sample nor in the larger
 statistics sample of generic-jet data. However, as shown in 
 Figure~\ref{fig:fig_5.11}, we discovered that the primary vertex of these 
 events has an asymmetric $z$-distribution ($z$ is the axis along the beam 
 line). Again, such an asymmetry is not observed in any of the
 large statistics data samples available.
 The binomial probability of observing an equal or larger asymmetry 
 due to a statistical fluctuation in the distribution of the
 event vertex is 1.1\%. Similar probabilities for the asymmetry 
 in several rapidity distributions are in the range between 1.5 to 10\%. 
 Since we know of no physics process that would produce such asymmetries,
 it is possible that an obscure detector problem, not seen in other
 samples, is responsible; or it may be that these asymmetries are due to a
 low probability statistical fluctuation.
\newpage
\begin{table}[p]
\begin{center}
\def\arraystretch{0.8}
\caption[]{Results of the K-S comparison between data and simulation.
           For each variable we list the observed K-S distance 
           $\delta^{0}$ and the probability
           $P$  of making an observation with a distance
           no smaller than $\delta^{0}$.}
\begin{tabular}{l c  c c c}
       & \multicolumn{2}{c}{ Events with a superjet} &
 \multicolumn{2}{c}{ Complementary sample} \\
 Variable           & $\delta^{0}$ & $P$ (\%)   & $\delta^{0}$  & $P$ (\%)  \\
  $E_T^{l}$               & 0.47   &  2.6       & 0.14          & 70.9 \\
  $\eta^{l}$              & 0.54   &  0.10      & 0.12          & 72.7 \\
  $E_T^{suj}$             & 0.38   & 11.1       & 0.15          & 43.0 \\
  $\eta^{suj}$            & 0.36   & 15.2       & 0.13          & 73.4 \\
  $E_T^{b}$               & 0.36   &  6.7       & 0.18          &  8.6 \\
  $\eta^{b}$              & 0.38   &  6.8       & 0.11          & 80.0 \\
  $E_T^{l+b+suj}$         & 0.39   &  2.5       & 0.17          & 18.8 \\
  $y^{l+b+suj}$           & 0.31   & 13.8       & 0.19          &  7.8 \\
  $\delta \phi^{l,b+suj}$ & 0.43   &  1.0       & 0.12          & 77.9 \\
  $Z_{vrtx}$              & 0.48   &  1.7       & 0.16          & 50.5 \\
\end{tabular}             
\label{tab:tab_5.2}
\end{center}
\end{table}

\clearpage
\newpage


\begin{figure}[htb]
\begin{center}
\leavevmode
\epsffile{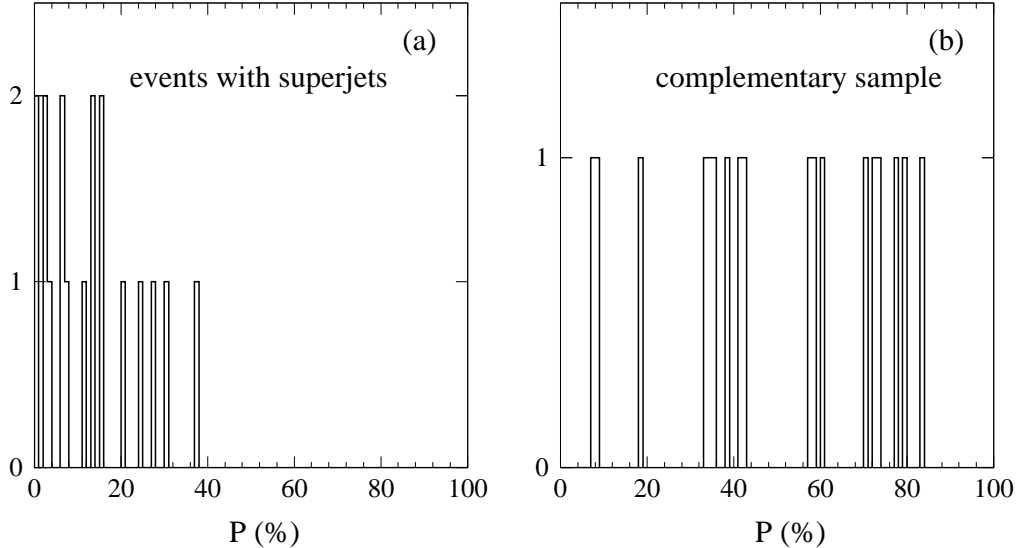}
\caption[]{Distribution of the probabilities $P$ that the 13 events with a
           superjet (a) and the complementary sample (b) are consistent with
           the SM prediction. The distribution (a) has a mean of 0.13 and
           a RMS of 0.11; the distribution (b) has a mean of 0.50 and
           a RMS of 0.24.}
\label{fig:fig_5.appb0}
\end{center}
\end{figure}

\clearpage
 The  set of 9 kinematic variables used to compare data and simulation is not 
 the only possible choice. We also looked at 9 complementary variables, and 
 Table~\ref{tab:tab_5.2tris} shows the result of the K-S test for this set of
 kinematic distributions: $\MET$, the corrected transverse missing energy; 
 $M_{T}^{W}$, the $\W$ transverse mass calculated using the primary
 lepton and $\MET$; $M^{b+suj}$, $y^{b+suj}$, and $E_T^{b+suj}$,
 the invariant mass, rapidity, and transverse energy of the system $b+suj$ 
 respectively; $M^{l+b+suj}$, the invariant mass of the system $l+b+suj$; 
 $\delta \theta^{b,suj}$ and $\delta \phi^{b,suj}$, the angle and the 
 azimuthal angle between the superjet and the $b$-jets, respectively; and 
 $\delta \theta^{l,b+suj}$, the angle between the primary lepton and the 
 system $b+suj$. The simulation correctly models these distributions for the
 complementary sample, while the probabilities for events with a superjet are
 systematically lower. However, the disagreement between events with a 
 superjet and their simulation is much reduced for this second set of variables.
 The probability distribution of the K-S comparisons for the 18 kinematic
 distributions is shown in Figure~\ref{fig:fig_5.appb0}.
\begin{table}[p]
\begin{center}
\def\arraystretch{0.8}
\caption[]{ K-S comparison of additional kinematical variables.
            For each variable we list the observed K-S distance 
            $\delta^{0}$ and the probability $P$ of making an observation
            with a distance no smaller than $\delta^{0}$.}
\begin{tabular}{l c  c c c}
  & \multicolumn{2}{c}{ Events with a superjet} &
    \multicolumn{2}{c}{ Complementary sample} \\
  Variable                  & $\delta^{0}$ & $P$ (\%) & $\delta^{0}$ & $P$ (\%)  \\
  $\MET$                    & 0.3 1        &  27.1    & 0.14         & 57.1 \\
  $M_T^W$                   & 0.36         &  13.1    & 0.16         & 38.2 \\
  $M^{b+suj}$               & 0.36         &   4.0    & 0.12         & 58.9 \\
  $y^{b+suj}$               & 0.35         &   7.1    & 0.14         & 34.9 \\
  $E_T^{b+suj}$             & 0.28         &  24.0    & 0.10         & 60.1 \\
  $M^{l+b+suj}$             & 0.31         &  21.0    & 0.15         & 33.6 \\
  $\delta \theta^{b,suj}$   & 0.26         &  30.1    & 0.15         & 41.1 \\
  $\delta \phi^{b,suj}$     & 0.31         &  15.3    & 0.10         & 83.8 \\
  $\delta \theta^{l,b+suj}$ & 0.25         &  37.3    & 0.16         & 35.7 \\
\end{tabular}             
\label{tab:tab_5.2tris}
\end{center}
\end{table}

\newpage
\vspace*{-1.0cm}
 As indicated by the figure, the probabilities of the complementary sample
 appear to be flatly distributed, as expected for a set of distributions
 consistent with the  simulation. In contrast, the probabilities of the
 superjet events cluster at low values. This indicates the difficulty of our 
 simulation to describe the kinematics of events with a superjet. 
 Given the {\em  a posteriori} selection of the 9 kinematic variables, the
 combined statistical significance of the observed discrepancies cannot be 
 unequivocally quantified. A thorough discussion of this issue is beyond 
 the goal of this paper, which is meant to present the basic measurements.
 We leave additional studies of these events and their possible interpretation
 to other publications. The characteristics of these events are listed in 
 Appendix~B.
%
\section{Check of the isolation and lifetime of the primary and soft leptons}
\label{sec:s-lepprop}
\begin{figure}[htb]
\begin{center}
\leavevmode
\epsfxsize \textwidth
\epsffile{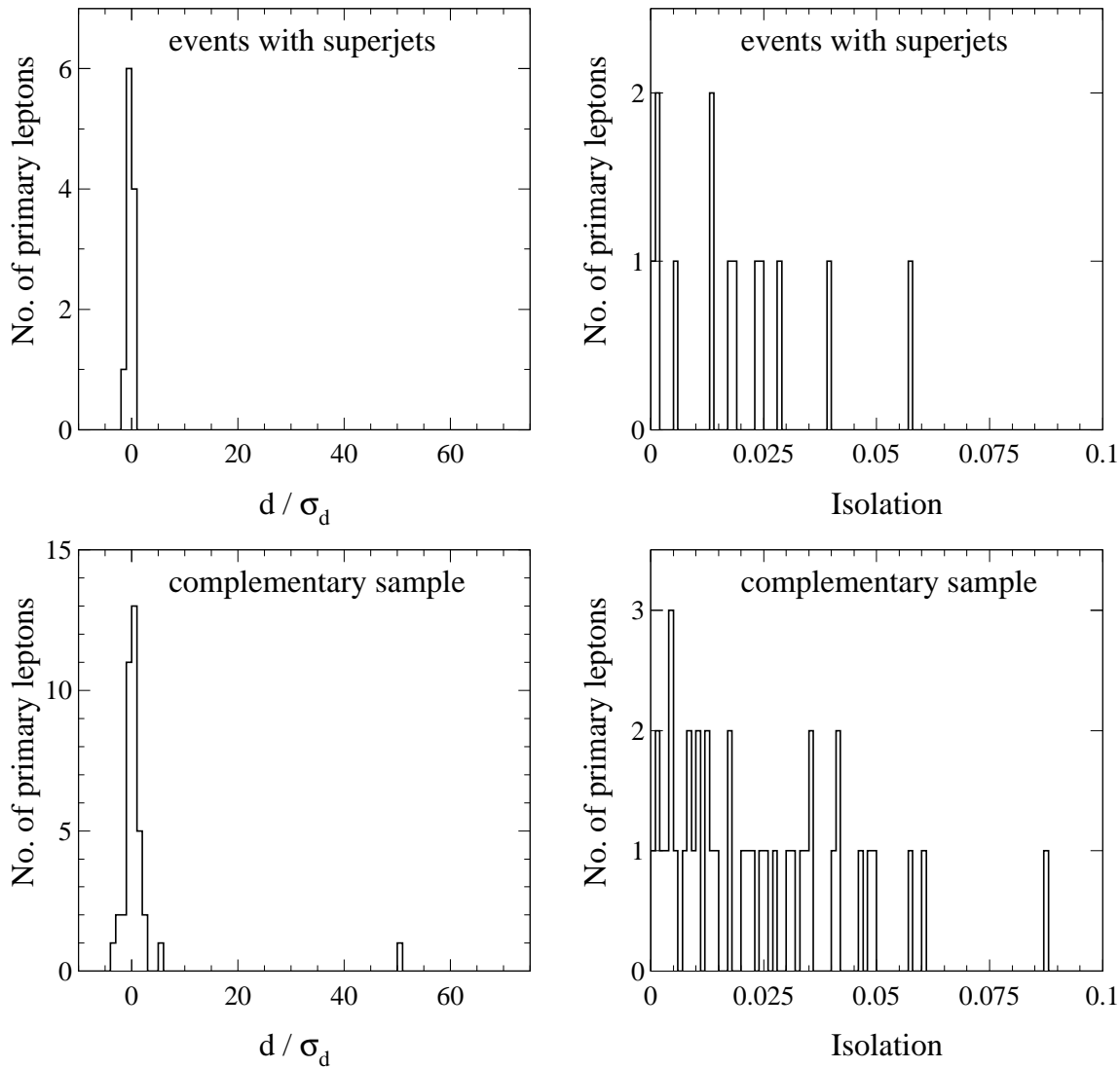}
\caption[]{Distributions of the signed impact parameter significance
           ($d/\sigma_{d}$) and of the isolation of primary leptons.}
\label{fig:fig_5.0}
\end{center}
\end{figure}

 The kinematics of the primary leptons in events with a superjet is poorly 
 described by the SM simulation, in which  they are mostly contributed
 from $\W$ decays. Therefore, we cross-check that the excess of events with
 a superjet is not due to a misestimate of the number of non-$\W$ events.
 According to the SM prediction, the small background of tagged non-$\W$ 
 events is due to semileptonic decays in $b\bar{b}$ and $c\bar{c}$ events.
 In such a case, the primary leptons are not isolated and have large
 impact parameters because of the long $b$ and $c$ quark lifetime.
 Figure~\ref{fig:fig_5.0} shows that  primary leptons in  the 13 events with a
 superjet are at least as well isolated as primary leptons in the complementary
 sample. Distributions of the signed impact parameter significance
 of the primary lepton track are also shown in Figure~\ref{fig:fig_5.0}.
 Tracks from long-lived decays usually have large ($\geq$ 3) impact parameter
 significance. The primary leptons in the 13 events are consistent with being
 prompt. One also notes that in the complementary sample two
 events have primary leptons with large positive impact parameter; this is 
 consistent with our estimate of 2.10 $\pm$ 0.05 non-$\W$ events (mostly from
 $b$-decays). 

\begin{figure}[htb]
\begin{center}
\leavevmode
\epsfxsize \textwidth
\epsffile{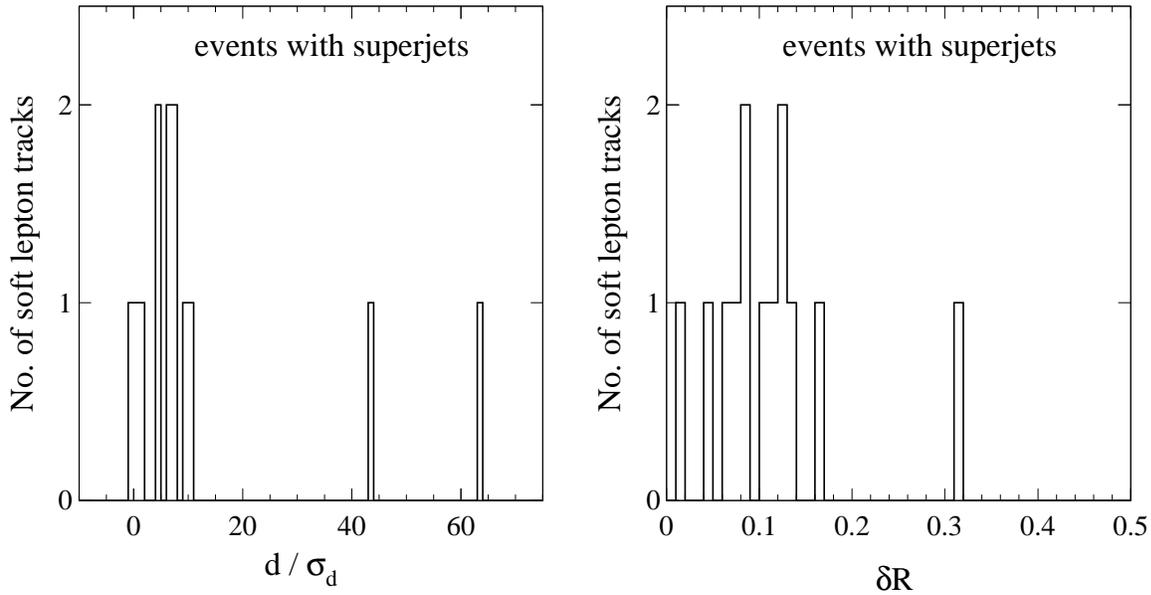}
\caption[]{Distributions of the signed impact parameter significance of soft
           lepton tracks and of their distance $\delta R=
           \sqrt{\delta \phi^{2}+ \delta \eta^{2} }$ from the superjet axis.}
\label{fig:fig_5.1}
\end{center}
\end{figure}

 Based on the SM expectation, the average transverse momenta of primary and 
 soft leptons are expected to differ by an order of magnitude (they are 
 selected with a 20 and 2 $\gevc$ transverse momentum requirement, 
 respectively). However, in the data the average transverse momenta 
 are 35 and 13 GeV/c, respectively. Since the $\W +\geq$ 1 jet sample has 
 been selected by removing all events containing a second lepton candidate
 with isolation $I \leq$ 0.15 and transverse momentum $p_T \geq$ 10 GeV/c,
 the superjets could be due to dilepton events which are not removed
 because the second lepton happens to be merged with a jet and is not isolated.
 We have removed only 16 dilepton candidate events tagged by SECVTX from the
 $\W +$ 2,3 jet sample.  From the simulation we expect that less than 0.5 events
 will have the second lepton randomly distributed in a cone of radius 0.4 
 around the axis of the jet tagged by SECVTX. Figure~\ref{fig:fig_5.1} shows
 that soft leptons are mostly found close to the superjet axis and are not
 uniformly distributed over the jet clustering cone of radius $R=0.4$. We have
 also looked at the distribution of the signed impact parameter significance
 of SLT tracks. Figure~\ref{fig:fig_5.1} shows that, in contrast with primary
 leptons, soft leptons inside a superjet are not prompt. As expected from the 
 simulation of heavy flavor decays, the soft lepton track
 is part of the SECVTX tag in 8 out of 13 superjets.
\vspace*{-1.0cm}
\clearpage
%
\section{Additional properties of the superjets}
\label{sec:s-sujprop}
 In this section we compare other properties of the superjets
 to the $\W+$ jet simulation to verify if, independent of the
 excess of soft lepton tags and the discrepancies found in Section VII, 
 they are otherwise compatible with being produced
 by semileptonic decays of $b$ and $c$ hadrons.
\subsection{Lifetime}
\label{sec:datlif}
 A measure of the lifetime of the hadron producing a secondary vertex is
 $$  {\rm pseudo-}\tau = \frac{L_{xy}}{c}
 \frac{M^{\rm SVX}}{p_T^{\rm SVX}}\; ,$$
 where $L_{xy}$ is the projection of the transverse displacement of the 
 secondary vertex on the jet-axis, $M^{\rm SVX}$ is the invariant mass and 
 $p_T^{\rm SVX}$ is the total transverse momentum of all tracks associated 
 with the secondary vertex. In this measurement, the Lorentz boost of the
 heavy flavor hadron is approximated with the Lorentz boost of the SECVTX tag.

\begin{figure}[htb]
\begin{center}
\leavevmode
\epsffile{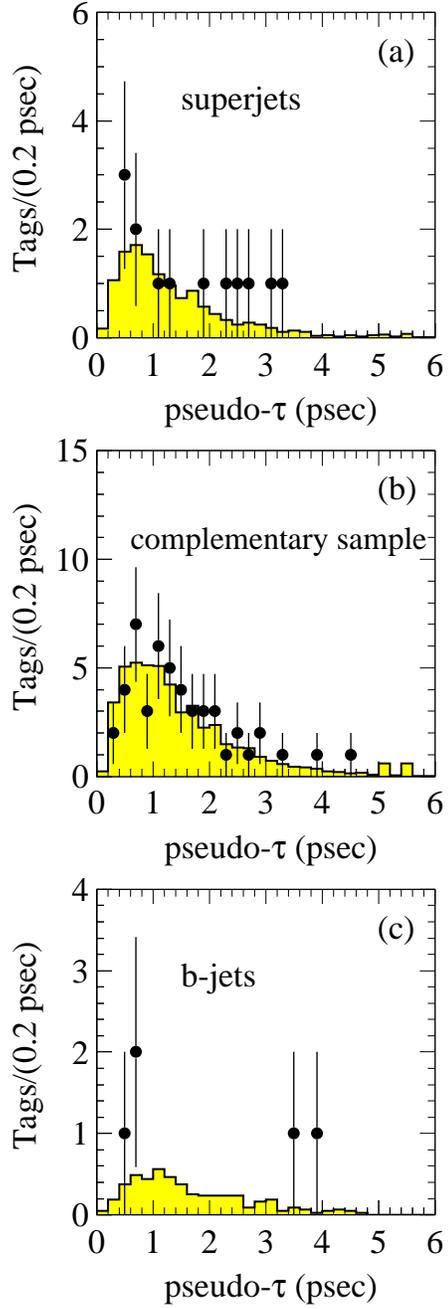}
\caption[]{Pseudo-$\tau$ distributions for superjets (a) and for tagged jets
           in the complementary sample (b) are compared to the simulation 
           (shaded histograms). The distribution for additional SECVTX tagged
           jets in superjet events (c) is compared to simulated $b$-jets.}
\label{fig:fig_9.0a}
\end{center}
\end{figure}

 Pseudo-$\tau$ distributions are compared in Figure~\ref{fig:fig_9.0a}
 to the simulation based on the sample compositions for the superjet and 
 complementary sample.
 The number of simulated superjets is rescaled to 13 events. One notes 
 that data and simulation have quite similar pseudo-$\tau$ distributions.
 The pseudo-$\tau$ calculation does not account for the neutral
 particles emitted in the heavy flavor decay. As a result a
 kinematic correction factor is needed to convert it into a lifetime 
 measurement. In the case of beauty or charmed mesons, this factor is
 approximately 1.1.

 A measure of the lifetime independent of the Lorentz boost is provided by
 $\tau_{ip} = {\displaystyle \frac{4}{\pi}} {\displaystyle \frac{<d_0 >}{c}}$,
 where $<d_0>$ is the error-weighted average impact parameter of all tracks 
 that form a SECVTX tag and have positive signed impact parameter.
 The distribution of the ratio
 $R_{\tau}={\displaystyle \frac {\tau_{ip}} { {\rm pseudo}-\tau} }$
 provides a check of the kinematic correction factor.

\begin{figure}[htb]
\begin{center}
\leavevmode
\epsffile{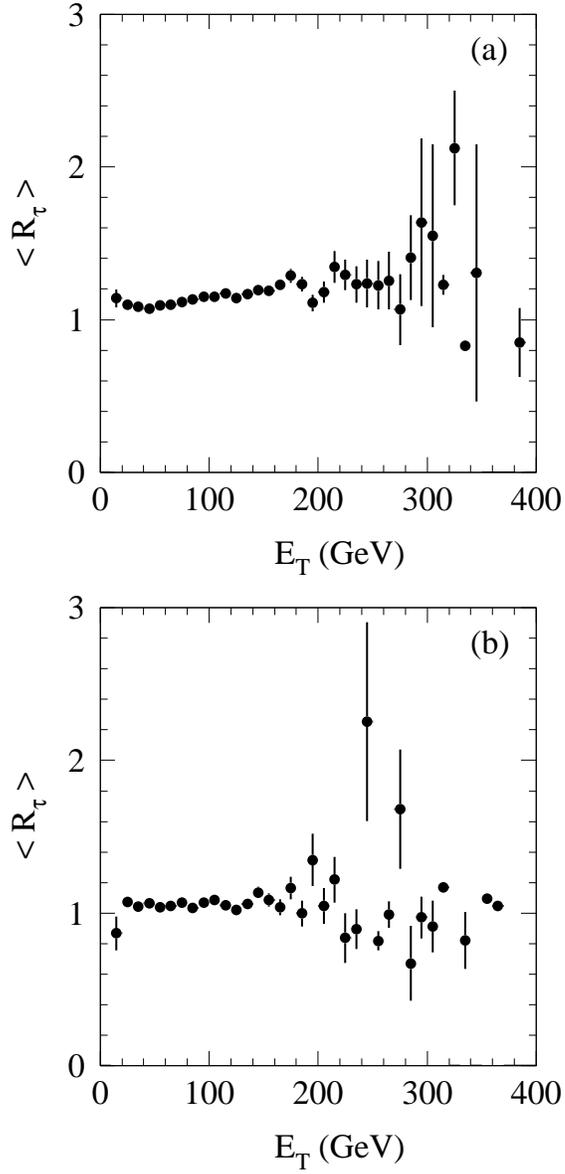}
\caption[]{ Yield of $<R_{\tau}>={\displaystyle <\frac {{\rm pseudo}-\tau} {\tau_{ip}} >}$
            as a function of the jet transverse energy in generic-jet data (a)
            and in the corresponding simulation (b).}
\label{fig:fig_9.2}
\end{center}
\end{figure}

 We first show that our simulation correctly models the correlation between 
 the lifetime measured with pseudo-$\tau$ and $\tau_{ip}$ by using the 
 generic-jet samples described in Appendix~\ref{sec:ss-jet}.  
 Figures~\ref{fig:fig_9.1} and~\ref{fig:fig_9.2} show that both methods yield
 consistent lifetime measurements in the data and in the simulation in which 
 SECVTX tags are produced by $b$ and $c$-hadrons. In this comparison, the  
 contribution of fake tags in jets without heavy flavor is removed by 
 subtracting the observed distribution of negative SECVTX tags 
 (see Section~\ref{sec:tag-alg}).

\begin{figure}[htb]
\begin{center}
\leavevmode
\epsffile{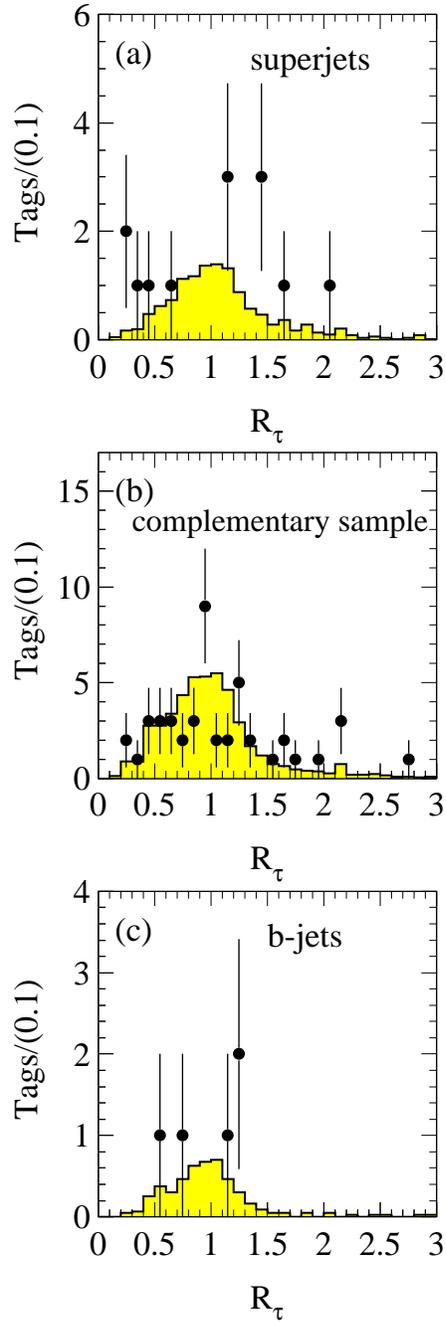}
\caption[]{Distributions of the variable $R_\tau$ (see text) for superjets (a)
           and for tagged jets in the complementary sample (b) are compared to
           the simulation (shaded histograms). The distribution for $b$-jets in
           superjet events (c) is compared to simulated $b$-jets.}
\label{fig:fig_9.0b}
\end{center}
\end{figure}

\begin{figure}[htb]
\begin{center}
\leavevmode
\epsfxsize\textwidth
\epsffile{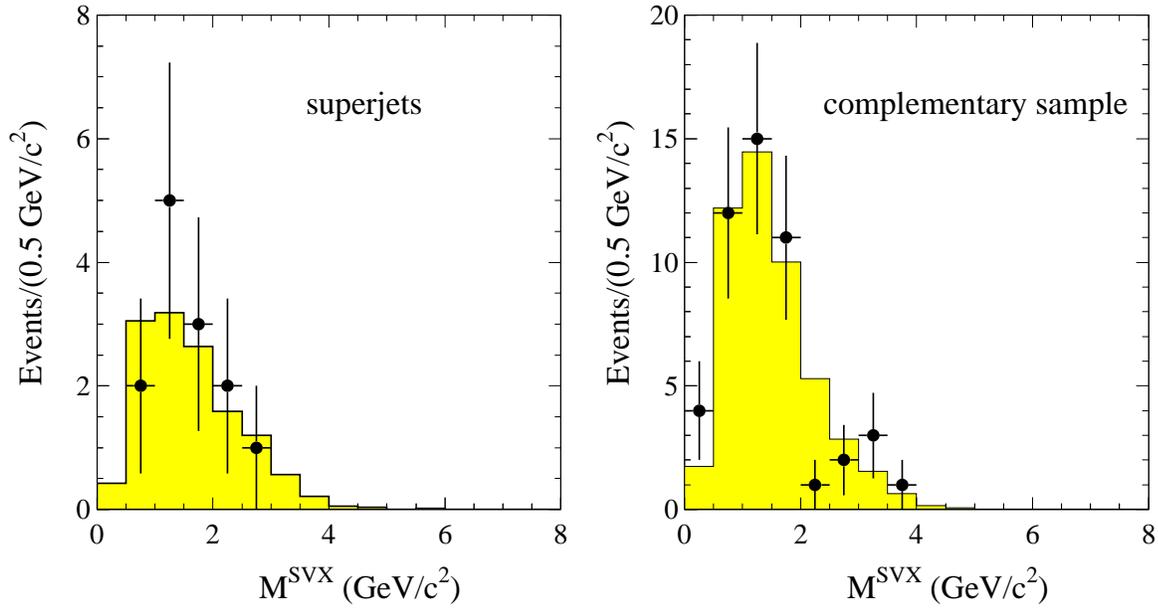}
\caption[]{Distributions of $M^{SVX}$, the invariant mass of the tracks
           associated with a secondary vertex, are compared to the simulation 
           (shaded histograms) normalized to the same number of events.}
\label{fig:fig_9.2bis}
\end{center}
\end{figure}

 Figure~\ref{fig:fig_9.0b}  presents the $R_{\tau}$ distributions
 in  superjet events and in the complementary sample.
 The result of the usual K-S comparisons (see Section~\ref{sec:ss-kolmo})
 between the data and the simulation are listed in Table~\ref{tab:tab_9.0}
 and indicate overall agreement. As shown in Figure~\ref{fig:fig_9.2bis}, 
 the distributions of the invariant mass $M^{SVX}$ are also correctly modeled
 by the simulation. The transverse momentum distribution of SECVTX tags is
 discussed in the next subsection.
\newpage
\begin{figure}[htb]
\begin{center}
\leavevmode
\epsffile{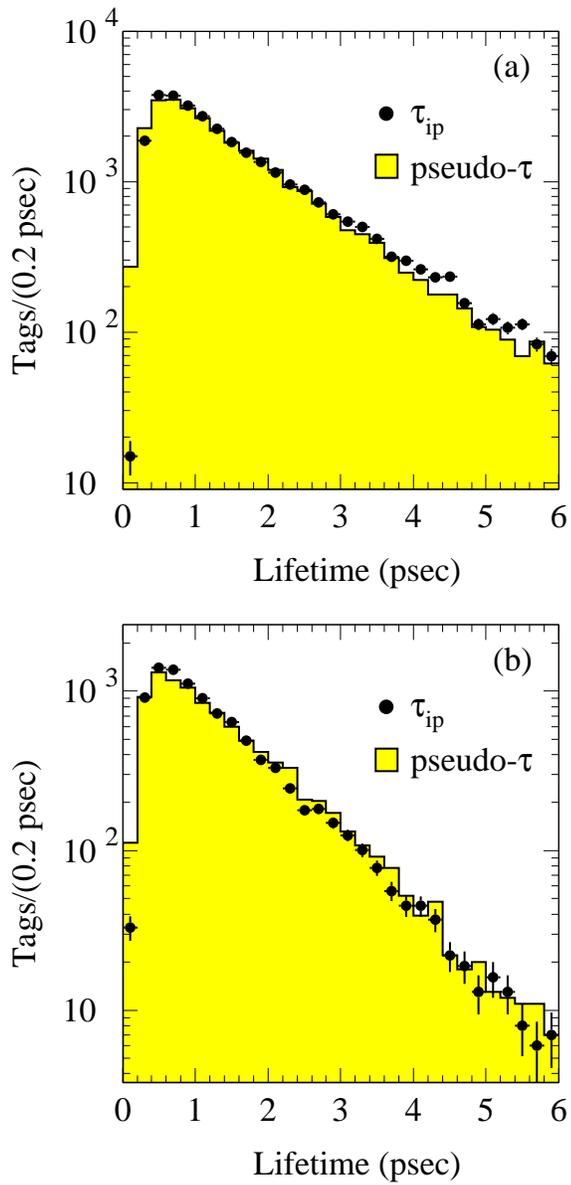}
\caption[]{Comparison of lifetime distributions using the pseudo-$\tau$
           and $\tau_{ip}$ method in generic-jet data (a) and in the
           corresponding simulation (b).}
\label{fig:fig_9.1}
\end{center}
\end{figure}
\newpage
\begin{table}[p]
\begin{center}
\def\arraystretch{0.8}
\caption[]{Result of K-S comparisons between data and simulation. For each 
           variable we list the observed K-S distance $\delta^{0}$ and the 
           probability $P$ of making an observation with a distance no 
           smaller than $\delta^{0}$.} 
\begin{tabular}{lcccc}
       & \multicolumn{2}{c}{ Events with a superjet} &
 \multicolumn{2}{c}{ Complementary sample} \\
 Variable            & $\delta^{0}$  & $P$ (\%)& $\delta^{0}$ & $P$ (\%) \\
  $R_{\tau}$ (superjets)    & 0.44   &   4.7   &    0.15      &  35.1 \\
  $R_{\tau}$ ($b$-jets)     & 0.44   &  39.0   &              &       \\
  $M^{SVX}$                 & 0.20   &  56.9   &    0.10      &  51.4 \\
  $p_T^{SLT}$               & 0.55   &   0.09  &              &       \\
  $p_T^{SVX}$               &        &         &    0.14      &  47.4 \\
\end{tabular}
\label{tab:tab_9.0}
\end{center}
\end{table}

\newpage
\subsection{Transverse momentum distribution of SLT tags}
\begin{figure}[htb]
\begin{center}
\leavevmode
\epsffile{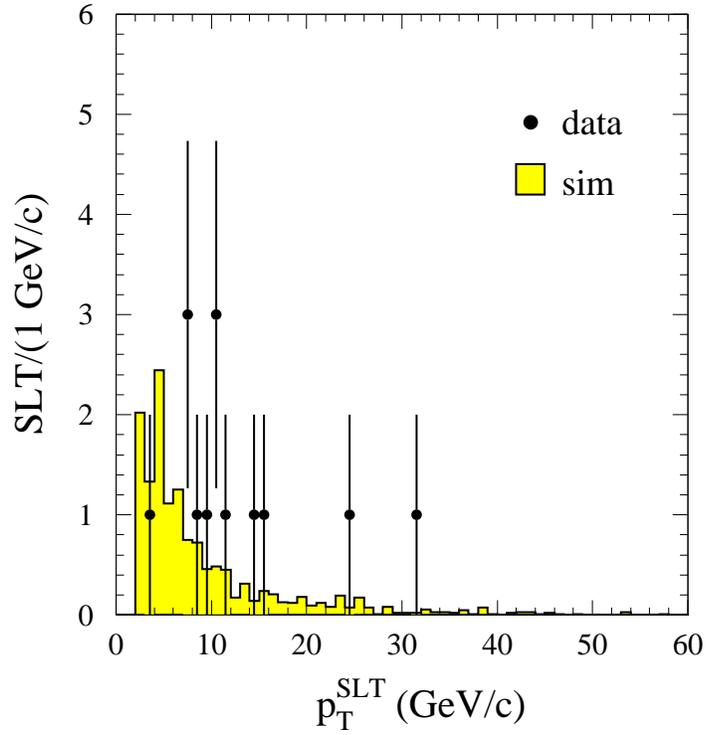}
\caption[]{The distribution of the transverse momentum of soft leptons in 
           superjet events is compared to the SM expectation normalized to the
           same number of tags and corrected for the superjet $E_T$ 
           distribution. One superjet contains two soft leptons.}
\label{fig:fig_9.3}
\end{center}
\end{figure}

\begin{figure}[htb]
\begin{center}
\leavevmode
\epsffile{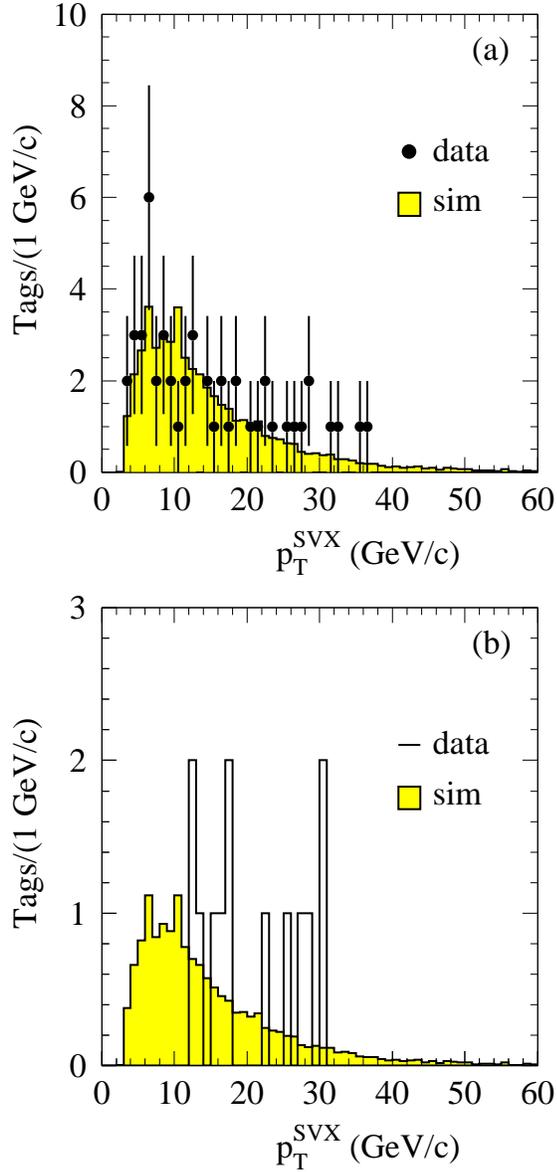}
\caption[]{Distributions of the transverse momentum of all tracks forming a 
           SECVTX tag in the complementary sample (a) and in superjets (b).}
\label{fig:fig_9.4}
\end{center}
\end{figure}

 Figure~\ref{fig:fig_9.3} compares the distribution of $p_T^{SLT}$, the soft
 lepton transverse momentum, in the 13 superjets to the simulation based on 
 the sample composition listed in Table~\ref{tab:tab_4.0}. The $p_T^{SLT}$ 
 spectrum depends on the jet transverse energy, and the superjet transverse 
 energy distribution in the data is stiffer than in the SM expectation (see
 Figure~\ref{fig:fig_5.4}). Therefore, we have corrected the transverse energy
 distribution of simulated superjets to make it look like the data. 
 Figure~\ref{fig:fig_9.3} shows that soft leptons in superjet events have 
 transverse momenta larger than what is expected for semileptonic decays of 
 $b$ and $c$-quarks. By construction the complementary sample does not contain
 soft lepton tags. However, $p_T^{SVX}$, the total transverse momentum of all
 tracks forming a SECVTX tag, is a useful analogue. If the difference between
 the transverse momentum of the soft lepton tag in the data and the simulation
 were due to inadequate modeling of the hadronization process, the $p_T^{SVX}$
 distribution in the complementary sample would also disagree with the 
 simulation. However, Figure~\ref{fig:fig_9.4}a shows agreement between the 
 complementary sample and the simulation\footnote{Since most of the SLT 
 tracks are associated with the secondary vertex, the $p_T^{SVX}$ distribution
 for superjets appears stiffer than in the complementary sample and in the
 simulation.}.
 The result of the K-S comparison of these distributions is also listed in
 Table~\ref{tab:tab_9.0}. The probability that the $p_T^{SVX}$ distribution in
 the complementary sample is produced according to the simulation is $P = 47$\%.
 The probability that the $p_T^{SLT}$ distribution in superjets is consistent
 with the SM simulation is $P = 0.1$\%.
%
\subsection{Comparison of $p_T^{SLT}$ and $p_T^{SVX}$ distributions 
            in generic-jet data to the simulation}
 We compare superjets in generic-jet data and in the corresponding simulation
 to check if the discrepancy between the observed and predicted transverse 
 momentum distribution of soft lepton tags is due to the modeling of 
 semileptonic decays in {\sc qq} or to the modeling of the hadronization in
 {\sc herwig}. The generic-jet data and simulation are described in 
 Appendix~\ref{sec:ss-jet}. The heavy flavor content of this sample is 
 similar to that of $\W+$ 2,3 jet events. We normalize data and simulation
 to the same number of events and in both we search for jets which contain
 positive and negative SECVTX tags. We then search for additional soft lepton
 tags in jets tagged by SECVTX. The data and simulation contain approximately
 the same number of supertags as a result of the calibration of the SLT  
 efficiency in the simulation (see Appendix~A). Fake SECVTX tags are evaluated
 and removed using the number of observed negative SECVTX tags in the data and
 the simulation. We do not remove the contribution of fake SLT tags from the 
 data but we add fake SLT tags to the simulation by weighting each track in a
 simulated jet with the same SLT fake probability normally used to evaluate 
 the rate of fake tags in the data.

\begin{figure}[htb]
\begin{center}
\leavevmode
\epsffile{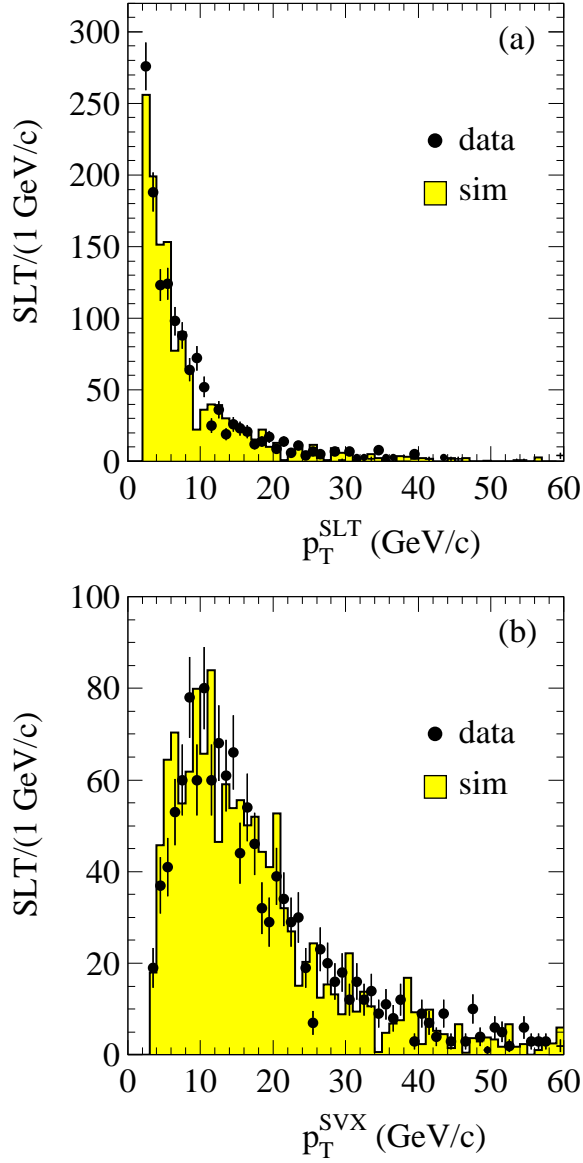}
\caption[]{ Distributions of the transverse momentum of soft leptons (a) and
            of all tracks forming the SECVTX tags (b) in superjets selected in
            generic-jet data and in the corresponding SM simulation. Data
            and simulation are normalized to the same number of events before
            tagging.}
\label{fig:fig_9.5}
\end{center}
\end{figure}

 In $5.5\times 10^{5}$ generic-jet events we find 1324 events with a supertag
 in the data and 1342 in the simulation. Distributions of the transverse
 momentum of soft lepton tags and of all tracks forming a SECVTX tag 
 are shown in Figure~\ref{fig:fig_9.5}. The agreement between data and 
 simulation provides evidence that we correctly model $b$ and $c$-jets.
%
\section{Additional cross-checks}
\label{sec:s-othersample} 
 The selection criteria used in this analysis were optimized for finding the 
 top quark~\cite{cdf-evidence}. The high-$p_T$ inclusive lepton data set, 
 from which we have selected the sample used in this study, consists of about
 82,000 events with one or more jets before making requirements on the 
 transverse momentum and isolation of the primary lepton and on the missing 
 transverse energy. Half of these events have primary leptons which are not
 well isolated ($I \geq0.2$). They are mostly due to multi-jet production with 
 one jet containing a fake lepton, but also include  a small amount of
 $b\bar{b} $ and $c\bar{c}$ production. The $p_T \geq 20\; \gevc$,
 $I \leq 0.1$ and $\MET \geq 20 \; \gev$ cuts reduce this data set to an 
 almost pure $\W +$ jet sample of about 11,000 events. In subsection~A, we
 investigate the rate of superjets in the kinematic regions removed in the 
 original selection of the $\W+ \geq$ 1 jet sample. This checks that events
 with a superjet are not the tail of a large unexpected background. In 
 subsection~B we look at the effect of removing the trigger requirement for
 primary muons and in subsection~C we extend our search to events with a
 primary electron in the plug calorimeter.
\subsection{Dependence on $\MET$, and  on the isolation and transverse 
            momentum of the primary lepton}
\begin{figure}[htb]
\begin{center}
\leavevmode
\epsffile{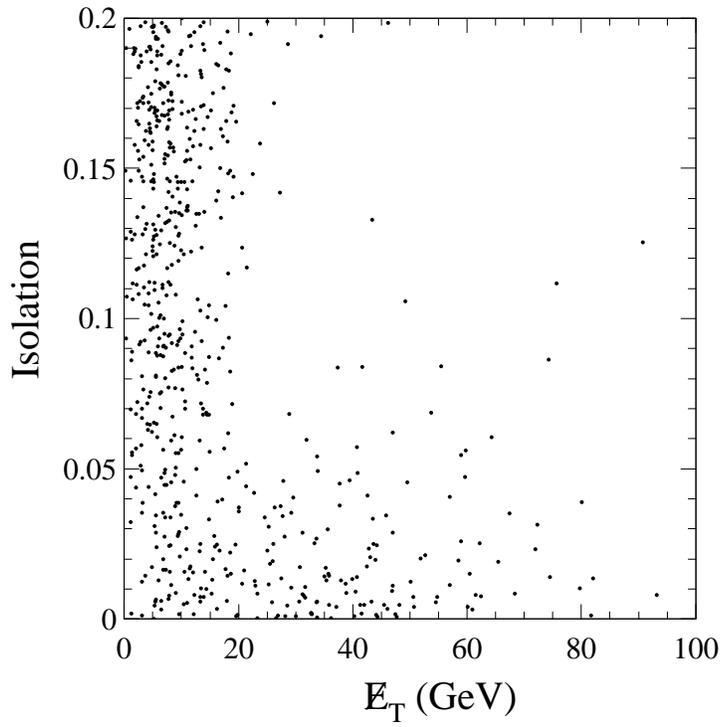}
\caption[]{Distribution of primary lepton isolation vs. $\MET$ for events
           containing one or more jets tagged by SECVTX. The primary lepton 
           transverse momentum is larger than 20 $\gevc$.}
\label{fig:fig_6.0}
\end{center}
\end{figure}

 There are 36,677 events with a primary lepton with $p_T \geq 20\; \gevc$ and
 $I \leq$ 0.2; 615 events have SECVTX tags (their $I$ vs. $\MET$ distribution
 is shown in Figure~\ref{fig:fig_6.0}). Using nominal cuts for selecting the 
 primary lepton, we first study  the rate of supertags in events tagged by
 SECVTX when  $\MET \leq 20\; \gev$. With the exception of non-$\W$ events, 
 which are the largest fraction, the relative contribution of all other SM
 processes does not depend on $\MET$. Since the ratios of supertags to SECVTX
 tags in non-$\W$ events and in the sum of the remaining processes are quite
 similar, in this case we predict the number of supertags in this sample by 
 multiplying the number of observed SECVTX tags by the predicted ratio of
 supertags to SECVTX tags for events with $\MET \geq$ 20 GeV. The observed 
 number agrees with the expectation as shown in Table~\ref{tab:tab_6.0}.

 In Table~\ref{tab:tab_6.1} we  compare rates of supertags in events
 tagged by SECVTX when the isolation of the primary lepton is large.
 These events are mostly contributed by $\bbbar$ production.
 The number of observed supertags in events with $\MET \geq 20\; \gev$ is  
 consistent with the prediction of the method  used to estimate the
 non-$\W$ background (we multiply the number of SECVTX tags in
 events with $\MET \geq 20\; \gev$ by the ratio of supertags to SECVTX tags
 in events with $\MET \leq 20\; \gev$).

 As shown in Figure~\ref{fig:fig_5.2}, many primary leptons in superjet events
 have transverse momentum close to the threshold used to select the sample.
 We have checked that we are not observing the tail of a distribution
 peaking at small transverse momenta by first removing the 20 $\gevc$ 
 transverse momentum cut on the primary lepton (the $p_T$ threshold of the
 L3 trigger is about 18 $\gevc$). Before tagging the size of the 
 $\W+$ jet sample increases by 20\%. As shown in Table~\ref{tab:tab_6.2},
 no additional events with a supertag are found.

 We then have searched for events with a superjet in the low-$p_T$ inclusive
 lepton sample collected during  the 1994-1995 collider run (Run~1B) using a
 L3 trigger threshold of 8 $\gevc$ (8 of the 13 events with a superjet were
 collected in  Run~1B). Because of the lower threshold, the trigger rate was 
 prescaled by a factor of 1.3. In this sample we find 7 events having a 
 primary lepton with $p_T \geq$ 10 $\gevc$ and $I \leq 0.1$, $\MET \geq 20$ 
 GeV, and containing a superjet and 1 or 2 additional jets. Six of the 7 
 events are the same events found in the high-$p_T$ inclusive lepton sample; 
 the additional event contains a primary electron with $E_T= 17.7$ GeV.
\newpage
\newpage
\mediumtext
\begin{table}[htb]
\begin{center}
\def\arraystretch{0.8}
\caption{ Number of tagged events as function of the jet multiplicity.
          The events are selected by requiring $\MET \leq$ 20 GeV and 
          a primary lepton with $p_T \geq 20 \; \gevc$ and $I \leq$ 0.1.
          The predicted number of supertags is based upon the observed number
          of SECVTX tags (see text).}
\begin{tabular}{lcccc}
 Tag type    &  1 jet & 2 jets &  3 jets & $\geq$ 4 jets \\
SECVTX       & 168    &   21   &     7   & 6 \\
Supertag     &  12    &    1   &     0   & 0  \\
 Prediction  & 10.2$\pm$1.3 & 1.2$\pm$0.2 &0.5$\pm$0.2 &0.5$\pm$0.2 \\
\end{tabular}
\label{tab:tab_6.0}
\end{center}
\end{table}
\widetext

\mediumtext
\begin{table}[htb]
\begin{center}
\def\arraystretch{0.9}
\caption{Yield of events with  supertags as function of the jet multiplicity.
         We select primary leptons with $p_T \geq 20 \; \gevc$ and
         isolation  $0.1 \leq I \leq 0.2$. The prediction of supertags in
         events  with $\MET \geq$ 20 GeV is derived using the ratio of
         supertags to SECVTX tags in events with $\MET \leq$ 20 GeV.}
\begin{tabular}{lcccc}
\multicolumn{5}{c}{ $\MET \leq$ 20 GeV }\\
 Tag type  &  1 jet & 2 jet &  3 jet & $\geq$ 4 jet \\
SECVTX     &   220  &   33  &   10   &      2 \\
Supertag   &    17  &    4  &    2   &      1 \\
\hline
\multicolumn{5}{c}{ $\MET \geq$ 20 GeV }\\
 Tag type  &  1 jet & 2 jet &  3 jet & $\geq$ 4 jet \\
SECVTX     &    8   &   3   &    5   &      0 \\
Supertag   &    2   &   0   &    1   &      0 \\
Prediction & 0.6 $\pm$ 0.1  & 0.4 $\pm$ 0.2  & 1.0$ \pm$ 0.7 &     0\\
\end{tabular}
\label{tab:tab_6.1}
\end{center}
\end{table}
\widetext

\mediumtext
\begin{table}[htb]
\begin{center}
\def\arraystretch{0.9}
\caption{Numbers of tagged $\W +$ jet events with $\MET \geq$ 20 GeV and  
         primary leptons with $I \leq$ 0.1 and $p_T \leq 20 \; \gevc$.}
\begin{tabular}{lcccc}
 Tag type  &   $\W+1 \,{\rm jet}$        &  $\W+2 \,{\rm jet}$        &  $\W+3 \,{\rm
 jet}$        & $\W+\geq4 \,{\rm jet}$     \\
 SECVTX       & 2 & 0 & 0 & 1  \\
 Supertag     & 0 & 0 & 0 & 0  \\
\end{tabular}
\label{tab:tab_6.2}
\end{center}
\end{table}
\widetext

\newpage
\subsection{Removal of the trigger requirement for primary muons}
 In  selecting the events used in this analysis, we require that the primary
 lepton has fired the appropriate second level (L2) trigger (see
 Section~\ref{sec:trigger}). The second level of the muon trigger requires a
 match between a CTC track reconstructed by a fast track processor~\cite{cft}
 and a track segment in the muon chambers, which fired the first level 
 trigger~\cite{cdf-evidence,cdf-tsig}. The L2 trigger efficiency for primary 
 muons is approximately 70\%~\cite{cdf-tsig}. Based on the observed 13 events
 with a superjet, we should have lost about two such events because the 
 primary muon failed the muon trigger (the detector has about the same 
 acceptance for electrons and muons). However, the original high-$p_T$ lepton
 data set contains also events triggered by other objects in the events.
 As shown in Figure~\ref{fig:fig_9.3}, 85\% of the superjets contain a soft 
 lepton with transverse momentum  comparable or larger than the 
 L2 trigger threshold. If the observed transverse momentum distribution
 of the soft leptons is  not a statistical fluctuation, we could
 find in the original data sample one or two additional events with a supertag
 in which the primary muon failed the trigger but the event was rescued by the
 soft muon. On the other hand, according to the SM simulation, only 9.6\% of 
 the $\W+$ jet events with a SLT tag contain a soft muon which passes the 
 trigger $p_T$-requirement. Using the predicted rates listed in 
 Table~\ref{tab:tab_3.3}, we estimate that: 31~$\W+$~1~jet events and 
 12~$\W+$~2,3~jet events with a primary muon have failed the trigger; 
 3~$\W+$~1~jet events and 1.1~$\W+$~2,3 jet events can be rescued by a soft
 muon. Of these events, 0.09~$\W +$~1~jet and 0.08~$\W+$~2,3 jet events are
 expected to contain a jet with a supertag.
 
 In the data, after removing the trigger requirement on the primary muon,
 we recover three $\W+$~1 jet events, none of which contains supertags.
 We also recover one $\W+$ 2 jet and one $\W+$ 3 jet event, both with a 
 supertag. No extra $\W+$ 4 jet event is found.
 The characteristics of these two events are listed in Appendix~B.
\subsection{Study of plug electrons}
\label{sec:plugel}
\begin{figure}[htb]
\begin{center}
\leavevmode
\epsfxsize \textwidth
\epsffile{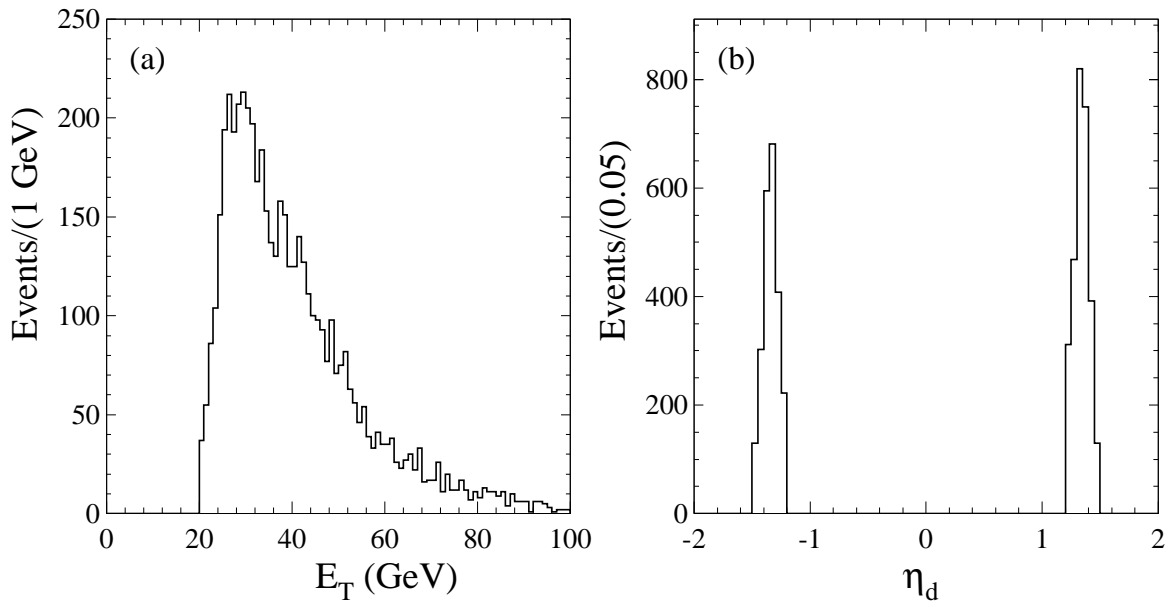}
\caption[]{Distributions of the transverse momentum and the pseudo-rapidity 
          with respect to the nominal interaction point of plug electrons.}
\label{fig:fig_6.1}
\end{center}
\end{figure}

 As shown in Figure~\ref{fig:fig_5.3}, the pseudo-rapidity distribution of 
 primary leptons in events with a superjet appears to rise at the end of the
 central detector acceptance ($|\eta| \simeq 1$). Motivated by this 
 observation, we have searched for events with a superjet using primary 
 electrons in the plug calorimeter. The pseudo-rapidity and transverse
 momentum distributions of plug electrons are shown in Figure~\ref{fig:fig_6.1}.
 We select $\W+$ jet events requiring an isolated plug electron with 
 $E_T \geq 20 \; \gev$ and $\MET \geq 20 \; \gev$.

 Table~\ref{tab:tab_6.4} lists rates of $\W+$ jet events with a primary plug 
 electron before and after tagging. We observe two additional $\W+$ 2,3 jet 
 events with a supertag, when 0.34 $\pm$ 0.04 events are expected from known 
 processes. The characteristics of these two additional events with a superjet
 are listed in Appendix~B.
\vspace*{-1.0cm}
\clearpage
\newpage
\begin{table}[htb]
\begin{center}
\def\arraystretch{0.8}
\caption{Number of events with an isolated plug electron and $\MET \geq$ 20 
         GeV before and after tagging. Since the relative contributions of
         different processes are not affected by the difference in the 
         pseudo-rapidity range covered by central leptons and plug electrons,
         the prediction of supertags is derived from Table~\ref{tab:tab_4.0}
         after normalizing to the same number of SECVTX tags.}
\begin{tabular}{lcccc}
 Source         &  $\W+1 \,{\rm jet}$        &  $\W+2 \,{\rm jet}$        &  $\W+3 \,{\rm jet}$        &
$\W+\geq4 \,{\rm jet}$     \\
\hline
 Data            & 1245 & 243 & 52 & 11\\
SECVTX tags        & 15 & 3 & 1 & 1 \\
Supertags    &  3          &  2          & 0         &      0  \\
SM prediction   & 0.9 $\pm$ 0.1 & 0.24 $\pm$ 0.03 & 0.10 $\pm$ 0.02 &0.10 $\pm$ 0.03 \\
\end{tabular}
\label{tab:tab_6.4}
\end{center}
\end{table}

%
 \section{Conclusions}
 \label{sec:s-concl}
 We have carried out a study of the heavy flavor content of jets produced in
 association with $\W$ bosons. Comparisons of the observed rates of SECVTX 
 (displaced vertex) and SLT (soft lepton) tags with standard model predictions,
 including NLO calculations of single and pair produced top quarks, are 
 generally in good agreement. However, we find an excess of events which have
 jets with both SECVTX and SLT heavy flavor tags. The standard model 
 expectation for these $\W+$ 2,3 jet events is $4.4 \pm 0.6$ events, while 13
 are observed. A detailed examination of the kinematic properties of these 
 events finds that they are statistically difficult to reconcile with a 
 simulation of standard model processes, which well reproduces closely related
 samples of data. Although obscure detector effects can never be ruled out,
 extensive studies of these events and investigations of larger statistics 
 samples of generic-jet data have not revealed any effects which indicate
 the existence of detector problems  or simulation deficiencies. We are not
 aware of any model for new physics which  incorporates the production and
 decay properties necessary to explain all features of these events. Work is
 continuing on studies of the present data. With much larger data samples from
 the Run II of the Tevatron, we will be able to explore in greater detail
 this class of events.
\acknowledgments
 We thank the Fermilab staff and the technical staff of the participating
 Institutions for their contributions. This work was supported by the 
 U.S.~Department of Energy and National Science Foundation; the Istituto 
 Nazionale di Fisica Nucleare; the Ministry of Education, Culture, Sports,
 Science and Technology of Japan; the Natural Science and Engineering Research
 Council of Canada; the National Science Council of the Republic of China; 
 the Swiss National Science Foundation; the A.P.~Sloan Foundation; 
 the Bundesministerium f\"{u}r Bildung und Forschung; the Korea Science and
 Engineering Foundation (KoSEF); the Korea Research Foundation; and the
 Comision Interministerial de Ciencia y Tecnologia, Spain. 
\appendix
\section{Comparison of rates of supertags in generic-jet data and in the
         corresponding simulation}
\label{sec:ss-jet}
 Table~\ref{tab:tab_4.1} lists rates of tags in generic-jet data and in the
 corresponding simulation. This comparison profits from the measurement of the
 heavy flavor composition of generic-jet data and of the calibration of the
 {\sc herwig} generator presented in Ref.~\cite{cdf-tsig}. A summary of that 
 study is provided here. Generic-jet data are events collected by requiring
 at least one jet with transverse energy above trigger threshold (i.e. a 20
 GeV threshold for JET~20 data). As usual we consider jets with $E_T \geq$ 15
 GeV and pseudo-rapidity $|\eta| \leq$ 2. We apply the additional requirement
 that at least one of the jets in the event contains two SVX tracks and is 
 therefore taggable by SECVTX or JPB. An equal number of $2 \rightarrow  2$ 
 hard-scattering events is simulated using option 1500 of the {\sc herwig} 
 generator and the MRS~(G) parton distribution functions~\cite{mrsg}. In the
 simulation, jets with heavy flavor come from heavy quarks in the initial or 
 final state of the hard scattering (flavor excitation and direct production)
 or from gluon splitting. A 13.2\% fraction of the simulated jets contains 
 heavy flavor (4.7\% due to $b$-hadrons and 8.5\% due to  $c$-hadrons).
 A 3.5\% fraction of the simulated jets contains heavy flavor and is tagged 
 by SECVTX (73\% of the tagged jets are initiated by a $b$-quark and 27\%
 by a $c$-quark).
 Jet-probability is more efficient than SECVTX in tagging $c$-jets.
 A 4.6\% fraction of the simulated jets contains heavy flavor and
 is tagged by jet-probability (55\% of the tagged jets are initiated by
 a $b$-quark and 45\% by a $c$-quark).

 The  heavy flavor production  cross sections calculated by {\sc herwig}
 have been tuned in Ref.~\cite{cdf-tsig} to reproduce the pattern of
 SECVTX and JPB  tags  observed in generic-jet data. {\sc herwig} gives a 
 good description of the data provided that the direct and flavor excitation
 production cross sections are increased by 1.10 $\pm$ 0.16 and the fraction
 of gluons branching to heavy quarks is increased by 1.36 $\pm$ 0.22. The
 accuracy of this calibration is limited by our understanding of the tagging
 efficiencies. The factors required to calibrate simulated rates of SECVTX or
 JPB tags are determined more accurately: 1.1 $\pm$ 0.1 for direct and flavor
 excitation production and 1.38 $\pm$ 0.09 for gluon splitting.

 Table~\ref{tab:tab_4.1} shows agreement also between the number of jets with
 heavy flavor tagged by the SLT algorithm in the data and simulation (the SLT
 algorithm was not used to calibrate the simulation). However the numbers of 
 SLT tags in the data have large errors because  the ratio of tags due to 
 heavy flavor to mistags is about 1/5. For jets with a supertag (SECVTX+SLT 
 or JPB+SLT) the ratio of tags due to heavy flavor to mistags  is about 2/1,
 and this allows a good calibration of the efficiency for finding supertags
 in the simulation. We compare ratios of supertags to SECVTX (JPB) tags in 
 the data and the simulation in order to cancel the contribution of the 
 uncertainty of the simulated SECVTX (JPB) algorithms.
 Efficiencies for finding SLT tags in jets already tagged by SECVTX or JPB
 are listed in Table~\ref{tab:tab_4.2}. We find that the efficiency for
 finding supertags in the data is (84 $\pm$ 5)\% of the simulated efficiency.
 The small differences in the tagging efficiency between data and simulation
 in Table~\ref{tab:tab_4.2} do not seem to be caused by a particular flavor
 type, because the relative fractions of $b$ and $c$-quarks are quite different
 in jets tagged by SECVTX and jet-probability. The uniformity of the
 data-to-simulation scale factor for finding supertags across the three 
 independent generic-jet samples also excludes any large dependence on the jet
 transverse energy. If we combine these three samples, we find that the 
 efficiency for finding supertags in the data is (85 $\pm$ 5)\% of the simulated
 efficiency for SECVTX tags and (86 $\pm$ 7)\% for JPB tags.
 Since the heavy flavor composition of generic-jet data with a SECVTX tag 
 (73\% $b$-quarks and 27\% $c$-quarks) is very similar to the composition
 of $\W +\geq$ 2,3 jet events with a SECVTX tag, the excess of $\W+$ 2,3 jet
 events with a supertag cannot be explained by correlations between the SLT
 and SECVTX algorithms unaccounted for by the simulation.
\newpage
\begin{table}[p]
\begin{center}
\def\arraystretch{0.9}
\caption[]{Number of tags due to heavy flavors observed in generic-jet data and
           in the simulation normalized to the same number of events before
           tagging. The amount of mistags removed from the data is indicated
           in parenthesis; errors include a 10\% uncertainty in the mistag  
           evaluation. The error of the number of simulated SLT tags includes 
           the 10\% uncertainty on the SLT tagging efficiency. This error is 
           not included for simulated  SECVTX+SLT and JPB+SLT tags as we
           intend to calibrate the simulation efficiency with the data.}
\begin{tabular}{lcc}
  \multicolumn{3}{c}{ JET 20 (194,009 events) }\\
 Tag type  & Data (removed fakes) & Simulation \\
SECVTX      & 4058$\pm$92 (616.0)        &  4052$\pm$143   \\
JPB         & 5542$\pm$295 (2801.0)      &  5573$\pm$173   \\
SLT         & 1032$\pm$402 (3962.0)      &  826$\pm$122    \\
SLT+SECVTX  & 219.8$\pm$20  (94.2)   &  263$\pm$29     \\
SLT+JPB     & 287.3$\pm$28  (166.7)  &  330$\pm$29     \\
\hline
  \multicolumn{3}{c}{ JET 50 (151,270 events) }\\
 Tag type  & Data (removed fakes) & Simulation \\
SECVTX      & 5176$\pm$158 (1360.0)       & 5314$\pm$142 \\
JPB         & 6833$\pm$482 (4700.0)       & 6740$\pm$171 \\
SLT         & 1167$\pm$530 (5241.0)       & 1116$\pm$111 \\
SLT+SECVTX  & 347$\pm$29 (169.0)          & 404$\pm$22  \\
SLT+JPB     & 427.5$\pm$42 (288.5)        & 490$\pm$32  \\
\hline
\multicolumn{3}{c}{ JET 100 (129,434 events) }\\
 Tag type  & Data (removed fakes) & Simulation \\
SECVTX      & 5455$\pm$239 (2227.0)       & 5889$\pm$176  \\
JPB         &  6871$\pm$659 (6494.0)      & 7263$\pm$202 \\
SLT         & 1116$\pm$642 (6367.0)       & 1160$\pm$168   \\
SLT+SECVTX  & 377.6$\pm$36 (243.4)  & 508$\pm$35     \\
SLT+JPB     & 451.8$\pm$55 (401.2)   & 563$\pm$34     \\
\end{tabular}
\label{tab:tab_4.1}
\end{center}
\end{table}

\newpage
\begin{table}[p]
\begin{center}
\caption[]{Fractions of SECVTX and JPB tags with a supertag in generic-jet 
           data and in the corresponding simulation. In the simulation the 
           fraction of supertags is slightly higher than in the data,
           independent of the jet transverse energy and the heavy flavor type.}
\def\arraystretch{1.2}
{\footnotesize
\begin{tabular}{lcccccc}
  & \multicolumn{2}{c}{ JET 20} & \multicolumn{2}{c}{ JET 50}
  & \multicolumn{2}{c}{ JET 100}\\
  & $\frac {{\rm SLT+SECVTX} }{{\rm SECVTX}}$& $\frac {{\rm SLT+JPB}}{{\rm JPB}} $ &
$\frac {{\rm SLT+SECVTX}}{{\rm SECVTX}}$& $\frac {{\rm SLT+JPB} }{{\rm JPB}} $ &
$\frac {{\rm SLT+SECVTX}}{{\rm SECVTX}}$& $\frac {{\rm SLT+JPB}}{{\rm JPB}} $\\
Data     & 0.054$\pm$0.005 &0.052$\pm$0.006 &0.067$\pm$0.006&0.063$\pm$0.008&
0.069$\pm$0.007 &0.066$\pm$0.010\\
Sim.      & 0.065$\pm$0.007& 0.059$\pm$0.005&0.076$\pm$0.004&0.073$\pm$0.005&
0.086$\pm$0.006&0.077$\pm$0.005\\
Data/Sim. &0.83$\pm$0.12&0.88$\pm$0.13&0.88$\pm$0.09&0.86$\pm$0.12
&0.80$\pm$0.10&0.86$\pm$0.14\\
\end{tabular}                              
}
\label{tab:tab_4.2}
\end{center}
\end{table}

\section{Characteristics of the events with a superjet}
 Tables~\ref{tab:tab_5.3} and~\ref{tab:tab_5.3bis} list the characteristics of
 the 13 events with a superjet. Four of these events are included in the data
 set used to measure the top quark mass~\cite{blusk}.

 Event 41540/127085 in Table~\ref{tab:tab_5.3} is classified in 
 Ref.~\cite{blusk} as a dilepton event. In the present analysis, which uses 
 tighter lepton selection criteria, the muon candidate appears to be due to 
 punch-through of a stiff track inside the jet with $E_T= 144.5$ GeV. The 
 fit of this event yields a top quark mass $M_{\rm top}$ = 158.8 $\gevcc$.

 The other three events (65581/322592, 67824/281883 and 56911/114159  in Table
 ~\ref{tab:tab_5.3bis}) contain an additional jet with $E_T\geq$ 8 GeV and
 $|\eta| \leq 2.4$. The fit of these events in Ref.~\cite{blusk} yields $M_{\rm
 top}$ = 152.7, 170.1 and 156.7 $\gevcc$, respectively.

 Table~\ref{tab:tab_6.3} lists the characteristics of the two events found by 
 removing the L2 trigger requirement for primary muons. The characteristics of
 the two additional events with a superjet and a primary plug electron are
 listed in Table~\ref{tab:tab_6.5}.
\newpage
\mediumtext
{\tightenlines \scriptsize
\begin{table}[p]
\begin{center}
\caption[]{Characteristics of $\W+$ 2 jet events with a superjet. Jets tagged 
           by the SECVTX (SLT) algorithm are labeled SECVTX (SLT). Jet energies
           are corrected for calorimeter non-linearities and out-of-cone losses;
           $\met$ is evaluated after these corrections are applied.}
\begin{tabular}{lcrcp{0.2cm}lcrc}
           & \multicolumn{1}{c}{$p_T$ ($\gevc$)}
           & \multicolumn{1}{c}{$\eta$}
           & \multicolumn{1}{c}{$\phi$ (rad)}
           &
           &
           & \multicolumn{1}{c}{$p_T$ ($\gevc$)}
           & \multicolumn{1}{c}{$\eta$}
           & \multicolumn{1}{c}{$\phi$ (rad)} \\
\hline
Run $46\,935$ event $266\,805$ &       &        &        & &   Run $41\,540$ event
$127\,085$ &       &       &      \\
electron (-)                   & 29.7  & -0.87  & 0.15   & &   electron (-)                  
& 22.2  & 0.84  & 0.57 \\
Jet 1                          & 49.6  & -0.61  & 5.46   & &   Jet 1  (SECVTX,SLT)              
&144.5  & 0.11  & 6.15 \\
Jet 2 (SECVTX,SLT)                & 41.1  &  0.43  & 2.70   & &   Jet 2                         
& 61.5  &-0.54  & 3.75 \\
$\MET$                         & 19.8  &        & 2.56   & &   $\MET$                        
& 92.1  &       & 3.05 \\
SLT ($\mu^-$)                  &\phantom{0}3.8  & 0.52   & 2.63   & &   SLT ($\mu^+$)                 
&  \phantom{0}8.8  & 0.18  & 6.14 \\
$Z_{\rm vrtx}$ (cm)                     &-20.71 &        &        & &   $Z_{\rm vrtx}$ (cm)                    
&-4.77  &       &      \\
Run $41\,627$ event $87\,219$  &       &        &        & &   Run $61\,167$ event
$368\,226$ &       &       &      \\
electron (-)                   & 78.5  & 0.90   & 4.56   & &   electron (+)                  
& 22.2  & 0.76  & 1.37 \\
Jet 1                          & 68.7  & 0.11   & 3.03   & &   Jet 1 (SECVTX,SLT)               
& 99.3  & -0.16 & 1.86 \\
Jet 2  (SECVTX,SLT)               & 58.0  & 0.50   & 1.23   & &   Jet 2 (SECVTX)                   
& 68.1  & 0.93  & 5.48 \\
$\MET$                         & 47.4  &        & 0.23   & &   $\MET$                        
& 36.0  &       & 3.61 \\
SLT ($\mu^-$)                  & 10.4  & 0.47   & 1.26   & &   SLT ($\mu^-$)                 
& 24.7  &-0.11  & 1.92 \\
$Z_{\rm vrtx}$ (cm)                     & -28.11&        &        & &   $Z_{\rm vrtx}$ (cm)                    
&-14.20 &       &      \\
Run $65\,384$ event $266\,051$ &       &       &         & &   Run $65\,741$ event
$654\,870$ &       &       &      \\
electron (-)                   & 21.9  & 0.68  & 0.65    & &   muon (+)                      
& 47.2  & 0.79  & 6.01 \\
Jet 1                          & 73.9  & 2.06  & 0.33    & &   Jet 1 (SECVTX,SLT)               
&109.4  & 0.63  & 4.58 \\
Jet 2 (SECVTX,SLT)                & 59.0  & 0.61  & 4.92    & &   Jet 2 (SECVTX)                   
& 63.9  & 0.31  & 2.87 \\
$\MET$                         & 96.2  &       & 3.02    & &   $\MET$                        
& 95.8  &       & 1.31 \\
SLT ($\mu^+$)                  & 10.9  & 0.61  & 4.80    & &   SLT ($e^+$)                   
& \phantom{0}7.1   & 0.76  & 4.61 \\
$Z_{\rm vrtx}$ (cm)                     &-24.24 &       &         & &   $Z_{\rm vrtx}$ (cm)                    
&- 14.20&       &      \\
Run $46\,357$ event $511\,399$ &       &        &        & &   Run $69\,520$ event
$136\,405$ &       &      &       \\
muon (-)                       & 22.2  & -0.82  & 5.64   & &   electron (-)                  
& 20.4  & 1.01 & 0.25  \\
Jet 1                          & 58.2  & -0.20  & 6.10   & &   Jet 1                         
& 44.2  &-0.61 & 5.57  \\
Jet 2 (SECVTX,SLT)                & 41.2  &  0.27  & 2.84   & &   Jet 2 (SECVTX,SLT)               
& 32.7  &-0.88 & 2.71  \\
$\MET$                         & 39.8  &        & 2.89   & &   $\MET$                        
& 27.5  &      & 2.42  \\
SLT ($\mu^+$)                  & 15.2  &  0.25  & 2.96   & &   SLT ($\mu^+$)                 
& 11.3  &-0.87 & 2.71  \\
SLT ($e^-$)                    &  \phantom{0}7.1  &  0.38  & 2.89   & &   $Z_{\rm vrtx}$ (cm)                    
&-12.36 &      &       \\
$Z_{\rm vrtx}$ (cm)                     &-24.13 &        &        & &                                 
&       &      &       \\
\end{tabular}
\label{tab:tab_5.3}
\end{center}
\end{table}
}
\widetext

\newpage
\mediumtext
{\tightenlines \scriptsize
\begin{table}[p]
\begin{center}
\caption[]{Characteristics of the $\W+$ 3 jet events with a superjet. Jets 
           tagged by the SECVTX (SLT) algorithm are labeled SECVTX (SLT). Jet
           energies are corrected for calorimeter non-linearities and 
           out-of-cone losses; $\met$ is evaluated after these corrections 
           are applied.}
\begin{tabular}{lcrcp{0.2cm}lcrc}
           & \multicolumn{1}{c}{$p_T$ ($\gevc$)}
           & \multicolumn{1}{c}{$\eta$}
           & \multicolumn{1}{c}{$\phi$ (rad)}
           &
           &
           & \multicolumn{1}{c}{$p_T$ ($\gevc$)}
           & \multicolumn{1}{c}{$\eta$}
           & \multicolumn{1}{c}{$\phi$ (rad)} \\
\hline
Run $56\,911$ event $114\,159$ &       &        &        & &   Run $61\,548$ event
$284\,898$ &       &      &       \\
electron (-)                   & 58.5  & 0.92   & 0.83   & &   muon (+)                      
& 20.3  & -0.54& 3.00  \\
Jet 1                          & 203.4 & -0.13  & 2.93   & &   Jet 1                         
& 72.4  &  0.55& 1.96  \\
Jet 2 (SECVTX,SLT)                & 65.5  & 0.82   & 5.80   & &   Jet 2 (SECVTX)                   
& 64.9  &  0.44& 3.94  \\
Jet 3                          & 24.1  & 0.60   & 0.00   & &   Jet 3 (SECVTX,SLT)               
& 58.7  &  0.07& 5.73  \\
$\MET$                         & 61.5  &        & 5.41   & &   $\MET$                        
& 38.8  &      & 0.02  \\
SLT ($\mu^+$)                  &\phantom{0}9.3  & 0.77   & 5.75   & &   SLT ($e^-$)                   
& 14.6  &  0.09& 5.83  \\
$Z_{\rm vrtx}$ (cm)                     &-13.89 &        &        & &   $Z_{\rm vrtx}$ (cm)                    
&16.38  &      &       \\
Run $65\,581$ event $322\,592$ &       &        &        & & Run $67\,824$ event
$281\,883$ &       &        &       \\
muon (-)                       & 21.4  & 0.57   & 6.00   & & electron (+)                  
& 52.3  & -0.16  & 3.64  \\
Jet 1 (SECVTX)                    & 146.3 & -0.56  & 1.21   & & Jet 1 (SECVTX)                   
& 78.8  & -0.49  & 0.90  \\
Jet 2 (SECVTX,SLT)                &  65.8 & 0.51   & 3.38   & & Jet 2                         
& 66.3  &  0.69  & 5.83  \\
Jet 3                          &  29.7 & 1.50   & 4.68   & & Jet 3 (SECVTX,SLT)               
& 55.8  &  0.68  & 2.09  \\
$\MET$                         &  70.2 &        & 3.78   & & $\MET$                        
& 57.6  &        & 4.30  \\
SLT ($\mu^-$)                  & 31.3  & 0.58   & 3.34   & & SLT ($\mu^-$)                 
&\phantom{0}7.2  &  0.88  & 1.97  \\
$Z_{\rm vrtx}$ (cm)                     & 5.54  &        &        & & $Z_{\rm vrtx}$ (cm)                    
&-10.56 &        &       \\
Run $46\,818$ event $221\,912$ &       &       &         & &  & & & \\
muon (-)                       & 48.2  & 1.02  & 2.36    & &  & & & \\
Jet 1  (SECVTX,SLT)               & 55.4  &-0.02  & 2.96    & &  & & & \\
Jet 2                          & 41.7  & 0.27  & 5.08    & &  & & & \\
Jet 3                          & 35.3  & 0.82  & 5.68    & &  & & & \\
$\MET$                         & 22.3  &       & 0.30    & &  & & & \\
SLT ($\mu^+$)                  & 10.5  & 0.06  & 2.93    & &  & & & \\
$Z_{\rm vrtx}$ (cm)                     &-17.28 &       &         & &  & & & \\

\end{tabular}
\label{tab:tab_5.3bis}
\end{center}
\end{table}
}
\widetext

\newpage
\narrowtext
{\tightenlines \small
\begin{table}[p]
\begin{center}
\caption[]{Characteristics of the  $\W+$ 2 jet events with a superjet
           rescued by removing the L2 trigger requirement.}
\begin{tabular}{lrrc}
           & \multicolumn{1}{c}{$p_T$ ($\gevc$)}
           & \multicolumn{1}{c}{$\eta$}
           & \multicolumn{1}{c}{$\phi$ (rad)} \\
\hline
 Run $61\,525$ event $116\,807$ &       &        &     \\
 muon (+)                       & 50.5  &  0.48  & 0.58\\
 Jet 1 (SECVTX,SLT)             & 66.3  &  0.10  & 4.45\\
 Jet 2                          & 36.8  & -0.71  & 1.87\\
 $\MET$                         & 22.2  &        & 4.30 \\
 SLT ($\mu^-$)                  & 11.2  & 0.11   & 4.36\\
 $Z_{\rm vrtx}$ (cm)            & 5.72  &        &     \\
 Run $68\,592$ event $250\,386$ &       &        &     \\
 muon (-)                       & 57.5  & -0.07  & 4.69\\
 Jet 1                          & 60.6  & -1.08  & 4.09\\
 Jet 2 (SECVTX,SLT)             & 42.5  & -0.17  & 1.44\\
 Jet 3                          & 32.5  &  1.58  & 0.97\\
 $\MET$                         & 36.1  &        & 1.12 \\
 SLT ($\mu^+$)                  &  7.9  & -0.21  & 1.42\\
 $Z_{\rm vrtx}$ (cm)            &14.48  &        &  \\
\end{tabular}
\label{tab:tab_6.3}
\end{center}
\end{table}
}
\widetext

\narrowtext
{\tightenlines \small
\begin{table}[hb]
\begin{center}
\caption[]{Characteristics of the  $\W+$ 2,3 jet events with a superjet
           found in the plug electron sample.}
{\small
\begin{tabular}{lrrc}
           & \multicolumn{1}{c}{$p_T$ ($\gevc$)}
           & \multicolumn{1}{c}{$\eta$}
           & \multicolumn{1}{c}{$\phi$ (rad)} \\
\hline
 Run $69\,941$ event $66\,919$ &       &        &     \\
 electron (-)                   & 43.4  &-1.33   & 0.77\\
 Jet 1 (SECVTX,SLT)             & 84.5  &-0.12   & 4.09\\
 Jet 2                          & 50.7  & 1.99   & 1.29\\
 $\MET$                         & 11.6  &        & 4.53 \\
 SLT ($\mu^+$)                  & 13.5  & -0.09  & 4.06\\
 $Z_{\rm vrtx}$ (cm)            &16.00  &        &     \\
 Run $58\,202$ event $109\,847$ &       &        &     \\
 electron (+)                   & 65.9  & 1.45   & 1.43\\
 Jet 1                          & 32.6  & 0.28   & 4.84\\
 Jet 2 (SECVTX,SLT)             & 30.8  &-0.75   & 4.38\\
 $\MET$                         & 12.5  &        & 4.73 \\
 SLT ($e-$)                     &  3.5  &-0.63   & 4.49\\
 $Z_{\rm vrtx}$ (cm)            &-18.08 &        &     \\
\end{tabular}
}
\label{tab:tab_6.5}
\end{center}
\end{table}
}
\widetext

\clearpage
\newpage

\begin{thebibliography}{99}
\label{bibliography}
\bibitem{SM} S.~L.~Glashow, Nucl.~Phys.~{\bf 22}, 579 (1961);
 S.~Weinberg, Phys.~Rev.~Lett.~{\bf 19}, 1264 (1967);
 A.~Salam, in {\it Elementary Particle Theory: Relativistic
 Groups and Analyticity (Nobel Symposium No.~8)}, edited by
 N.~Svartholm (Almqvist and Wiksell, Sweden, 1968), p.~367;
 M.~Gell-Mann, in {\it QCD, 20 years later}, edited by 
 P.~M.~Zervas and H.~A.~Kastrup (World Scientific, 1993). 

\bibitem{cdf-wjet}
 F.~Abe {\it et al.,} Phys.~Rev.~Lett.~{\bf 70}, 4042 (1993);
 Phys.~Rev.~Lett.~{\bf 73}, 2296 (1994);
 Phys.~Rev.~Lett.~{\bf 76}, 3070 (1996);
 Phys.~Rev.~Lett.~ {\bf 81}, 1367 (1998).

\bibitem{tot_xsec} F.~Abe {\it et al.,} Phys.~Rev.~{\bf D50}, 5550 (1994);
                   D.~Cronin-Hennessy {\it et al.,} Nucl.~ Inst.~and
                   Methods.~{\bf A443}, 37 (2000).\\
  The luminosity is derived  using the total $p\bar{p}$ cross section
  value 80.03 $\pm$ 2.24 mb.

\bibitem{cdf-evidence}
 F.~Abe {\it et al.,} Phys. Rev.~{\bf D50}, 2966 (1994);
 Phys.~Rev.~Lett.~{\bf 73}, 225 (1994).

\bibitem{cdf-discovery}
 F.~Abe {\it et al.,} Phys.~Rev.~Lett.~{\bf 74}, 2626 (1995).

\bibitem{xsec} F.~Abe {\it et al.,} Phys.~Rev.~Lett.~{\bf 80}, 2773 (1998).

\bibitem{cdf-tsig} T.~Affolder {\it et al.,} Phys.~Rev.~{\bf D64}, 032002 (2001).

\bibitem{tsig-pred} F.~Bonciani {\it et al.,} Nucl.~Phys.~{\bf B529}, 450 (1998);
            E.~Berger and H.~Contopanagos, Phys.~Rev.~{\bf D54}, 3035 (1996);
            S.~Catani {\it et al.,} Phys.~Lett.~{\bf B378}, 329 (1996);
            E.~Laenen {\it et al.,} Phys.~Lett.~{\bf B321}, 254 (1994).

\bibitem{d0} S.~Abachi {\it et al.,} Phys.~Rev.~Lett.~{\bf 79}, 12003 (1997).

\bibitem{higgs} J.~Gunion {\it et al.,} {\it The Higgs Hunter's Guide} (Addison-Wesley, 1990).

\bibitem{techni} E.~Eichten and K.~Lane, Phys.~~Lett.~{\bf B388}, 803 (1996).

\bibitem{hf-sem} Review of Particle Physics, D.~E.~Groom {\it et al.,} 
                 Eur.~Phys.~J. {\bf C15}, 1 (2000).

\bibitem{cdf-det} D.~Amidei {\it et al.,} Nucl.~ Inst.~and Methods, {\bf A271},
                  387 (1988); Fermilab Report 94/024-E (1994). 

\bibitem{cdf-svx} D.~Amidei {\it et al.,} Nucl.~Inst.~and Methods~Phys.~Res.,
                  Sect. {\bf A350}, 73 (1994); 
                  P.~Azzi {\it et al.,} Nucl.~Inst.~and Methods~{\bf A360},
                  137 (1994).

\bibitem{jet_clus} F.~Abe {\it et al.,} Phys.~Rev.~D {\bf 45}, 1448 (1992).

\bibitem{clus_err} F.~Abe {\it et al.,} Phys.~Rev.~Lett.~{\bf 70}, 1376 (1993).

\bibitem{top_mass} F.~Abe {\it et al.,} Phys.~Rev.~Lett.~{\bf 80}, 2767 (1998).

\bibitem{blusk} T.~Affolder {\it et al.,} Phys.~Rev.~{\bf D63}, 032003 (2001).

\bibitem{pythia}
            T.~Sj\"{o}strand, Computer Physics Commun. {\bf 39}, 347 (1986) ;
            T.~Sj\"{o}strand and M. Bengtsson, Computer Physics Commun. 
            {\bf 43}, 367 (1987);
            Computer Physics Commun. {\bf 46}, 43 (1987).

\bibitem{herwig} G.~Marchesini and B.~R.~Webber, Nucl.~Phys.~{\bf B310}, 461
 (1988); G.~Marchesini {\it et al.,} Comput.~Phys.~Comm.~{\bf 67}, 465 (1992).

\bibitem{vecbos} F.~A.~Berends, W.~T.~Giele, H.~Kuijf and B.~Tausk,
                 Nucl.~Phys.~{\bf B357}, 32 (1991).
\bibitem{heprt} J.~Benlloch, Proceedings of the 1992 DPF Meeting, 10-14 Nov.,
                1992, Batavia, IL, ed. C.~H.~Albright {\it et al.,} 
                World Scientific, (1993) p.~1091.
\bibitem{mrsd0} A.~D.~Martin, R.~G.~Roberts and W.~J.~Stirling, 
                Phys.~Lett.~{\bf B306}, 145 (1993);  {\bf B309}, 492(E) (1993).
\bibitem{lep}   A.~Ballestrero {\it et al.,} hep-ph/0006259 (2000);
                M.~Mangano,  hep-ph/9911256 (2000).
\bibitem{gbb}   M.~Seymour, Nucl.~Phys.~{\bf B436}, 163 (1995);
                M.~Mangano and P.~Nason, Phys.~Lett.~{\bf B285}, 160 (1992). 
\bibitem{cleo} P.~Avery, K.~Read, G.~Trahern, Cornell Internal Note
               CSN-212, March 25, 1985 (unpublished). We use Version 
               9\_1 of the CLEO simulation and our own lifetime database.

\bibitem{dibos} J.~Ohnemus and J.~F.~Owens, Phys.~Rev.~{\bf D44}, 1403 (1991);
                Phys.~Rev.~{\bf D44}, 3477 (1991); 
                Phys.~Rev.~{\bf D43}, 3626 (1991);
                B.~Mele {\it et al.,} Nucl.~Phys. {\bf B357}, 409 (1991);
                S.~Frixione {\it et al.,} Nucl.~Phys.~{\bf B383}, 3 (1992);
                S.~Frixione, Nucl.~Phys.~{\bf B410}, 280 (1993)).

\bibitem{single_top} T.~Stelzer, Z.~Sullivan and S.~Willenbrock, 
                     Phys.~Rev.~{\bf D56}, 5919 (1997);
                     M.~Smith and S.~Willenbrock, 
                     Phys.~Rev.~{\bf D54}, 6696 (1996).

\bibitem{mlm}  M.~Mangano, Nucl.~Phys.~{\bf B405}, 536 (1993).

\bibitem{prob} The probability is estimated with Monte Carlo pseudo-experiments
               which include both Poisson fluctuations and Gaussian 
               uncertainties in the prediction. Using these experiments
               we derive the probability of observing a likelihood
 ${\cal L} =\displaystyle {\prod_{i=1}^{4}} [\mu_i^{n_i} \exp^{-\mu_i}/n_i !] $,
               where $n_i$ and $\mu_i$ are the observed and predicted numbers
               of tags in the $i$-th jet bin, no larger than that of the data.

\bibitem{kuiper} N.~H.~Kuiper, {\it Proceeedings of the Koninklijke Nederlandse
                 Akademie van Wetenschappen}, ser.~A, 28 (1962). 

\bibitem{sadoulet} W.~T.~Eadie,~D.~Dryard,~F.~E.~James,~M.~Roos,~and~B~Sadoulet,
                   {\it Statistical Methods in Experimental Physics} 
                   (American Elsevier, 1971).

\bibitem{numrep}	W.~H.~Press, S.~A.~Teutolsky, W.~T.~Vetterling, 
                  B.~P.~Flannery, {\it Numerical Recipes}
                  (Cambridge University Press, 1992).

\bibitem{cft} The central fast tracker  (CFT) is a hardware processor which 
              uses fast timing information from the CTC as input and provides
              a list of $r-\phi$ tracks to the second level trigger (L2). The
              L2 muon trigger required CFT tracks with $p_T \geq 9.2\; \gevc$ 
              in the first part of the collider run and $p_T \geq 12\; \gevc$ 
              in the remaing 80\% of the data taking period.

\bibitem{mrsg} A.~D.~Martin, R.~G.~Roberts and W.~J.~Stirling, 
               Phys.~Lett.~{\bf B354}, 155 (1995). 

\end{thebibliography}
\end{document}